\documentclass[iop,letterpaper]{emulateapj}
\usepackage{times}
\usepackage{courier}
\usepackage{chngpage}
\usepackage{color}
\usepackage{float}
\usepackage{rotating}
\usepackage[hidelinks]{hyperref}
\usepackage[fleqn]{amsmath}
\usepackage{relsize}
\usepackage{amssymb}
\usepackage{multirow}

\slugcomment{Accepted 2015 April 15 to The Astronomical Journal}

\shorttitle{Asteroid Lightcurves in PTF}
\shortauthors{Waszczak et al.}

\begin{document}

\title{Asteroid lightcurves from the Palomar Transient Factory survey: \\
Rotation periods and phase functions from sparse photometry}

\author{Adam Waszczak$^1$, Chan-Kao Chang$^2$, Eran O. Ofek$^3$, Russ Laher$^4$, Frank Masci$^5$, David Levitan$^6$, Jason Surace$^4$, \\
Yu-Chi Cheng$^2$, Wing-Huen Ip$^2$, Daisuke Kinoshita$^2$, George Helou$^5$, Thomas A. Prince$^6$, Shrinivas Kulkarni$^6$\\}

\affil{\\$^1$Division of Geological and Planetary Sciences, California Institute of Technology, Pasadena, CA 91125, USA: \href{mailto:waszczak@caltech.edu}{\color{blue}{waszczak@caltech.edu}}\\
         $^2$Institute of Astronomy, National Central University, Jhongli, Taiwan\\         
         $^3$Benoziyo Center for Astrophysics, Weizmann Institute of Science, Rehovot, Israel\\
         $^4$Spitzer Science Center, California Institute of Technology, Pasadena, CA 91125, USA\\
         $^5$Infrared Processing and Analysis Center, California Institute of Technology, Pasadena, CA 91125, USA\\
         $^6$Division of Physics, Mathematics and Astronomy, California Institute of Technology, Pasadena, CA 91125, USA
         }

\begin{abstract}

We fit 54,296 sparsely-sampled asteroid lightcurves in the Palomar Transient Factory (PTF) survey to a combined rotation plus phase-function model. Each lightcurve consists of 20 or more observations acquired in a single opposition. Using 805 asteroids in our sample that have reference periods in the literature, we find the reliability of our fitted periods is a complicated function of the period, amplitude, apparent magnitude and other lightcurve attributes. Using the 805-asteroid ground-truth sample, we train an automated classifier to estimate (along with manual inspection) the validity of the remaining $\sim$53,000 fitted periods. By this method we find 9,033 of our lightcurves (of $\sim$8,300 unique asteroids) have `reliable' periods. Subsequent consideration of asteroids with multiple lightcurve fits indicate a 4\% contamination in these `reliable' periods. For 3,902 lightcurves with sufficient phase-angle coverage and either a reliably-fit period or low amplitude, we examine the distribution of several phase-function parameters, none of which are bimodal though all correlate with the bond albedo and with visible-band colors. Comparing the theoretical maximal spin rate of a fluid body with our amplitude versus spin-rate distribution suggests that, if held together only by self-gravity, most asteroids are in general less dense than $\sim$2 g/cm$^3$, while C types have a lower limit of between 1 and 2 g/cm$^3$. These results are in agreement with previous density estimates. For 5--20 km diameters, S types rotate faster and have lower amplitudes than C types. If both populations share the same angular momentum, this may indicate the two types' differing ability to deform under rotational stress. Lastly, we compare our absolute magnitudes (and apparent-magnitude residuals) to those of the Minor Planet Center's nominal ($G=0.15$, rotation-neglecting) model; our phase-function plus Fourier-series fitting reduces asteroid photometric RMS scatter by a factor $\sim$3.

\end{abstract}


\section{Introduction}

In this work we model an asteroid's apparent visual magnitude $V$ (log flux) as

\begin{equation}
V=H+\delta+5\log_{10}(r\Delta)-2.5\log_{10}[\phi(\alpha)],
\end{equation}

\noindent where $H$ is the absolute magnitude (a constant), $\delta$ is a periodic variability term due to rotation ({\it e.g.}, if the object is spinning and has some asymmetry in shape or albedo), $r$ and $\Delta$ are the heliocentric and geocentric distances (in AU), and $\phi =\phi(\alpha)$ is the \emph{phase function}, which varies with the solar phase angle $\alpha$ (the Sun-asteroid-Earth angle). When $\alpha=0$ ({\it i.e.}, at opposition), $\phi=1$ by definition, while in general $0<\phi<1$ for $\alpha>0$ (with $\phi$ decreasing as $\alpha$ increases).

A key feature of our approach is the simultaneous fitting of both the phase function $\phi$ and the rotation term $\delta$. The detailed forms of $\phi$ and $\delta$, as well as the algorithm underlying our fitting procedure, are motivated by a variety of prior work in this area, as described in the following sections. 

\subsection{Asteroid rotation}

Building upon the work of \cite{kaa01}, \cite{han12} discuss the inversion of asteroid lightcurve data taken over several oppositions to obtain a 3D shape solution. The form of $\delta$ (cf. Equation [1]) in this case consists of a large number of free parameters (several tens to hundreds). Results from inversion agree well with those from stellar occultations, adaptive optics imaging, and in-situ spacecraft imagery \citep{han13}. Knowledge of the detailed irregular shapes of asteroids improves our ability to constrain models of their internal structure, as well the magnitude and timescale of spin and orbital evolution due to solar-radiation and thermal emission, including the Yarkovsky and YORP effects (see \citealp{bot06} and references therein).

A simpler model for $\delta$---suitable for fitting to data sparser than that required for most inversion methods---is a Jacobi ellipsoid \citep{cha69} in its principal-axis spin state. The lightcurve of such an ellipsoid is a double-peaked sinusoid, given by a simple expression depending solely (assuming constant surface albedo) on the axes ratio, and angle between the line of sight and spin axis. The fitted amplitude thus yields a lower-bound elongation estimate for the asteroid.

The predicted distribution of the rotation frequencies of a collisionally-equilibrated system of particles has long been claimed to be a Maxwellian function \citep{sal87}, which---as reviewed by \cite{pra02}---very well approximates the observed distribution of several hundred of the brightest ($\sim$40-km or larger) asteroids, but breaks down for smaller objects, among which an excess of slow and fast rotators appear to exist. \cite{ste15} more recently argue that collision instead leads to a L\'{e}vy distribution, and that a significant primordial spin component remains in the present observed population. Some studies that have examined the spin distribution of small objects are \cite{pra08}, \cite{pol09}, the Thousand Asteroid Lightcurve Survey \citep{mas09}, and two brief observing runs conducted within the PTF survey (\citealp{pol12}; \citealp{cha14a}).

\cite{war09} describe the Lightcurve Database (LCDB), which compiles several thousand densely-sampled lightcurves of asteroids targeted by dedicated observing teams. Lightcurves in the LCDB have the following features:

\begin{enumerate}
\item LCDB lightcurves' dense sampling generally permits fitting of Fourier series with many harmonic terms,
\item LCDB lightcurves are often sampled over the shortest time window necessary to measure the period, and therefore generally do not require large or uncertain corrections due to phase angle effects,
\item LCDB lightcurves' fitted periods are assigned integer quality codes by a human reviewer (from 1 = poor to 3 = confident).
\end{enumerate}

All three of the above features are either impractical or infeasible when the set of lightcurves is very large and the data sparsely sampled, as is the case for PTF. In this work we adopt the following modified approaches when fitting lightcurves:

\begin{enumerate}
\item We truncate the rotation curve's Fourier-series fit after the 2nd harmonic, a simplification broadly justified by \cite{har14} and the assumption of an ellipsoidal shape (cf. Section 3.1.2),
\item We simultaneously fit a phase-function model with the rotational part,
\item We use a machine-learned classifier to objectively aid in estimating the validity of each fitted period. The classifier is trained using all fitted lightcurves that have previously (and confidently) measured LCDB periods and takes into account the accuracy with which the true period was retrieved along with 20 lightcurve metrics (fitted period, amplitude, ratio of peaks, $\chi^2$ per degree of freedom of fit, number of data points, and more). 
\end{enumerate}
	
Use of a machine classifier in asteroid lightcurve period quality assessment is entirely novel and inspired in part by work done by PTF collaborators in extragalactic transient science \citep{blo12} and variable star science (\citealp{masc14}; \citealp{mil14}), as well as Waszczak et al. (in prep)'s work on detection techniques for streaking NEOs. Among the advantages of using a machine-classified quality score is that, via cross-validation with the known-period sample, one estimates the completeness and contamination, {\it i.e.}, the true-positive and false-positive rates with respect to identifying an accurately-fit period, as a function of, {\it e.g.}, the period, amplitude, etc. The resulting true- and false-positive rates may then be used to de-bias the classifier-filtered period distribution.

\begin{figure}
\centering
\includegraphics[scale=0.8]{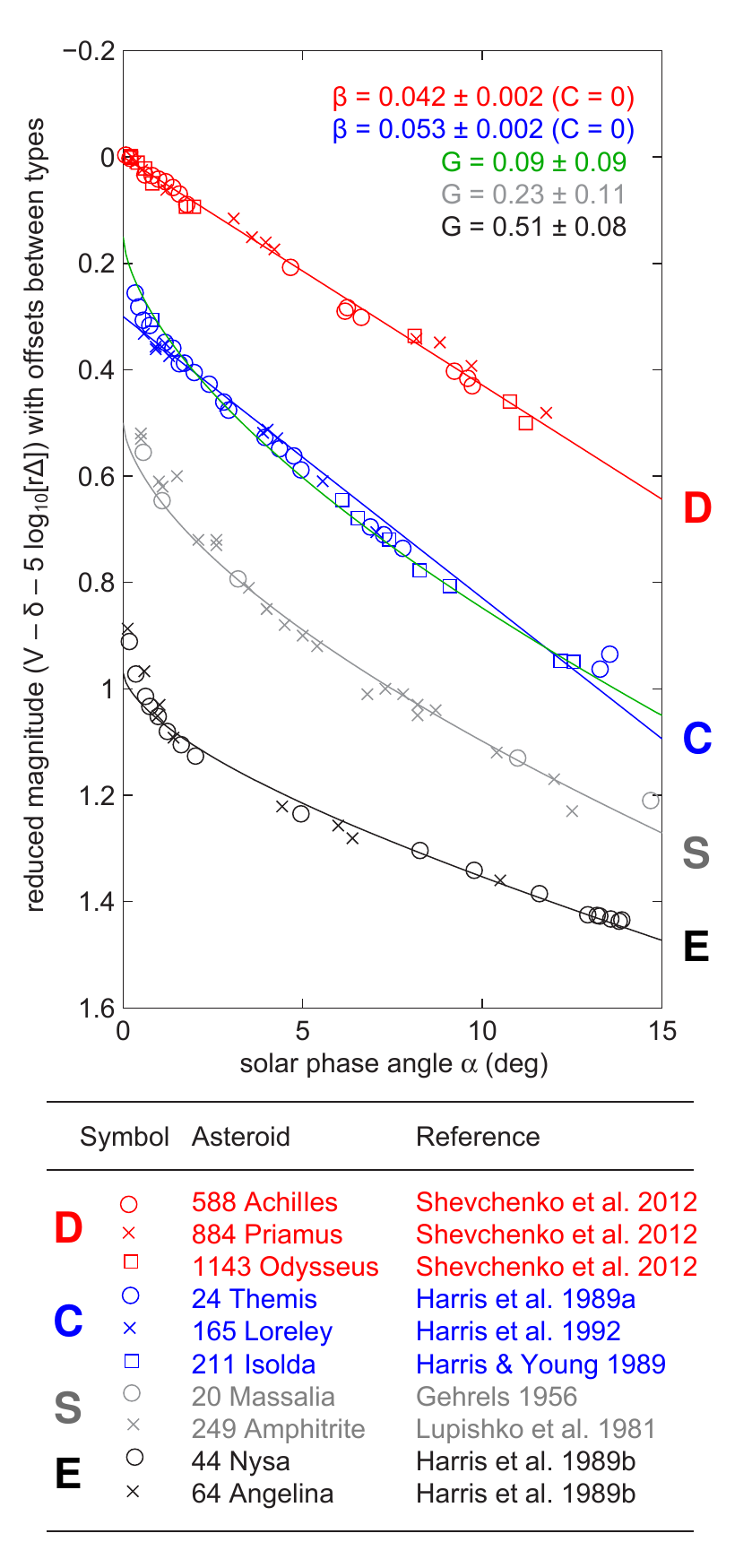}
\caption{Phase curves (from the literature) containing densely-sampled, rotation-corrected photometry of asteroids in four taxonomic classes. Colored lines are our original fits to the data using various single-parameter $\phi$ models (cf. Section 3.2).}
\vspace{10pt}
\end{figure}

\subsection{Asteroid phase functions}

The analytic phase function of an ideal Lambertian-scattering sphere fits well to featureless, atmospheric planets like Venus, but quite poorly to airless bodies (see Figure 3.9 of \citealp{sea10} for a comparison). In later sections we describe several $\phi$ models that have been derived for (or empirically fit to) asteroids. Qualitatively, asteroids show an approximately linearly decreasing $\phi$ out to $\alpha\approx 100$ deg, modified by a surge (increase in slope) at low phase angles ($\alpha\lesssim5$ deg), known as the \emph{opposition effect} (see Figure 1).

Early work ({\it e.g.} \citealp{bow89} and refs. therein) on a small sample of well-observed asteroids, suggested that different asteroid spectral types display distinct behavior in $\phi$. Figure 1 compares example phase curve data for D, C, S and E types\footnote{\cite{bus02} review these and other asteroid taxonomic classes, which are defined on the basis of low-resolution ($R\approx100$) visible reflectance spectra.}, incorporating photometry from various sources. We emphasize the fact that all of the data points in Figure 1 have been corrected for rotational modulation (the $\delta$ in Equation [1]) through dense sampling of each asteroid's lightcurve at each phase angle (equivalently, each epoch).

Using a large corpus of low-precision photometry from the MPC\footnote{IAU Minor Planet Center, \href{http://minorplanetcenter.net}{\color{blue}{http://minorplanetcenter.net}}}, Oszkiewicz et al. (\citeyear{osz11}, \citeyear{osz12}) showed that a fitted parameter of one particular $\phi$ model correlates well with an asteroid's SDSS visible color. While they were unable to correct for rotational variation ($\delta$-term in Equation [1]), the Oszkiewicz et al. work nevertheless demonstrates a solid trend between $\phi$ and a compositional attribute (color). 

\begin{figure*}
\centering
\includegraphics[scale=0.65]{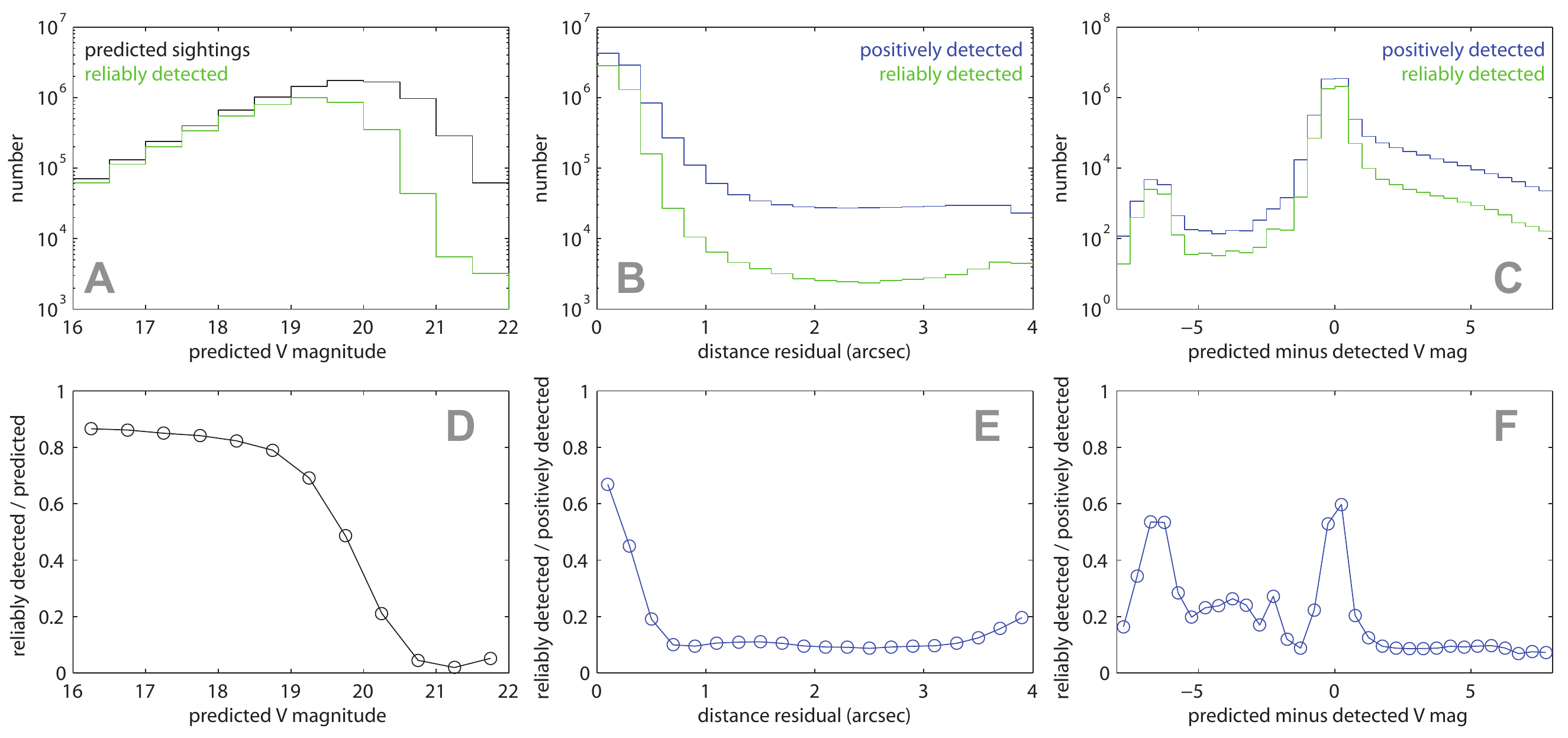}
\caption{Comparison of predicted asteroid sightings against positive and `reliable' asteroid detections. We define a `reliable' detection as any positive detection which (1) lacks any \emph{catalogued} background sources within a 4$''$ radius, (2) has a calibrated magnitude uncertainty of less than 0.1 mag, (3) lacks any processing warning flags. As suggested by the middle and right column of plots, this definition of 'reliable' still contains some small contamination (at the $<$1\% level) from uncatalogued background sources and/or noise, as indicated by detections with distance residuals greater than $\sim$1 arcsecond or magnitude residuals of greater than $\sim$1 mag. In panel D, the less than 100\% completeness at the bright end reflects the non-negligible probability that any asteroid will fall within 4$''$ of a catalogued background source (regardless of the magnitude of either the asteroid or the background source).}
\end{figure*}

These prior works motivate several defining aspects of this work's phase-function analysis:

\begin{enumerate}
\item We fit multiple phase function models to each lightcurve, both for compatibility with the literature and to explore how the fitted parameters are related,
\item We simultaneously fit the rotational component with the phase-function part,
\item We introduce a single colorimetric index for quantifying C-type vs. S-type taxonomic classification, based on the compilation of several visible-band-color asteroid datasets (see Appendix), and examine the variation in phase-function parameters as a function of this color index.
\end{enumerate}

\section{Observations}

\subsection{Overview of the PTF survey}

The \emph{Palomar Transient Factory\footnote{\href{http://ptf.caltech.edu}{\color{blue}{http://ptf.caltech.edu}}}} (PTF) is a synoptic survey designed primarily to discover extragalactic transients (\citealp{law09}; \citealp{rau09}). The PTF camera, mounted on Palomar Observatory's 1.2-m Oschin Schmidt Telescope, uses 11 CCDs (each 2K $\times$ 4K) to image 7.3 deg$^2$ of sky at a time at $1.0''$/pixel resolution. Most exposures ($\sim$85\%) use a Mould-$R$ filter\footnote{The Mould-$R$ filter is very similar to the SDSS-$r$ filter; see \cite{ofe12a} for its transmission curve.} (hereafter ``$R$''). The remaining broadband images acquired use a Gunn $g$-band filter. Nearly all broadband PTF images are 60-second integrations, regardless of filter. About 15\% of nights (near full moon) are devoted to a narrowband (H$\alpha$) imaging survey of the full Northern Sky.

Science operations began in March 2009, with a nominal one- to five-day cadence for supernova discovery and typical twice-per-night imaging of fields. Median seeing is $2''$ with a limiting magnitude $R\approx20.5$ (for 5$\sigma$ point-source detections), while dark conditions routinely yield $R\approx21.0$ \citep{law10}.

The PTF survey is ongoing and expected to continue through mid-2016. In January 2013 the PTF project formally entered a second phase called the \emph{intermediate PTF} (`iPTF'; \citealp{kul13}). In this paper we simply use `PTF' to mean the entire survey, from 2009 through the present (2015). The iPTF program accommodates more varied `sub-surveys' as opposed to a predominantly extragalactic program, including variable star and solar system science. Images are still acquired with the same telescope/camera/filters with 60s exposures, and are processed by the same reduction pipeline.

\cite{lah14} describe the PTF data reduction and archiving pipelines, hosted at the Infrared Processing and Analysis Center (IPAC) at Caltech. Processing at IPAC includes bias and flat-field corrections, astrometric calibration against UCAC3 \citep{zac10}, astrometric verification against 2MASS \citep{skr06}, creation of source catalogs with Source Extractor \citep{ber96}, and production of reference images (stacks of $\sim$20--30 PTF images that reach $V\approx 22$).

Ofek et al. (\citeyear{ofe12a}, \citeyear{ofe12b}) describe the PTF survey's absolute photometric calibration method, which relies on source matching with SDSS DR7 \citep{aba09}, and thus requires at least partial overlap of PTF with SDSS each night. A separate, \emph{relative} photometric calibration (based on lightcurves of non-variable field stars) also exists for PTF data and is described by \cite{lev11} and in the appendix of \cite{ofe11}. In this work we utilize all $R$-band and $g$-band PTF data accumulated from the survey's start (March 2009) through July 2014. The asteroid magnitudes reported in this work use relative photometric zeropoints when available (which as of this writing applies to $\sim$85\% of PTF images) and absolute photometric zeropoints otherwise.

The PTF's robotic survey program and processing pipeline, as well as our data aggregation and analysis in this work, make use of many functions from the MATLAB package for astronomy and astrophysics \citep{ofe14}.

\begin{table*}
\caption{Description of the PTF asteroid database. Includes PTF data acquired from March 2009 through July 2014, excluding H$\alpha$ survey data.}
\begin{tabular}{lrl}
\hline
table & \# rows & example columns (not necessarily comprehensive)\\
\hline
PTF tiles              & 11,169         & R.A., Dec., tile ID \\
exposures         & 304,982         & epoch, filter, exposure time, absolute photometric zeropoint, tile ID, exposure ID\\
CCD images & 3,305,426 & CCD ID, corners RA \& Dec, seeing, limiting mag., relative phot. zeropoint, \# of sources, exposure ID, image ID\\
asteroids & 401,810 & name, orbital elements, color data ({\it e.g.}, SDSS), IR data ({\it e.g.}, WISE), known rotation period, asteroid ID (number)\\
predicted sightings &  17,929,274   & R.A., Dec., rates, helio- \& geocentric range, phase \& elong. angle, pred. $V$ mag., image ID, asteroid ID, prediction ID\\
positive detections &  8,842,305   & R.A., Dec., instrumental mag., local zeropoint, shape data, quality flags, prediction ID, lightcurve ID, detection ID\\
reliable detections$^1$ &  4,392,395  &  detection ID\\
lightcurves$^2$    &  587,466 & \# of constituent detections, filter, opposition year, median mag., asteroid ID, lightcurve ID\\
lightcurve fits$^3$    &   54,296  & fitted lightcurve parameters, human-assigned quality code, machine-classified quality index, lightcurve ID, fit ID\\
reliable-period fits$^4$    &   9,033  & fit ID\\
reliable-$G_{12}$ fits$^5$    &   3,902  & fit ID\\
\hline
\end{tabular}
\smallskip
\\$^1$`Reliable' detections are those free from possible background-source or bright star contamination, magnitude errors $>$ 0.1 mag, and certain SExtractor flags.
\\$^2$A lightcurve is here defined as a set of positive detections of a given asteroid in a single filter and opposition.
\\$^3$Lightcurve fits only exist for lightcurves which contain at least 20 reliable detections and converged to a solution during the lightcurve-fitting process.
\\$^4$Fits have reliable rotation periods if a human screener labels the period reliable \emph{and} the machine classifier rates it above a certain quality threshold (see text).
\\$^5$Fits have reliable $G_{12}$ phase-function parameter if (1) amplitude $<$ 0.1 mag \emph{or} period is reliable, (2) fit has sufficient phase angle coverage (see section 6.3).
\vspace{15pt}
\end{table*}

\subsection{This work's data set}

\cite{was13} used a custom spatial indexing algorithm to search the set of all PTF single-epoch transient detections (through July 2012) for detections of all asteroids with orbits known as of August 2012. That search procedure first generated uniformly-spaced ephemerides for each asteroid using JPL's online service (HORIZONS; \citealp{gio96}). Each asteroid's ephemeris defines a 3D-curve (two sky coordinates plus one time); the intersection of each curve with the 3D kd-tree of transient detections was then computed and positive detections within a 4$''$ matching radius saved.

In this work we use a modified version of the \cite{was13} algorithm. The updates/changes are as follows.

Firstly, in terms of content, we now search all PTF ($R$ and $g$-band) data from 01-March-2009 through 18-July-2014 for all \emph{numbered} asteroids as of 12-July-2014 (401,810 objects). We now exclude unnumbered objects as the positional uncertainty of these objects can be very large, and as they tend to be very faint their lightcurves will not in general be of high quality.

Secondly, in place of a single-step matching of a 3D transient-detection kd-tree against 3D ephemeris curves, we now divide the search into two main steps. We first perform a 2D spatial matching that exploits the natural indexing of PTF exposures into tiles ({\it i.e.}, the grid of evenly spaced boresights or `fields' on the sky). Each 2D ephemeris curve's intersection with the 2D PTF survey footprint is computed, the object's position cubically-interpolated to all epochs of exposures possibly containing the object, and the object's precisely-computed positon is then compared to the precise image boundaries of candidate exposures. Matching of predicted positions against actual detections takes place subsequently as source catalogs are then loaded into memory (as needed and in parallel). This method is faster than the original \cite{was13} method and enables separate logging of predicted and positive detections.

The results of the known-asteroid search, as well as the derived lightcurve data (described later) are stored in a relational database, the size and contents of which are summarized in Table 1. Out of $\sim$18 million predicted single-epoch asteroid sightings (including predicted magnitudes as dim as $V\approx 23$, well below PTF's sensitivity), there were 8.8 million positive detections (within a 4$''$ radius). Of these, we define 4.3 million detections as `reliable' as they (1) lack any \emph{catalogued} background sources within the 4$''$ radius, (2) have a calibrated magnitude uncertainty of less than 0.1 mag, (3) lack any processing flags indicative of contamination. Figure 2 compares predicted, positive and `reliable' detections; the middle and right panels of Figure 2 show that our definition of `reliable' seems to include a small fraction of likely bad observations ($<$1\% contamination, note the vertical log scale), namely those which have distance residuals greater than $\sim$$1''$ or magnitude residuals greater than $\sim$1 mag. Because these reliable detections are the subset of observations which we input into our lightcurve fitting model (Section 4), the fitting algorithm includes logic designed to remove isolated data points that have very large residuals, either with respect to the median lightcurve value or relative to their uncertainty.

\section{Lightcurve model}

Equation (1) presents the overall form and notation of our asteroid lightcurve model. In this section we describe the detailed parameterization and assumptions of the model.

\subsection{Rotation component}

\subsubsection{Intra-opposition constraint}

The most important parameter in the rotation component (the $\delta$ in Equation [1]) is the synodic spin period $P$, a constant which satisfies

\begin{equation}
\delta(\tau)=\delta(\tau+nP),
\end{equation}

\noindent where $\tau\equiv t-\Delta/c$ is the light-time-corrected observation timestamp, $\Delta=\Delta(t)$ is the asteroid's geocentric distance, $c$ is the speed of light, and $n$ is any integer satisfying

\begin{equation}
|n|\ll P_\text{orb}/P,
\end{equation}

\noindent where $P_\text{orb}$ is the synodic \emph{orbital} period, 

\begin{equation}
P_\text{orb}=\left(\frac{1}{\text{yr}}-\frac{1}{T_\text{orb}}\right)^{-1}=\left(\frac{1}{\text{yr}}-\frac{\sqrt{G M_\odot}}{2\pi a_\text{orb}^{3/2}}\right)^{-1},
\end{equation}

\noindent where $T_\text{orb}$ is the asteroid's \emph{sidereal} orbital period and $a_\text{orb}$ is its orbital semi-major axis (related by Kepler's third law). $P_\text{orb}$ is the time elapsed between the asteroid's consecutive oppositions. Pursuant to this restriction, we constrain each $\delta$ solution using observations from within the same opposition---{\it i.e.}, for most asteroids, within a 1.1- to 1.6-year interval centered on the date of locally minimally observed $\alpha$.

The intra-opposition restriction is important given that our data set (described in the next section) spans $\sim$5 years. For an asteroid with a zero inclination circular orbit and spin axis perpendicular to its orbital plane, we can relax Equation (3) to allow $n$ to be any integer, in which case $\delta$ can be constrained using observations spanning many years. In general however, Equation (2) must be modified to accommodate a varying viewing geometry with respect to the spin axis:

\begin{equation}
\delta(\tau)=F(\tau)\delta(\tau+nP),
\end{equation}

\noindent where $F$ is some unknown periodic function satisfying $F(t)=F(t+mT_\text{orb})$, where $m$ is any integer and $T_\text{orb}$ is the sidereal orbital period. Provided the amplitude of $F$ is not large relative to that of $\delta$, and provided the spin vector is not changing with respect to the orbital plane ({\it i.e.}, precessing\footnote{Principal-axis rotation (a stable equilibrium state) is assumed for most planetary bodies. \cite{bur73} discuss the relevant timescales of spin evolution.}) on a timescale comparable to $P_\text{orb}$, we are justified in assuming Equation (2) (with the Equation [3] restriction) applies.

\subsubsection{Second-order Fourier series}

Any $\delta$ satisfying Equation (2) can be approximated to arbitrary precision using a Fourier series. \cite{har14} discuss why, from a geometric standpoint,  the second harmonic tends to dominate an asteroid's fitted $\delta$. As noted earlier (section 1.1), most large asteroids approximately resemble triaxial prolate ellipsoids ({\it e.g.}, Jacobi ellipsoids), having equatorial axis ratios of at most $\sim$3:1 (corresponding to a $\delta_\text{max}-\delta_\text{min}$ amplitude of $\sim$1.2 mag). For less extreme axis ratios (specifically, those producing a $\sim$0.4 mag or smaller second-harmonic amplitude), other harmonics related to shape or albedo asymmetries may contribute comparable coefficients to the Fourier approximation of $\delta$.

The PTF survey program has---on a few rare occasions---conducted high-cadence ($\sim$10-minute spaced) observations of low ecliptic latitude fields. These runs produced a set of $\sim$1,000 densely-sampled main-belt asteroid rotation curves, which have already been analyzed and published (\citealp{pol12}; \citealp{cha14a}). These high-cadence ``pilot studies'' are relevant to our present work in that they demonstrate (1) the quality of the PTF survey's photometric calibration for asteroids with unambiguously valid $\delta$ solutions, and (2) the above-described prevalence of a dominant second-harmonic in most of the objects sampled.

Following these pilot studies, we adopt a second-order Fourier series model:

\begin{equation}
\delta\equiv \sum_{k=1,2} A_{1,k} \sin\left(\frac{2\pi k \tau}{P}\right) + A_{2,k} \cos\left(\frac{2\pi k \tau}{P}\right),
\end{equation}

\noindent where $\tau$ is the light-time corrected epoch (cf. Equation [2]). In the pilot studies, most of the fitted $\delta$ solutions qualitatively resemble a simple sine or cosine function. Such a solution can be represented by \emph{either} a:

\begin{enumerate}
\item first harmonic with period $P=P_1$ (with $A_{i,1}\ne0$ and $A_{i,2}=0$), \emph{or}
\item second harmonic of period $P=\frac{1}{2}P_1$ (with $A_{i,1}=0$ and $A_{i,2}\ne0$).
\end{enumerate}

\noindent Given the prolate ellipsoid model, choice (2) is more realistic and hence preferred. However, again recognizing that other harmonics can have a non-negligible contribution, in fitting $\delta$ to our lightcurve sample we allow the first-harmonic coefficients $A_{i,1}$ to be non-zero, but introduce logic into the fitting algorithm (cf. Section 4) which checks for double-period solutions satisfying certain criteria and iterates accordingly.

\subsection{Phase-function component}

In this work we simultaneously fit each lightcurve's phase function $\phi$ along with its rotation curve $\delta$ (cf. Equation [1]). This approach is intermediate in complexity between some of the simpler, two-parameter ($\delta$-neglecting) models that have been applied to very large data sets ({\it e.g.}, \citealp{wil12}; \citealp{osz12}), and the more complex, shape plus pole-orientation models (\citealp{kaa04}; \citealp{cel09}; \citealp{han12}) which can involve tens of parameters and require data spanning multiple oppositions.

Regarding the former class of models, we note that there is a formal statistical problem associated with neglecting $\delta$ when fitting $\phi$. If modeling the observations $M$ by $V'\equiv V-\delta=H+5\log_{10}(r\Delta)-2.5 \log_{10}(\phi)$, then the distribution of residuals $M-V'$ is \emph{not} Gaussian.  Assuming $\delta$ is a sinusoid with amplitude $A$, for observations $M$ sampling the lightcurve at random times, the residual probability density function $p=p(M-V')$ has a local minimum value $p_\text{min}$ at $M-V'=0$ and maximum value $p_\text{max}$ near $M-V'=\pm A$. Thus $p$ is bimodal and roughly bowl-shaped---not at all Gaussian-shaped. The uncertainty in $\phi$ produced by a standard $\chi^2$ minimization---which assumes Gaussian-distributed errors---is thus inaccurate. However, since $p$ is symmetric about $M-V'=0$, for densely-sampled data the fitted phase function $\phi$ remains unaffected by neglecting $\delta$; in such a case the only effect is an underestimated uncertainty.

We obtain three separate fits for each lightcurve, each using a different phase-function ($\phi$) and allowing for unique solutions for $H$ and $\delta$ in Equation (1). The three phase-function models are:

\begin{enumerate}
\item the two-parameter model of \cite{she97},
\item the one-parameter $G$ model \citep{bow89},
\item the one-parameter $G_{12}$ model \citep{mui10}.
\end{enumerate}

In this section we review and motivate the application of each of these $\phi$ models.

\subsubsection{Two-parameter Shevchenko model}

\cite{she97} introduced a phase function dependent on two parameters; in terms of Equation (1) the model is\footnote{In Shevchenko's original notation, $\beta$ is denoted $b$ and $C$ is denoted $a$. Moreover, in the original notation, $\phi(0)=-a$; we here added a constant term $+a$ to make $\phi(0)=1$, following convention with other phase functions.}

\begin{equation}
-2.5\log_{10}[\phi(\alpha)]\equiv\beta\alpha-C\frac{\alpha}{1+\alpha},
\end{equation}

\noindent where $\beta$ has units of mag/deg and $C$ is the amplitude of the opposition surge (units of mag). This model was subsequently considered in-depth by \cite{bel00}, hereafter B\&S, who compiled the most complete (to date) set of high-precision, targeted phase curve observations of main-belt asteroids from various data sets spanning several decades. 

Though in practice Shevchenko's model is the least commonly used phase function out of the three we consider, it is by far the simplest to express mathematically, and is the only model for $\phi$ whose parameters have linear dependence in Equation (1).

Furthermore, this model's parameters are the most straightforward to associate with physical asteroid properties. B\&S highlighted a robust relationship between an asteroid's ($\beta,C$) phase-function parameters and its geometric albedo\footnote{Also known as the \emph{visible} albedo or the \emph{physical} albedo.}. As we later explore a similar relationship in the present work, we here review the basis of this observation. 

The geometric albedo $p_V$ is formally \emph{defined} in terms of the phase function $\phi$:

\begin{equation}
p_V\equiv\frac{A_\text{bond}}{2}\left(\int_0^\pi \phi(\alpha)\sin(\alpha)\;d\alpha\right)^{-1}\equiv\frac{A_\text{bond}}{q},
\end{equation}

\noindent where $A_\text{bond}$ is the (visible) bond albedo, defined as the total visible light energy reflected or scattered by the asteroid (in all directions) divided by the total visible light energy incident upon the asteroid (from the Sun). We also here define the phase integral $q$.

B\&S showed that, in the range of $\beta$ observed from S-type to C-type asteroids, $\beta$ and $C$ are empirically correlated, in a relation that we approximate here as

\begin{equation}
C\approx(\text{0.9 mag})-(\text{17 deg})\beta\;\;\text{for}\;\;0.03<\frac{\beta}{\text{mag}/\text{deg}}<0.05.
\end{equation}

\noindent Using Equation (9) to substitute for $C$ in Equation (7), inserting the result into Equation (8) and numerically evaluating the integral gives

\begin{equation}
p_V\approx A_\text{bond}\left(0.4-\frac{2.2\beta}{\text{mag}/\text{deg}}\right)\;\;\text{for}\;\;0.03<\frac{\beta}{\text{mag}/\text{deg}}<0.05.
\end{equation}

B\&S saw a negative correlation between $p_V$ and $\beta$ in the data\footnote{B\&S actually stated the correlation in terms of $\log p_V$ vs. $\beta$, though the range in $\beta$ is sufficiently small that $p_V$ vs. $\beta$ is essentially valid as well.}, consistent with Equation (10) \emph{only if} either $A_\text{bond}$ is assumed constant among different asteroid types (not a reasonable assumption) \emph{or} if $A_\text{bond}$ negatively correlates with $\beta$, which B\&S did not explicitly show.

The bond albedo $A_\text{bond}$ can be thought of as an intrinsic, bulk-compositional characteristic of an asteroid's surface\footnote{More accurately, the single-scattering albedo $w$, which is the analog of $A_\text{bond}$ for a ``point-source'' particle, more fundamentally embodies this bulk-compositional attribute. \cite{hap12} details how $A_\text{bond}$ is solely a function of $w$ for an asteroid whose surface consists of isotropic scatterers; we here use $A_\text{bond}$ as a proxy for $w$.}, much like an asteroid's color, whereas $\beta$ and $C$ relate (in part) to the textural, particulate, and macroscopic roughness of the asteroid's surface. B\&S and other authors separately associate $\beta$ with the \emph{shadow-hiding} effect and $C$ with the \emph{coherent backscatter} effect. Both of these physical phenomena are understood from a theoretical standpoint ({\it e.g.}, \citealp{hel89}; \citealp{hap12}) to be functions of $A_\text{bond}$, with $\beta$ negatively related to $A_\text{bond}$ and $C$ positively related. This is consistent with Equation (9), and renders Equation (10) consistent with B\&S's noted $p_V$-vs.-$\beta$ correlation. Other properties such as particle size, particle geometry and regolith porosity also have predicted (and laboratory-measured) contributions to the observed phase function (\citealp{hap12} and refs. therein); these properties can conceivably vary independently of $A_\text{bond}$.

In short, our interpretation of the S-type and C-type asteroid data reviewed by B\&S is that a compositional indicator ($A_\text{bond}$) correlates with indicators of two independent phenomena ($\beta$ and $C$) that contribute to how light scatters from an asteroid's surface. This statement intentionally makes no mention of $p_V$, since Equation (8) tells us $p_V$ by definition varies with $\beta$ (in a non-obvious way) and with $A_\text{bond}$, the latter being a more basic compositional attribute.

As stated above, the phase function can be related to properties other than $A_\text{bond}$, such as regolith porosity. Many of these other properties in theory and experiment contribute to effects involving \emph{multiply}-scattered light, and therefore do not alter the effect of shadow-hiding ($\beta$-term in Equation [7]), which is dominated by \emph{singly}-scattered light \citep{hap12}. In contrast, the coherent backscatter effect ($C$-term) \emph{does} involve multiply-scattered light. B\&S saw non-monotonic behavior in $C$ as a function of $p_V$ when including the rarer, high-$p_V$ E-type asteroids in the same plot as C and S types. E types do conform however to the same negative monotonic trend in $p_V$-vs.-$\beta$ satisfied by the C and S types, consistent with the hypothesis that $\beta$ is adequately expressed as a function of $A_\text{bond}$ alone, yet E types have a lower-than-predicted $C$ value based on extrapolation of Equation (9).

One possibility is that Equation (9) is not valid for all asteroids, but must be replaced by some unknown non-monotonic relationship, possibly because $C$ depends non-monotonically on $A_\text{bond}$ and/or has comparable dependence on other properties ({\it e.g.}, porosity or grain size). Assuming Equation (7) is a sufficiently general model for $\phi$, and lacking knowledge of a good model for $C$, it follows that $\beta$ and $C$ should in practice always be fit separately. Another possibility is that Equation (7) is an incorrect or incomplete model, however B\&S described no instances wherein their model was unable to adequately fit the data for a particular asteroid or class of asteroids.

\subsubsection{Lumme-Bowell $G$ model}

The next phase function model we consider is the Lumme-Bowell model \citep{bow89}, also known as the ($H$,$G$) or IAU phase function:

\begin{equation}
\left\{
\begin{split}
\phi&\equiv(1-G)\phi_1+G\phi_2\\
\phi_1&\equiv\exp(-3.33\tan^{0.63}[\alpha/2]) \\
\phi_2&\equiv\exp(-1.87\tan^{1.22}[\alpha/2])
\end{split}
\right.
\end{equation}

Like Shevchenko's model, this model includes two terms (the basis functions $\phi_1$ and $\phi_2$) representing two physically-distinct contributions to the observed $\phi$. As detailed in \cite{bow89}, this model is \emph{semi}-empirical in that it was derived from basic principles of radiative transfer theory with certain assumptions, and at various stages tailored to match existing laboratory and astronomical observations. That the two basis functions' coefficients are related to a single parameter $G$ bears resemblance to the $\beta$-vs.-$C$ correlation described by Equation (9).

\cite{mar86} marked the IAU's adoption of this phase function as a standard model for predicting an asteroid's brightness. Since then this model has seen widespread application, and is often used with the assumption $G=0.15$ ({\it e.g.}, in the ephemeris computation services offered by the MPC and JPL). \cite{hary88} present mean values of $G$ for several of the major asteroid taxonomic classes (based on a sample of $\sim$80 asteroids), with $G=0.15$ being an average between the C types ($G\approx 0.08$) and the S types ($G\approx 0.23$). The $G$-model fails to accurately fit the rarer D types (which have linear phase curves) and E types (which have very sharp opposition spikes), whereas the Shevchenko model can properly accommodate these rarer types.

Use of the Lumme-Bowell $\phi$ in our lightcurve model (Equation [1]) introduces a second non-linear parameter ($G$) into the model, the period $P$ being the other non-linear parameter. This complicates the fitting algorithm somewhat, as described in Section 4.

\subsubsection{Muinonen et al. $G_{12}$ model}

The third phase function model we consider, introduced by \cite{mui10}, bears resemblance to the $G$-model but includes a second free parameter and a third basis function:

\begin{equation}
\phi\equiv G_1\phi_1+G_2\phi_2+(1-G_1-G_2)\phi_3
\end{equation}

As opposed to the analytic trigonometric basis functions of the $G$-model, here $\phi_1$, $\phi_2$ and $\phi_3$ (all functions of $\alpha$ alone) are defined in terms of cubic splines (see \citealp{mui10} for the exact numerical definitions). Assuming the coefficients $G_1$ and $G_2$ are constrained independently, these basis functions were designed to provide the most accurate fits to the phase functions of all major asteroid taxonomic types, including the rarer D types and E types.

For situations where fitting $G_1$ and $G_2$ separately is infeasible, \cite{mui10} specialized their above model to make it a function of a single parameter, $G_{12}$, which parameterizes $G_1$ and $G_2$ using piecewise functions:

\begin{equation}
\begin{aligned}
G_1 &=\left\{
\begin{split}
&0.7527 G_{12} +0.06164\;\;\;\text{if}\;G_{12}<0.2;\\
&0.9529 G_{12} +0.02162\;\;\;\text{otherwise};
\end{split}
\right.\\
\\
G_2 &=\left\{
\begin{split}
&-0.9612 G_{12} +0.6270\;\;\;\text{if}\;G_{12}<0.2;\\
&-0.6125 G_{12} +0.5572\;\;\;\text{otherwise};
\end{split}
\right.\\
\end{aligned}
\end{equation}

In this work we use this single-parameter $G_{12}$ form of the Muinonen et al. model, making it analogous to the $G$-model in terms of implementation, including the complication associated with a non-linear parameter.

\subsubsection{Multi-parameter Hapke model}

Just as we commented on the more rigorous means of fitting a rotation curve via 3D shape modeling with multi-opposition data, for completeness we note that a more rigorous model (than the three presented above) exists for phase functions. Given better-sampled lightcurves and more computational power, future modeling of large photometric datasets would benefit from applying the more theoretically-motivated model of \cite{hap12}, an abbreviated form of which is

\begin{equation}
\phi = \frac{B_C K}{p_V}\left[\left(\frac{w}{8}(B_S g-1)+\frac{r_0-r_0^2}{2}\right)h+\frac{2}{3}r_0^2 \phi_L\right]
\end{equation}

Here $w$ is the single-scattering albedo (cf. Footnote 9), of which $r_0$ is solely a function. The remaining factors all are functions of phase angle ($\alpha$). Each opposition-surge term ($B_S$ and $B_C$) has two free parameters (width and amplitude). $K$ depends on the mean topographic roughness (a function of one free parameter); $g$ is the single-scattering angular distribution function (typically includes one parameter); $h$ is a function of $\alpha$ only; and $\phi_L$ is the phase function of an ideal Lambertian-scattering sphere (a simple function of $\alpha$).

With its $\phi\propto p_V^{-1}$ dependence, the Hapke model (Equation [14]) can conveniently eliminate both $p_V$ and $H$ from the modeling process. Inserting Equation (14) into Equation (1), and using the common relation\footnote{Rather that attributing it to any specific author(s), we note that Equation (15) may be derived directly using Equation (8) and the following definition of the bond albedo, which we stated in words immediately after Equation (8): \begin{equation*}
A_\text{bond}\equiv\frac{\int_0^\pi 10^{-V(\alpha)/2.5}\sin(\alpha)d\alpha}{(10^{-M_\text{Sun}/2.5}/4\pi \text{AU}^2)\times\pi(D/2)^2}
\end{equation*} where $V(\alpha)=H-2.5\log_{10}\phi(\alpha)$ is Equation (1) evaluated at $\delta=0$ and $r=\Delta=1$ AU.}

\begin{equation}
H=-5\log_{10}\left(\frac{D\sqrt{p_V}}{\text{1329 km}}\right),
\end{equation}

\noindent where $H$ is the absolute visual magnitude, $D$ is the asteroid's effective diameter and 1329 km is a constant (set by the arbitrarily-defined magnitude of the Sun), produces a model with many physically meaningful parameters and free of both $H$ and $p_V$.

\section{Lightcurve-fitting algorithm}

We solve Equation (1) using a custom \emph{linear least squares} (LLSq) method. A basic review of LLSq can be found in \cite{hog10}. Each fitted asteroid lightcurve contains $N_\text{obs}\ge20$ observations, with measured apparent magnitudes $m_i$ and measurement uncertainties $\sigma_i$. All instrumental magnitudes are elliptical aperture \citep{kro80} measurements (SExtractor's \texttt{MAG\_AUTO}) calibrated with a local zeropoint ({\it i.e.}, the `ZPVM' correction of \citealp{ofe12a}). The uncertainties contain a Poisson-noise component (SExtractor's \texttt{MAGERR\_AUTO}) as well as systematic error from the calibration. For images lacking a relative photometric solution, the relevant systematic error is the \texttt{APBSRMS} parameter in the PTF database; for images having a relative photometric solution, the systematic error is a combination of the \texttt{sysErr} and \texttt{zeroPointErr} database quantities (added in quadrature).

In all cases, our model (Equation [1]) is non-linear in at least one parameter (the period $P$, or equivalently the frequency $f=1/P$). We test $N_\text{frq}$  evenly-spaced frequencies between $f=0$ (infinite rotation period) and $f=12$ day$^{-1}$, {\it i.e.}, up to the $\sim$2-hour spin barrier. 

Asteroids rotating \emph{faster} than the $\sim$2-hour spin barrier are likely monolithic objects and---particularly if larger than $\sim$150 m---are interesting in their own right (cf. the discussion in \citealp{pra02}). However, given the apparent observed rarity of such super-fast rotators (SFRs) and the large interval in frequency space that must be searched to discover them; we impose 2 hours = 12 cycles per day as our upper limit on fitted frequency in order to make computational time reasonable without sacrificing sensitivity to the majority of asteroids' spin rates. \cite{cha14a} presents preliminary results of an independent, ongoing effort to use PTF data (or at least specific subsets thereof) to search for SFRs, with at least one SFR having been discovered and confirmed \citep{cha14b}.

We use a frequency spacing $\Delta f=1/(4\Delta t)$, where $\Delta t$ is the time interval between the first and last observation in the lightcurve. Formally $\Delta t$ can be as long as 1.1 to 1.6 yr for most asteroids (cf. Section 3.1.1); however the median value of $\Delta t$ (among lightcurves that ultimately acquired fits) is $\sim$45 days, with 16$^\text{th}$ and 84$^\text{th}$ percentiles of 13 and 106 days, respectively.

In addition to the non-linear parameter $f$, the lightcurve model in general has $N_\text{lin}$ linear parameters. We seek to solve the following tensor equation for $X$:

\begin{equation}
m_i=\sum_{j,k}L_{ijk}X_{jk}\;\;\left\{\begin{array}{l}
i=1,2,...,N_\text{obs}\\
j=1,2,...,N_\text{frq}\\
k=1,2,...,N_\text{lin}
\end{array}
\right.
\end{equation}

\noindent where $m_i$ is the $i^\text{th}$ observation, $L$ is the `design matrix' (a 3D array of size $N_\text{obs}\times N_\text{frq}\times N_\text{lin}$) and $X$ is the linear-parameter matrix ($N_\text{frq}\times N_\text{lin}$) containing the linear-parameter solutions as a function of frequency.

\subsection{Linear phase-function parameters}

For the particular case wherein we use Shevchenko's model (Equation [7]) for the phase function $\phi$, the design matrix is

\begin{equation}
L_{ij}
=\left(
\begin{array}{c}
1\\
\sin(2\pi f_j \tau_i) \\
\cos(2\pi f_j \tau_i) \\
\sin(4\pi f_j \tau_i) \\
\cos(4\pi f_j \tau_i) \\
\alpha_i \\
\alpha_i/(1+\alpha_i)
\end{array}
\right)
\end{equation}

\noindent where the $k$-index has been omitted with the convention that $k=1$ is the $1^\text{st}$ row of the above column vector, $k=2$ is the second row, etc. Here $\tau_i$ and $\alpha_i$ are the time and phase angle of the $i^\text{th}$ observation, $f_j$ is the $j^\text{th}$ frequency, etc. Likewise, the linear-parameter matrix $X$ in this case is

\begin{equation}
X_{j}
=\left(
\begin{array}{c}
H_j\\
(A_{1,1})_j \\
(A_{2,1})_j \\
(A_{1,2})_j \\
(A_{2,2})_j \\
\beta_j \\
C_j
\end{array}
\right)
\end{equation}

\noindent where $H_j$ is the fitted absolute magnitude for the $j^\text{th}$ frequency, etc.

The general LLSq solution to Equation (16) is

\begin{equation}
X_{jk} = \sum_{\ell,n,p}S_{jk\ell}L_{n j \ell} (B^{-1})_{np}m_p,
\end{equation}

\noindent where $B^{-1}$ is the inverse of the data-covariance matrix $B$:

\begin{equation}
B=\left(
\begin{array}{cccc}
\sigma_1^2 & 0          & \cdots & 0 \\
0          & \sigma_2^2 & \cdots & 0 \\
\vdots     & \vdots     & \ddots & \vdots \\
0          & 0          & \cdots & \sigma_{N_\text{obs}}^2
\end{array}
\right),
\end{equation}

\noindent and $S_{jk\ell}$ is the parameter-covariance matrix, given by

\begin{equation}
S_{jk\ell}=[(s_j)^{-1}]_{k\ell},
\end{equation}

\noindent where in the above definition we invert each of the $N_\text{frq}$ matrices $s_j$, these being defined by

\begin{equation}
(s_j)_{k\ell} \equiv \sum_{n,p} L_{njk}(B^{-1})_{np}L_{pj\ell}.
\end{equation}

\noindent The elements of the parameter-covariance matrix $S$ are the variances and covariances of the fitted parameters (as a function of frequency). The fit's residuals (as a function of frequency) are:

\begin{equation}
R_{ij} = m_i-\sum_{k}L_{ijk}X_{jk},
\end{equation}

\noindent and the fit's chi-squared (as a function of frequency) is:

\begin{equation}
(\chi^2)_{j} = \sum_{\ell,n}R_{\ell j}(B^{-1})_{\ell n}R_{nj}.
\end{equation}

\begin{figure*}
\centering
\includegraphics[scale=0.515]{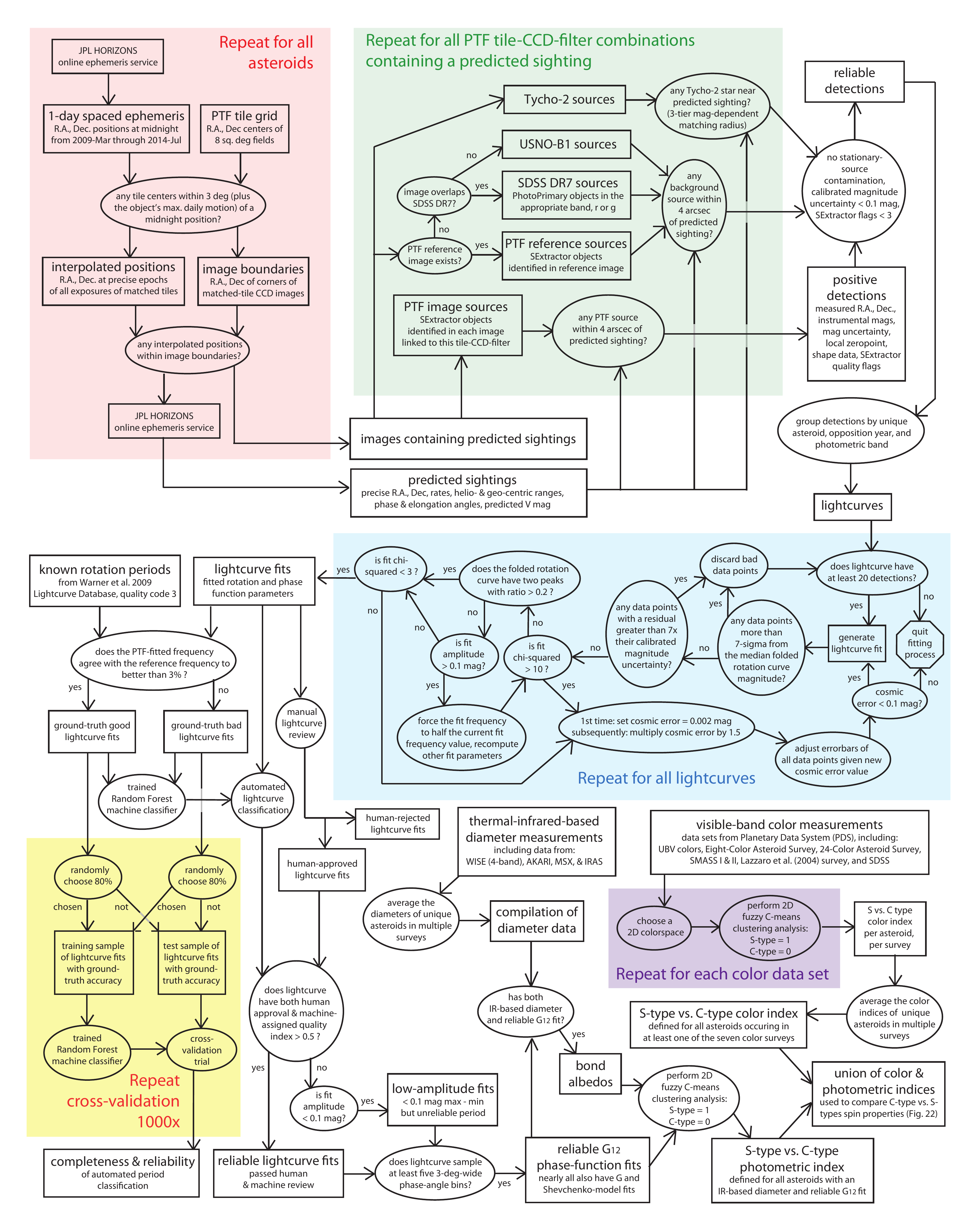}
\caption{Diagram detailing the logic of this work's data reduction and analysis. Includes mining the survey for known-asteroid observations, aggregation of the data into lightcurves, vetting of the lightcurves and an application wherein phase functions are compared to color-derived asteroid taxonomy. See text for details.}
\end{figure*}

The frequency-dependent chi-squared $(\chi^2)_j$ is also known as the \emph{periodogram}. Formally, the best-fit rotation frequency corresponds to the minimal value of $(\chi^2)_j$, but this may differ from the \emph{preferred} frequency solution if the lightcurve is contaminated by other systematic periodic signals, if the data suffer from underestimated measurement uncertainties, or if the best-fit frequency corresponds to a dominant first harmonic (as opposed to a preferred dominant second harmonic, cf. Section 3.1.2).

Figure 3 details our iterative lightcurve-fitting algorithm's logic. Fitting commences as long as 20 or more `reliable' data points (cf. Section 2.2 and Figure 2) are associated with a lightcurve. Irrevocably-bad data points are discarded in the first round of iterations, these include detections with 7$\sigma$ or greater residuals from the initial solution. Examples of detections with such high residuals include contamination from background sources missing in the reference catalog, bad detector pixels that were not flagged by the pipeline, or spurious zeropoint solutions.

In the next stage of iterations, the fit's $\chi^2$ per degree of freedom is reduced to $\sim$1 (formally, it is reduced until it is less than 3, cf. Figure 3) by gradually inflating the observations' errorbars through addition of a `cosmic error', so-named because it encompasses contamination from possible errors (in all the `cosmos'). In general the cosmic error represents the same diverse contaminating phenomena responsible for the $>$7$\sigma$ deviations seen in the initial iterations (cf. previous paragraph) just to a lesser extent.

Separately, this errorbar inflation compensates for our model's inability to fit each asteroid's precise periodic structure using only two harmonic terms in the Fourier series. In the limit of infinite observations and sufficiently many Fourier terms, we would ideally expect our data's errorbars to reflect true Gaussian variance. However, by truncating the series at two harmonics and using sufficiently precisely-calibrated photometry, we are in effect choosing to sacrifice (downsample) some of our photometric precision to obtain a formally better fit at the coarser resolution limit of the model. 

To illustrate use of the cosmic error, consider the example of an eclipsing binary lightcurve, {\it i.e.}, a rotation curve which is effectively sinusoidal \emph{except} for a small interval around the phase of minimum flux, when it dips to a lower-than-predicted brightness. Examples from our dataset appear in Figure 10. Observations acquired during such eclipses will have systematic negative deviations greater in absolute value than would be explained by Gaussian variance alone. Increasing the errorbars of these observations will decrease the fits' $\chi^2$ without altering the value of the fitted frequency. The fitted parameters' uncertainties (both for frequency and the linear parameters) are accordingly inflated as a penalty, and the fitted amplitude will be underestimated. As detailed in Figure 3, the initial cosmic error used is 0.002 mag, and each iteration it is multiplied by a factor 1.5 until the $\chi^2$ is sufficiently low. If the cosmic error exceeds 0.1 mag, the fitting is aborted. If the $\chi^2$ (per degree of freedom) drops below 3 while the cosmic error is still below 0.1 mag, the fitting process concludes `successfully' (see Figure 3).

Concurrently, each iteration includes a test for the presence of double peaks in the folded rotation curve (only if the fitted amplitude is at least 0.1 mag). In particular, if there exist two maxima and two minima in the folded lightcurve, we demand that the ratio of these peaks be greater than 0.2. Such a solution is preferred (cf. Section 3.1.2) given our ellipsoidal shape assumption, as described by \cite{har14}.

Denote as $f_\text{best\_global}$ the frequency yielding the absolute minimum $\chi^2$ per degree of freedom value, denoted $\chi^2_\text{min\_global}$ (after the cosmic error has been tuned). If the folded lightcurve is single-peaked (or has only a relatively small secondary peak), then another deep minimum usually exists at the harmonic frequency $f_\text{best\_harmonic}=0.5\times f_\text{best\_global}$, the local minimum $\chi^2$ value of which we denote $\chi^2_\text{min\_harmonic}$). For cases wherein $\chi^2_\text{min\_harmonic}<\chi^2_\text{min\_global}+\text{inv-}\chi^2\text{-cdf}(0.95,7)$, where $\text{inv-}\chi^2\text{-cdf}(p,N)$ is the inverse of the $\chi^2$ cumulative distribution function for $N$ free parameters evaluated at $p$, then we instead choose $f_\text{best\_harmonic}$ rather than $f_\text{best\_global}$. The 1$\sigma$ uncertainty interval for the best-fit frequency is then found by computing the upper and lower intersections between $\chi^2_\text{min}+\text{inv-}\chi^2\text{-cdf}(0.68,7)$ and the periodogram in the vicinity of $f_\text{best}$. Note that we used $N=7$ free parameters in this case, {\it i.e.}, the number of elements of $X_j$ (Equation 18).

\subsection{Nonlinear phase-function parameters}

Modeling the phase function $\phi$ with either the $G$ or $G_{12}$ model (Equations [11] and [12]), introduces a second non-linear parameter (after the frequency $f$) and so we must modify the equations of the previous section accordingly. We sample $N_\text{pha}=200$ evenly-spaced phase-function parameter values. In particular, for $G$ we test the interval $-0.3\le G\le 0.7$ in steps of $\Delta G=0.005$, and for $G_{12}$ we test the interval $0\le G_{12} \le 1$ in steps of $\Delta G_{12}=0.005$. 

Our approach is to modify the left-hand side of Equation (16) by defining a new matrix $m'_{iq}$ which contains all possible phase-function-corrected observed magnitudes:

\begin{equation}
m'_{iq}\equiv m_i-\Phi_{iq}=\sum_{j,k}L_{ijk}X_{jkq}\;\left\{\begin{array}{l}
i=1,2,...,N_\text{obs}\\
j=1,2,...,N_\text{frq}\\
k=1,2,...,N_\text{lin}\\
q=1,2,...,N_\text{pha}
\end{array}
\right.
\end{equation}

\noindent where, {\it e.g.}, for the case of the $G$-model (Equation [11]),

\begin{equation}
\begin{aligned}
\Phi_{iq}&\equiv-2.5\log_{10}[\phi(\alpha_i,G_q)] \\
&=-2.5\log_{10}[(1-G_q)\phi_1(\alpha_i)+G_q\phi_2(\alpha_i)]\\
\end{aligned}
\end{equation}

The linear-parameter-solution array $X$ now has an extra index $q$, reflecting the fact that we are now solving for each linear parameter as a function of the two non-linear parameters. The design matrix has the same number of indices as before (but fewer rows):

\begin{equation}
L_{ij}
=\left(
\begin{array}{c}
1\\
\sin(2\pi f_j \tau_i) \\
\cos(2\pi f_j \tau_i) \\
\sin(4\pi f_j \tau_i) \\
\cos(4\pi f_j \tau_i)
\end{array}
\right),
\end{equation}

\noindent while the linear-parameter matrix $X$ is now

\begin{equation}
X_{jq}
=\left(
\begin{array}{c}
H_{jq}\\
(A_{1,1})_{jq} \\
(A_{2,1})_{jq} \\
(A_{1,2})_{jq} \\
(A_{2,2})_{jq}
\end{array}
\right).
\end{equation}

The appeal in adopting the above approach is that the general solution is only slightly modified:

\begin{equation}
X_{jkq} = \sum_{\ell,n,p}S_{jk\ell}L_{n j \ell} (B^{-1})_{np}m'_{pq},
\end{equation}

\noindent where the only difference between equations (19) and (29) are the $q$ indices appended to $X$ and $m$ (and the latter being redefined as $m'$).

\begin{figure*}
\centering
\includegraphics[scale=0.54]{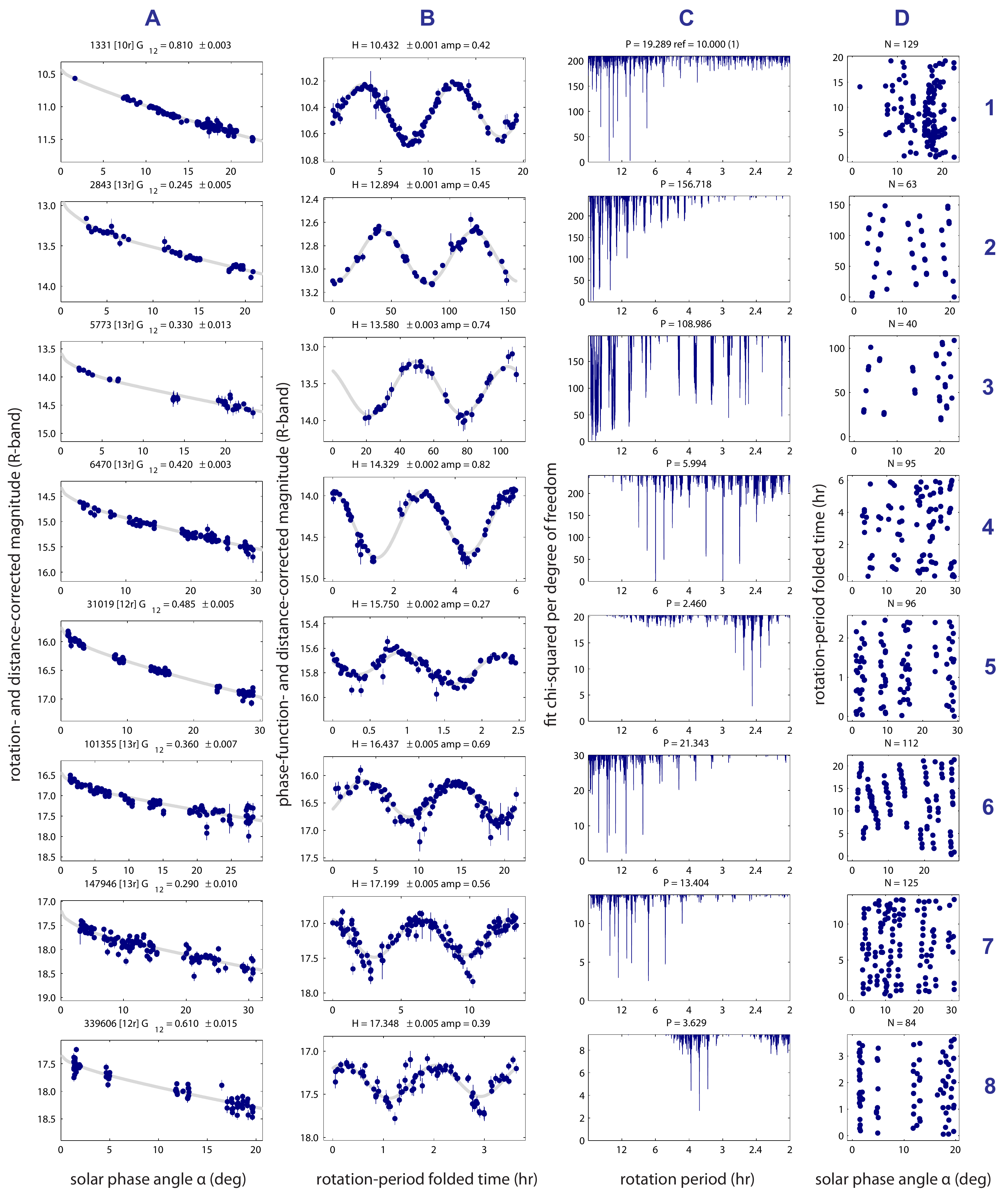}
\caption{Examples of lightcurves having both well-sampled rotation \emph{and} phase-function components. Each row corresponds to a different asteroid. These example asteroids are sorted vertically by their physical diameter (assuming 7\% albedo); the top object is $\sim$45 km and the bottom object is $\sim$2 km. Column A shows the phase curve (corrected for rotation); Column B shows the rotation curve (corrected for phase-function); Column C shows the periodogram; Column D shows the distribution of the observations in rotational phase vs. solar phase angle. Above each plot is additional information depending on the column: (A) the asteroid number, followed by (in square brackets) the opposition year (most are 2013) and filter (in all cases `r') followed by the fitted $G_{12}$ parameter; (B) the fitted absolute magnitude and amplitude; (C) the fitted period (in hours); (D) the number of data points included (and shown) in the fit.}
\end{figure*}

The fit's residuals $R$ are now a function of frequency \emph{and} phase-function parameter:

\begin{equation}
R_{ijq} = m'_{iq}-\sum_{k}L_{ijk}X_{jkq},
\end{equation}

\noindent as is the fit's chi-squared:

\begin{equation}
(\chi^2)_{jq} = \sum_{\ell,n}R_{\ell jq}(B^{-1})_{\ell n}R_{njq}.
\end{equation}

As a function of any of the \emph{linear} parameters, the fit's $\chi^2$ varies precisely quadratically, whereas as a function of frequency it has an intricate spectral structure with many local minima. As a function of a non-linear phase parameter ($G$ or $G_{12}$), the $\chi^2$ tends to have a single minimum (on the range we evaluate): in this sense $G$ and $G_{12}$ are more similar to the linear parameters than they are to frequency. However, the generally asymmetric shape of the phase parameter' $\chi^2$ dependence necessitates its grid-based numerical treatment---particularly to ensure accurate estimation of the phase parameter's uncertainty.

The two-dimensional $\chi^2$ surface given by Equation (31), which is defined on a $N_\text{freq}\times N_\text{pha}$ grid, can be reduced to a one-dimensional $\chi^2$ function by choosing, for each frequency index $j$, the phase-parameter index $q$ that minimizes the $\chi^2$. The result is a one-dimensional periodogram, as in Equation (24). Once the fitted frequency is identified, we compute the uncertainty in the fitted $f$ by the method described in the previous section using the inv-$\chi^2$-cdf() function. We then likewise numerically compute the uncertainty in the phase parameter by again collapsing $(\chi^2)_{jq}$ to a one-dimensional vector, this time as a function of the phase parameter with the frequency fixed at the fitted value ($j$-index), and use the inv-$\chi^2$-cdf() function to estimate the uncertainty in the phase parameter.

As noted in Table 1, a total of 587,466 lightcurves exist in PTF, where each lightcurve by definition consists of all reliable observations of a unique asteroid observed in a single opposition in a single photometric band. Of these, only $\sim$10\% (59,072 lightcurves) have at least 20 observations and therefore qualified for fitting with our algorithm. A total of 54,296 lightcurves actually produced a fit---the remaining $\sim$5,000 lightcurves failed to produce a fit either because some observations were discarded and the total fell below 20 data points, or because the fitted cosmic error grew to exceed 0.1 mag. 

Figure 4 shows several examples of lightcurves fitted with the algorithm described in this section. In the third column (column C) of Figure 4, we show the periodograms of each lightcurve. Note that although the periodogram's horizontal axes are labeled with the \emph{period} (for easier interpretation), the chi-squared (per degree of freedom) values are actually plotted linearly with respect to \emph{frequency}. This is because, as described earlier, our sampling is uniform with respect to frequency, and the harmonics are more easily seen with constant frequency spacing. Column (D) shows the data sampling in rotational phase versus solar phase angle, a useful plot to ensure there is no obvious correlation between the two (which could lead to an erroneous fit, {\it e.g.}, for long periods, large amplitudes and/or few data points).

\subsection{Comments on implementation}

Each iteration in the fitting of each asteroid lightcurve involves evaluating the arrays and tensor-products in either Equation (19) or (29). This includes inverting the data-covariance matrix $B$ (Equation [20]) and inverting the $N_\text{frq}$ matrices $s_j$ (Equation [22]). The arrays $L$, $m'$, $X$ and $R$ can have a relatively large number of elements, making them and their relevant products potentially taxing with respect to computational memory.

Our particular implementation of this algorithm leverages the efficient array-manipulation capabilities of MATLAB, especially its ability to perform fast matrix multiplication and matrix inversion utilizing BLAS calls\footnote{\href{http://www.netlib.org/blas}{\color{blue}{http://www.netlib.org/blas}}} and OpenMP multi-threaded C loop code\footnote{\href{http://openmp.org}{\color{blue}{http://openmp.org}}}. Given typical numbers of observations and frequency sampling, each of our lightcurve fits (including the multiple iterations) takes on average several tens of seconds to run on an eight-core machine (multi-threading enabled), and typically consumes less than $\sim$4 GB of memory using single-precision computation.

In the online supplementary material we provide our custom MATLAB function used for fitting the $G$-parameter version of the lightcurve model (\texttt{asteroid\_lc\_fit\_G.m}). Analogous versions exist for the Shevchenko and $G_{12}$ models. This function takes as input an asteroid's apparent magnitudes, magnitude uncertainties, observed epochs, phase angles, geocentric and heliocentric distances. Its outputs include the linear-parameter-solution array (Equation 28), residuals (Equation 30), chi-squared array (Equation 31), and additional information about each lightcurve solution such as the amplitude and peak ratios.

\section{Reliability of fitted rotation periods}

A primary concern in the quality assessment of our fitted lightcurve parameters is the validity of our derived rotation periods. In this section we describe several methods of estimating the reliability of these periods, beginning with comparison to a ground-truth subsample of known-period asteroids and followed by a full vetting of our entire sample using a combination of machine-learning and manual classification.

The fitted period may differ (slightly or significantly) between the fits using the different phase function models. In this section for simplicity we consider only the period value obtained when fitting with the $G_{12}$ phase-function model (Section 3.2.3). In subsequent sections we will again consider all three $\phi$ models.

\begin{figure}
\centering
\includegraphics[scale=0.75]{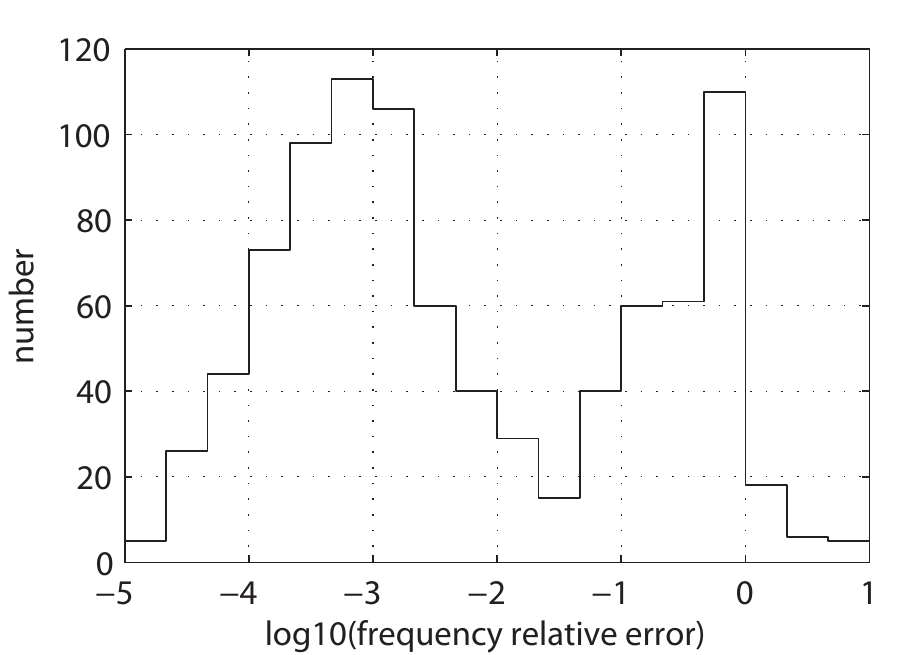}
\caption{For the 927 lightcurves (805 unique asteroids) having a quality code 3 period in the Lightcurve Database of \citealp{war09} \emph{and} an original fit in this work, we plot the distribution of the relative error in our fitted rotation frequencies with respect to the literature-referenced frequencies. The distribution is bimodal, with the left-hand mode corresponding to those fits having better than $\sim$3\% agreement.}
\vspace{10pt}
\end{figure}

\begin{figure*}
\centering
\includegraphics[scale=0.54]{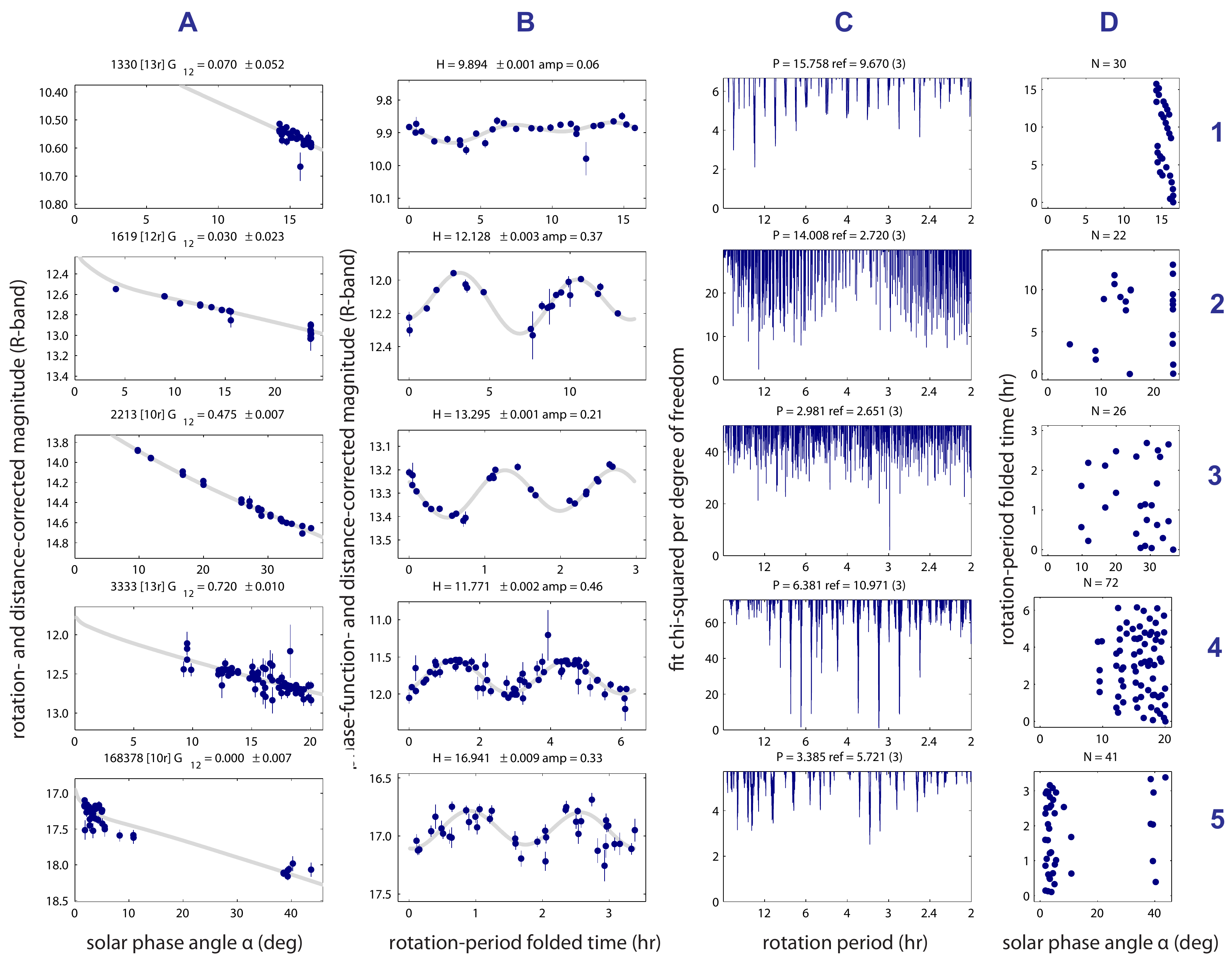}
\caption{Examples of lightcurves whose fitted frequency differs from the reference frequency by more than 3\%, so that they fall in the right mode in the histogram shown in Figure 5 and are formally defined as inaccurate fits. \emph{Row 1}: Low-amplitude rotator. \emph{Row 2}: Incorrect period (too few observations?). \emph{Row 3}: A fitted frequency that differs from the reference frequency by 12\%. \emph{Row 4}: period that differs by a non-integer multiple, despite looking reasonable. \emph{Row 5}: Folded lightcurve appears to be fitting noise in the data.}
\end{figure*}

\begin{figure*}[t!]
\centering
\includegraphics[scale=0.56]{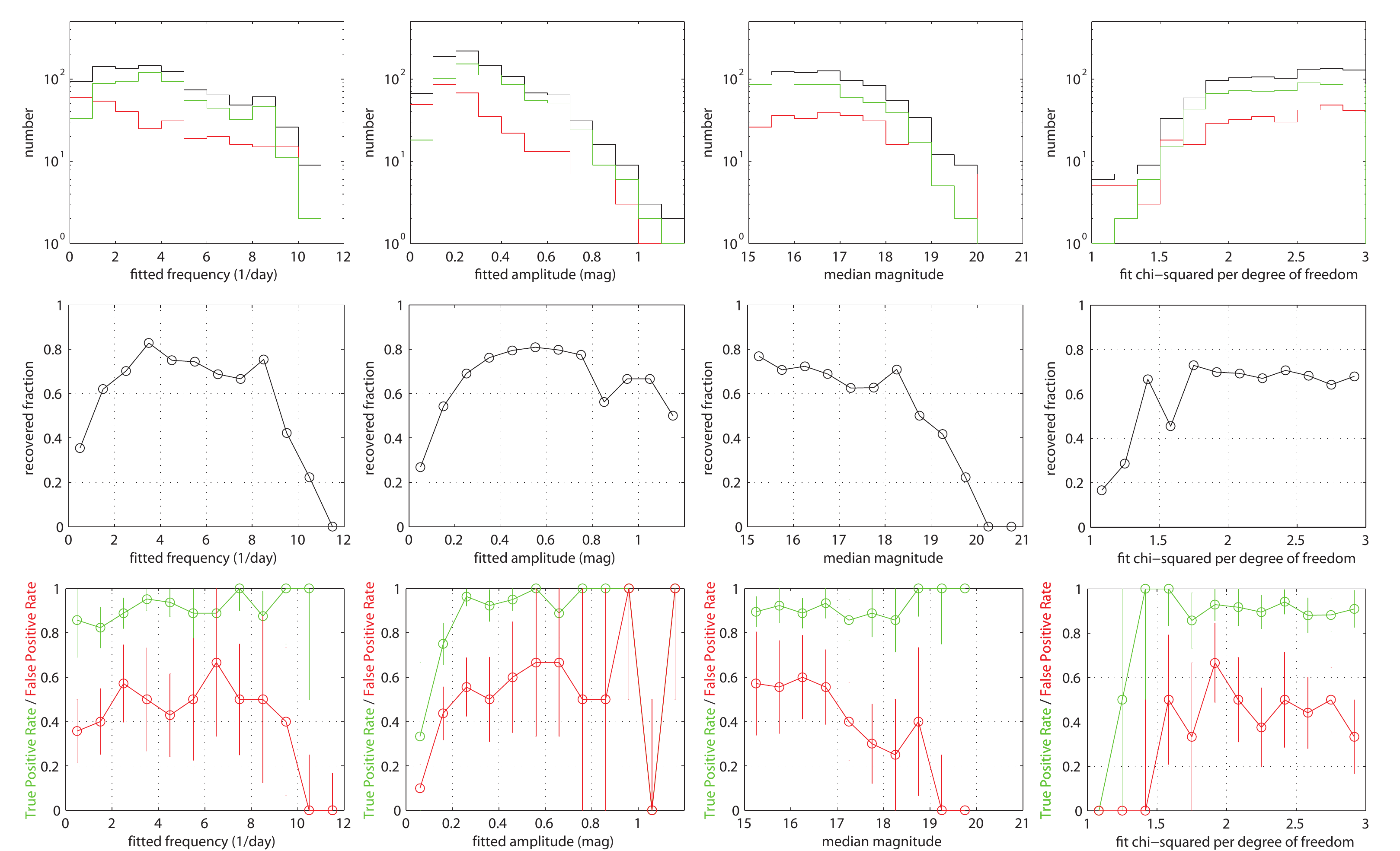}
\caption{\emph{Top row}: The 927-lightcurve known-period sample (black), divided into the accurately-fitted (green) and inaccurately-fitted (red) subgroups. \emph{Middle row}: Ratio of the green to black histograms. \emph{Bottom row}: Results of cross-validation of the machine-classifier (see Section 5.2.2).}
\vspace{10pt}
\end{figure*}

\begin{figure*}[!]
\centering
\includegraphics[scale=0.56]{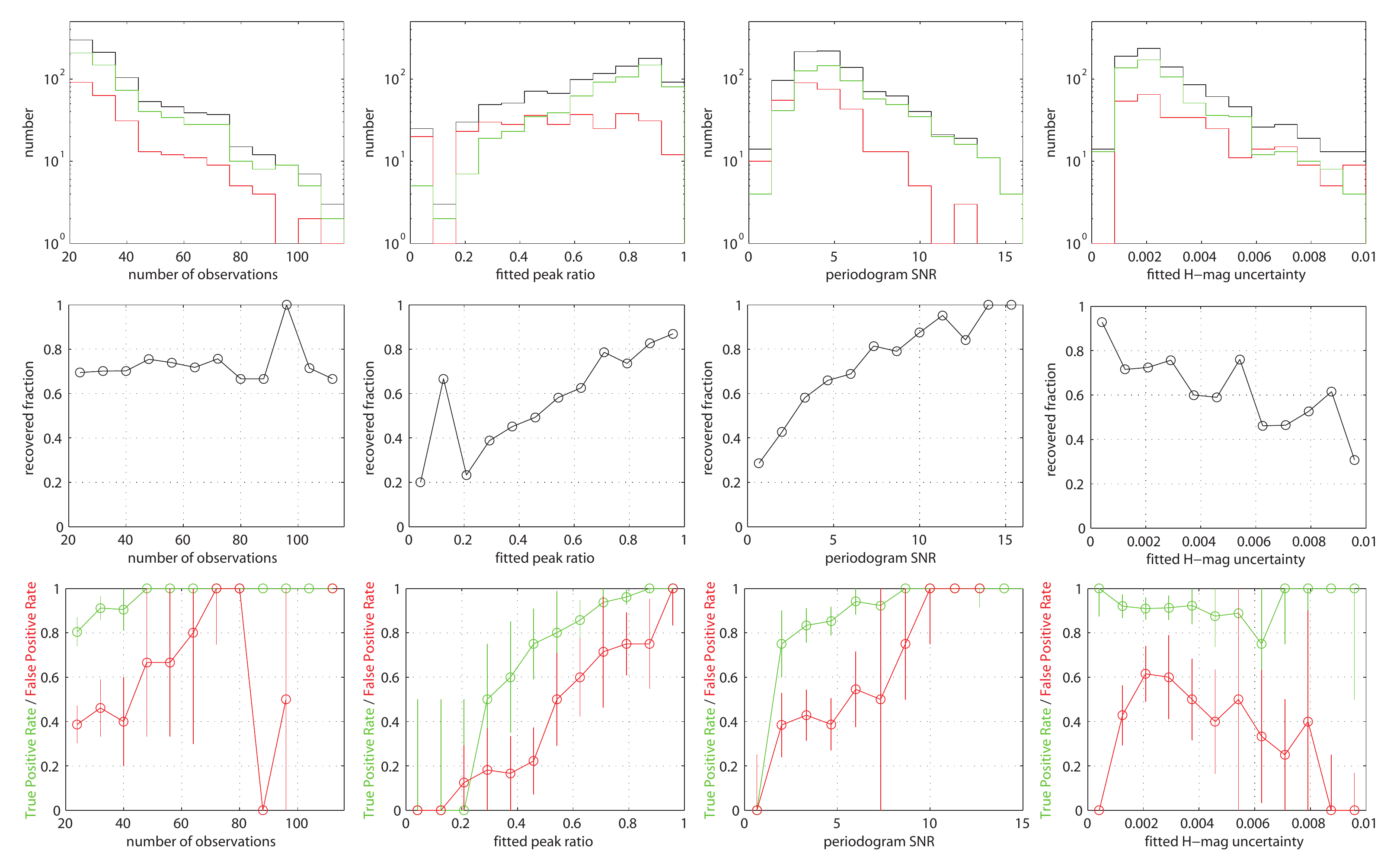}
\caption{\emph{Top row}: The 927-lightcurve known-period sample (black), divided into the accurately-fitted (green) and inaccurately-fitted (red) subgroups. \emph{Middle row}: Ratio of the green to black histograms. \emph{Bottom row}: Results of cross-validation of the machine-classifier (see Section 5.2.2).}
\vspace{10pt}
\end{figure*}

\subsection{Known-period subsample}

A total of 927 ($\sim$2\%) of our fitted lightcurves belong to 805 unique asteroids having a previously-measured period listed in the Lightcurve Database (LCBD) of \cite{war09}. This includes only asteroids having a quality code of 3 (highest quality) in the LCDB.

Figure 5 shows that the distribution of relative errors on our fitted frequencies is bimodal, with the left mode corresponding to periods having better than $\sim$3\% agreement with the reference period, and the right mode corresponding to periods in disagreement with the reference period. These disagreeing fits include lightcurves which differ from the reference value by a harmonic (half = relative error 0.5, double = relative error 1.0), as well as frequencies that do not differ by a factor of two or any integer multiple. About 1/3 of the lightcurves in Figure 5 fall into the right mode and are thus considered disagreeing fits.

Figure 6 shows some examples of these disagreeing fits. Row 1 shows an apparent low-amplitude rotator, whose fitted period of 15.7 hr differs from the reference value of 9.7 hr. Row 2 is an object whose periodogram contains a great deal of noise, divided into two broad forests of frequency minima. The left forest appears to have been selected by our fitting algorithm while the right forest seems associated with the true period of $\sim$2.7 hr. Row 3 contains an object whose 12\% relative frequency error exceeds the 3\%-accuracy threshold we have defined, and so despite appearing to be a good fit it is formally categorized as inaccurate. Row 4 also looks like a reasonable fit at 6.4 hr, but disagrees with the reference period of 11.0 hr (though the latter does have a perceptible local minimum in the periodogram). Finally, Row 5 includes a likely example of the algorithm fitting noise in the photometry of a faint asteroid.

In Figures 7 and 8 (top and middle rows) we detail the distribution of the accurately-recovered-period and inaccurately-recovered-period subgroups in terms of eight different lightcurve parameters. Some basic observations from these histograms are:

\begin{enumerate}
\item fitted periods are far less reliable if longer than $\sim$1 day or shorter than $\sim$2.7 hours,
\item fitted amplitudes of less than 0.1 mag correspond to the least reliably fit periods,
\item lightcurves consisting of observations dimmer than $\sim$18.5 mag are much less reliable than brighter lightcurves (though they are also far less numerous in the known-period sample),
\item fit $\chi^2$ (per degree of freedom) values of less than $\sim$1.7 correlate with less reliable periods (though they are also far less numerous in the known-period sample). Note that, in the fitting process, growth of the cosmic error term ceased once the $\chi^2$ (per degree of freedom) fell below 3 (cf. Figure 3).
\item the number of observations in a lightcurve is \emph{not} directly correlated to the reliability of the fitted period,
\item the ratio of the folded lightcurve's two peaks, the signal-to-noise ratio of the periodogram's chosen minimum, and the uncertainty in the absolute magnitude parameter are all strong indicators of the reliability of the fitted period.
\end{enumerate}

The above comments reflect consideration of the \emph{one-dimensional} distributions in Figure 7 and 8; however we can easily imagine there are correlations in more dimensions not evident from these plots alone. An obvious example would be the two-dimensional distribution in amplitude versus median magnitude: reliability is presumably greater for bright asteroids having amplitudes $<$0.1 mag than it is for dim asteroids having amplitudes $<$0.1 mag. Period versus amplitude is also likely an insightful distribution (and was considered for example by \citealp{mas09}). The number of observations possibly \emph{does} correlate with reliability if we were to restrict another parameter or parameters to some specific interval.

Rather than manually examining the period-fitting reliability as a function of all possible multi-dimensional combinations of the eight lightcurve parameters detailed in Figures 7 and 8, we can take a more general approach of considering the reliability to be a single function defined on the multi-dimensional parameter space in which all of the lightcurves reside. We hypothesize that accurately-fit lightcurves and inaccurately-fit lightcurves occupy distinct regions in this multi-dimensional volume. As these volumes can overlap to some extent, we can at least estimate the \emph{probability} that a lightcurve with that particular vector of parameters corresponds to an accurately-recovered (or inaccurately-recovered) period when obtained by the fitting algorithm of Section 4.

There are two general ways of accomplishing this goal. One way is to produce a large number of synthetic lightcurves filling out the multidimensional lightcurve-parameter space, subject these synthetic lightcurves to our fitting algorithm, and thereby map out {\it e.g.}, by binning and interpolation, the fit reliability throughout the multi-dimensional volume. This method requires us to accurately simulate all sorts of varying sampling cadence as well as measurement uncertainties, including contributions from both systematics and noise, and it requires significant extra computing time to actually subject the synthetic data to our fitting procedure. The second method---the approach we take in this work---uses a ground-truth sample (the known-period lightcurves already described in this section) to train a machine classifier to discriminate reliable versus unreliable fits within the multi-dimensional lightcurve-parameter space.

\subsection{Machine learning}

\begin{table*}
\caption{Summary of the 20 lightcurve parameters (features) used by our period-quality classifier. See text for a discussion of the cross-validation-derived importance value (Section 5.2.2).}
\begin{tabular}{lcl}
\hline
\multirow{2}{*}{feature} & importa-& \multirow{2}{*}{description} \\
        &  nce (\%) &  \\
\hline
\texttt{peakRatio}&	11.1& Ratio of the fitted lightcurve's two peaks ($=\text{max} - \text{min}$). Zero if only one peak, one if exactly the same height.\\
\texttt{amplitude}&	10.2& Fitted amplitude of the folded lightcurve. Equivalent to the height ($\text{max} - \text{min}$) of the larger of the two peaks. \\
\texttt{periodFit}&	8.6& Rotation period value obtained using this work's data and fitting algorithm. \\
\texttt{freqSNR}&	8.4& Signal-to-noise of the fitted (minimum) frequency in periodogram = $2\times|\text{min} - \text{median}|/(84^\text{th}\text{-percentile} - 16^\text{th}\text{-percentile})$\\
\texttt{hMagErr}&	5.8& Uncertainty in the fitted $H$-magnitude ({\it i.e.}, error in the fitted absolute magnitude) \\
\texttt{a12Coeff}	&4.3& Fourier coefficient $A_{12}$\\
\texttt{a22Coeff} 	&4.2& Fourier coefficient $A_{22}$\\
\texttt{numObsFit}&4.1& Number of observations in the final fitted lightcurve, after discarding any bad observations\\
\texttt{medMag}	&4.1& Median calibrated magnitude (in the photometric band specific to the lightcurve, either $R$ or $g$)\\
\texttt{chisq}	&4.1& Reduced chi-squared of the fit {\it i.e.}, $\chi^2$ per degree of freedom)\\
\texttt{a21Coeff}	&4.0& Fourier coefficient $A_{21}$ \\
\texttt{a11Coeff}	&3.8& Fourier coefficient $A_{11}$\\
\texttt{rmsFit}	&3.8& Root-mean-squared residual of the fit\\
\texttt{hMagRef}	&3.8& Reference $H$-magnitude ({\it i.e.}, absolute magnitude of the asteroid in $V$-band as listed by the MPC)\\
\texttt{kIndex}	&3.7& Stetson's $K$-index (a measure of kurtosis in the magnitude distribution of a folded lightcurve, introduced by \citealp{ste96}.)\\
\texttt{freqResol}	&3.7& Resolution of the periodogram: $\Delta f=1/(4\Delta t)$ where $\Delta t$ is the time between the first and last observations in the lightcurve\\
\texttt{hMagResid}&3.7& Difference between the reference absolute magnitude (\texttt{hMagRef}) and the fitted $H$-magnitude \\
\texttt{cuspIndex}&3.6& `Cusp index': Median squared residual of the dimmest 10\% points divided by the median squared residual of all other points \\
\texttt{numObsRem}&2.9& Number of observations removed during the  fitting process (due to $>$7-sigma residuals with respect to preliminary fits)\\
\texttt{cosmicErr}&2.1& Final `cosmic error' value at end of fitting process ($<$0.1 mag in all cases)\\
\hline
\end{tabular}
\end{table*}

\begin{figure*}
\centering
\includegraphics[scale=0.68]{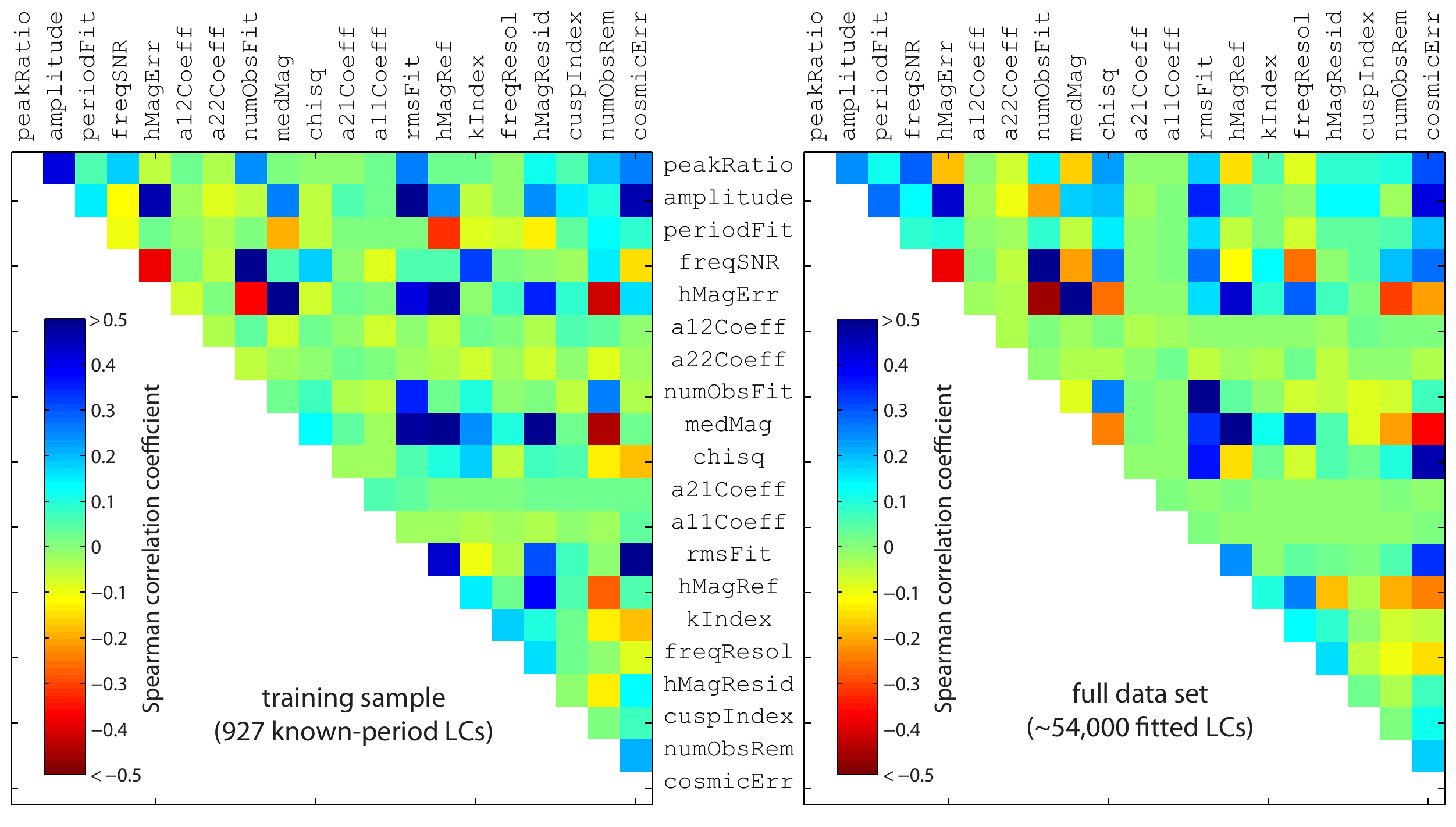}
\caption{Correlation matrices (Spearman's $\rho$ coefficient) for the 20 lightcurve features (Table 2) in the training sample (left) and in the full data set (right).}
\vspace{10pt}
\end{figure*}

\begin{figure*}
\centering
\includegraphics[scale=0.54]{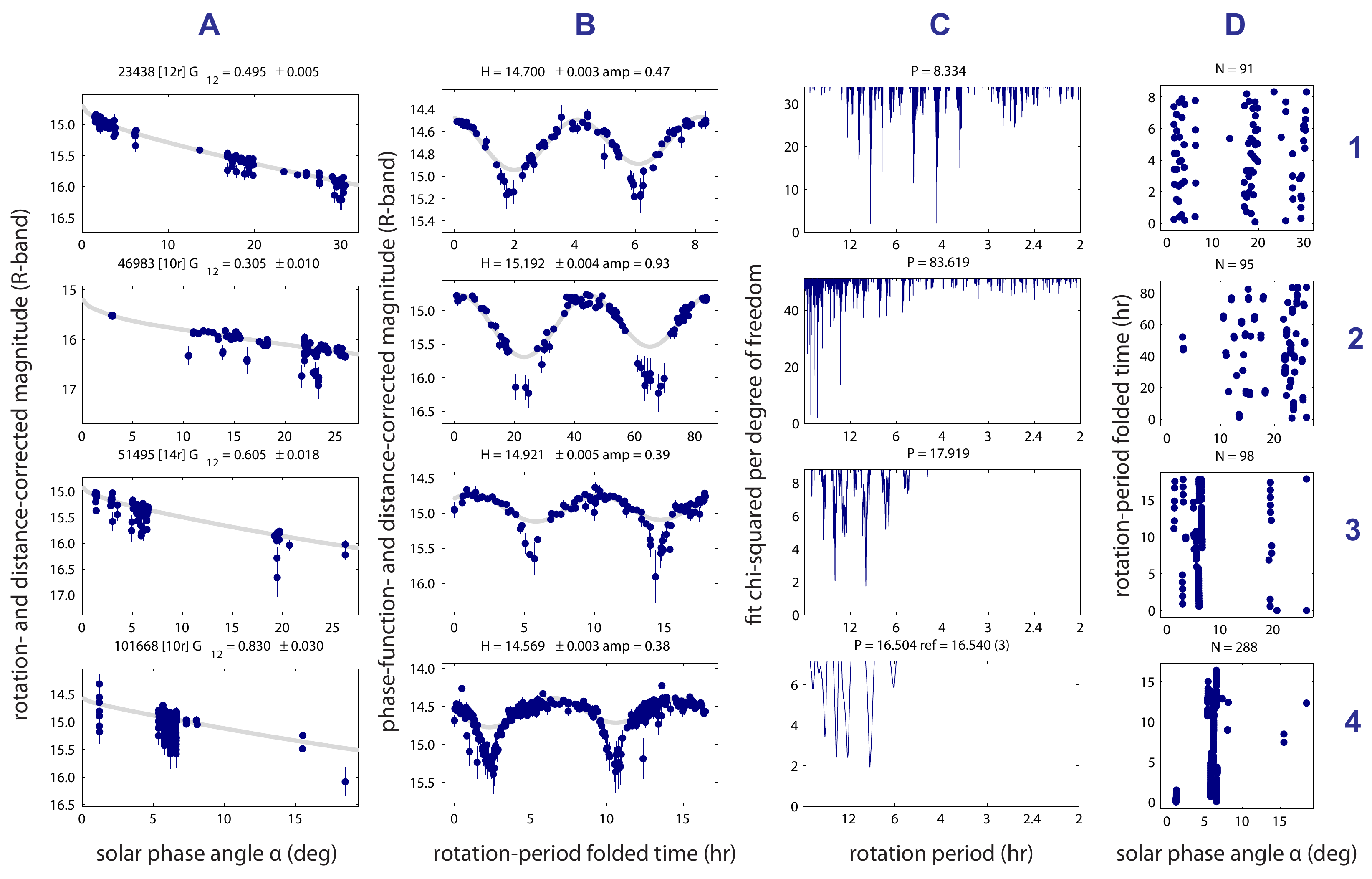}
\caption{Examples of reliable lightcurves whose folded rotation curve include cusp-like minima (systematic negative deviation from the 2nd-order Fourier fit at minimum brightness), suggestive of a binary system. Many more examples exist in our lightcurves, however in this work we have not specifically flagged such lightcurves. Future works will more carefully label and analyze this particular class of objects. }
\vspace{10pt}
\end{figure*}

We adopt a supervised ensemble-method approach for classification, originally popularized by \cite{bre84}, specifically the \emph{random forest} (RF) method \citep{bre01}. RF classification has extensive and diverse applications in many fields ({\it e.g.}, economics, bioinformatics, sociology). Within astronomy in particular RF classification is one of the more widely-employed methods of machine-learning, though many alternatives exist. For example, \cite{masc14} use the RF method for variable-star lightcurve classification, while others have approached this problem via the use of, {\it e.g.}, support vector machines \citep{woz04}, Kohonen self-organizing maps \citep{bre04}, Bayesian networks and mixture-models \citep{mah08}, principle component analysis \citep{deb09}, multivariate Bayesian and Gaussian mixture models \citep{blo11}, and thick-pen transform methods \citep{par13}.

For general descriptions of RF training and classification, we refer the reader to \cite{bre01}, \cite{brem04}, and the many references cited by \cite{masc14}. Our use of a RF classifier is particularly motivated by its already-proven application to the discovery and classification of astrophysical transients in the same PTF survey data \citep{blo12}, as well as streaking near-Earth asteroid discovery in PTF data (Waszczak et al. in prep.).

Machine-learning application generally consists of three stages: training, cross-validation, and classification. In the training stage of building a machine classifier, the multi-dimensional parameter space is hierarchically divided into subspaces called \emph{nodes}, these nodes collectively comprise a \emph{decision tree}. The smallest node---also known as a \emph{leaf}---is simply an individual datapoint (in our case, a single lightcurve). Given a set of leaves with class labels, one can build an ensemble of trees (called a \emph{forest}), each tree representing a unique partitioning of the feature space, wherein the nodes are split with respect to different randomly-chosen subsets of the parameter list. Each node splitting attempts to maximize the separation of classes between the sub-nodes. Serving as a model, in the subsequent classification stage the forest allows one to assign a probability that a given vector of features belongs to a given class. During cross validation (an essential early stage in this process), the training and classification steps are repeated many times, each time using different subsamples (of labeled data) as the training data and testing data. Cross validation evaluates the classifier's performance and ensures it is not overfitting the training data.

For our lightcurves, we are interested in a binary classification, {\it i.e.}, whether the fitted period is accurate (`real') or inaccurate (`bogus'). \cite{blo12} coined the term \texttt{realBogus} to describe this binary classification probability in the context of extragalactic transient identification. In the present work we are essentially adapting Bloom et al.'s \texttt{realBogus} concept to the problem of lightcurve-period reliability assessment.

We employ a MATLAB-based Random Forest classifier\footnote{\href{https://code.google.com/p/randomforest-matlab}{\color{blue}{https://code.google.com/p/randomforest-matlab}}} which is a port of the original RF software (originally implemented in R). This software includes two main functions, which perform the training and classification steps separately.

\subsubsection{Classifier training}

Our training data consist of the known-period lightcurves (cf. the previous section) belonging to the two classes under consideration: 618 lightcurves having accurately-fit rotation periods and 309 lightcurves having inaccurately-fit periods. Membership in one class versus the other depends on our arguably arbitrary $3\%$ relative accuracy threshold, though we claim the clearly bimodal shape of the distribution in Figure 5 justifies this 3\% criterion. We note also that the classifier ultimately only provides a \emph{probability} that a given lightcurve belongs to one class or the other, so that objects very near to the 3\% cutoff may conceivably correspond to classification probabilities close to 0.5.

An important point is that the `ground-truth' reference periods we have taken from the database of \cite{war09} may include some number of inaccurate periods. Such periods may be the product of erroneous fitting on the part of any one of its many different contributors, each of whom may employ a different fitting procedure and/or adhere to different confidence criteria. For the sake of this work however we consider all quality code 3 periods to be accurate---any actual inaccuracy will contribute to decreased classifier performance.

Besides ground-truth periods that are simply inaccurate, we also in principle risk contamination from reference periods that are \emph{no-longer accurate}. We assume that the majority of asteroids' periods are not changing with time, at least not at levels measureable with our data. For instance, direct measurement of the YORP mechanism in at least one asteroid \citep{low07} reveal a relative rotation period change of several parts per million over several years. Any \emph{measureable} period changes would likely be due to recent collisional events. The case of asteroid 596 Scheila \citep{bod11} demonstrates that detectable collisional events among main-belt asteroids do occur on a relatively regular basis, though even this robustly-detected collision imparted no measurable change in the asteroid's spin rate \citep{she13}.

Although Figures 7 and 8 detail the period-fitting reliability as a function of only eight lightcurve parameters, we construct our classifier using 12 additional parameters, for a total of twenty lightcurve parameters. In the context of machine-learning these parameters are known as \emph{features}. The twenty features we use were chosen on the basis of their availability (most are output directly by the fitting process and do not require additional computation) as well as their actual importance (as computed during the cross-validation tests described in the next section).

Our twenty lightcurve features are listed in Table 2, in order of decreasing importance. Most of these quantities we have discussed already in previous sections in the context of our model and fitting procedure. The list also includes two features characterizing the magnitude distribution of the folded lightcurve: (1) Stetson's $K$-index, a measure of the kurtosis borrowed from variable star lightcurve analysis \citep{ste96}, and (2) a `cusp index' which quantifies the extent to which the dimmest 10\% of the data points in the folded lightcurve deviate from the best fit relative to the other 90\% of the data points. We designed the cusp index to potentially identify eclipsing systems which are poorly fit by the two-term Fourier approximation but nonetheless may have accurately-fit periods (examples of lightcurves with such cusp-like minima appear in Figure 10). Eclipsing binaries would be most properly treated with a different model entirely, as would tumbling asteroids (which we also did not systematically try to identify in the data, and probably lack reliable lightcurve solutions when subjected to this work's algorithm).

\begin{figure}
\centering
\includegraphics[scale=0.9]{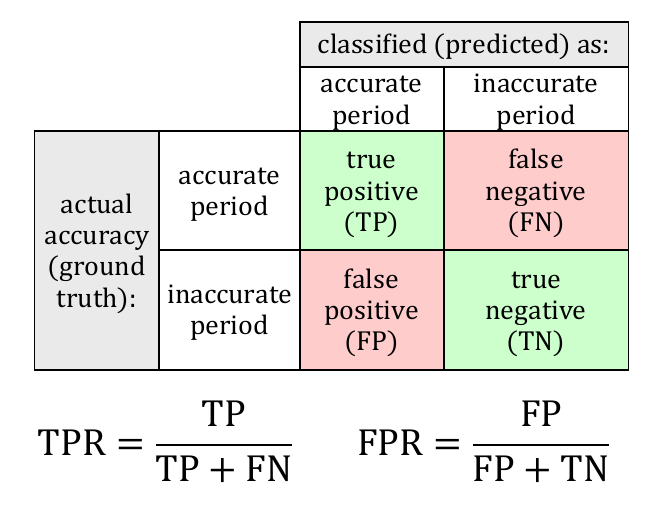}
\caption{Definitions of true vs. false and positive vs. negative labels. True-positive rate (TPR) is sometimes called the \emph{completeness} or \emph{sensitivity}, while false-positive rate (FPR) is otherwise known as the \emph{false-alarm rate}, one minus the \emph{reliability}, or one minus the \emph{specificity}.}
\vspace{1.5pt}
\end{figure}

\begin{figure}
\centering
\includegraphics[scale=0.75]{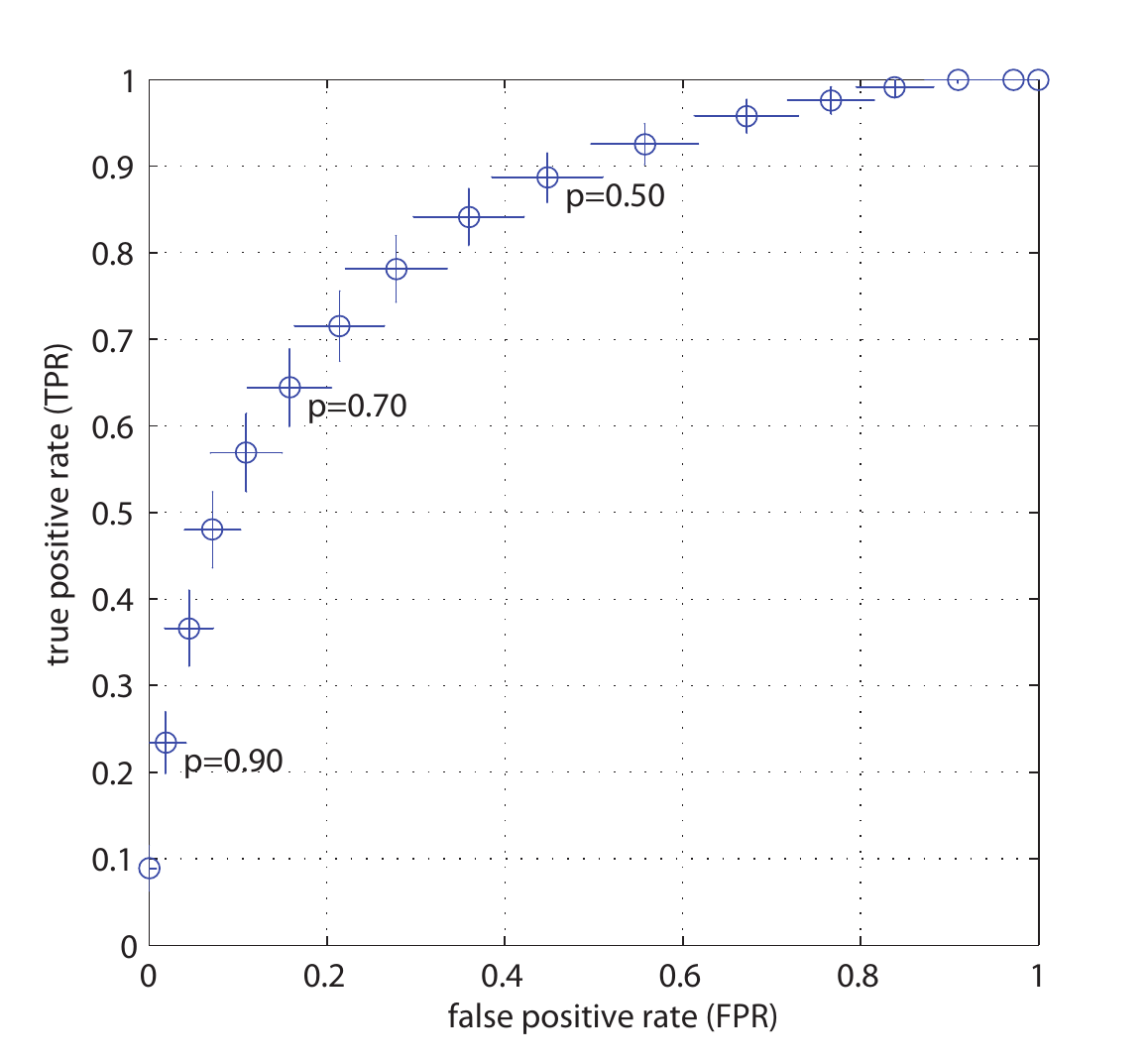}
\caption{True-positive versus false-positive rates for the cross-validation trials. Such as plot is sometimes referred to as a \emph{receiver operating characteristic} (\emph{ROC}) curve. Each trial trains the classifier using a randomly-chosen 80\% of the known accurate fits and 80\% of the known inaccurate fits among the 927-lightcurves that have reference periods. The 20\% remaining lightcurves serve as the test sample. Moving along the hyperbolic locus of points in this plot is equivalent to tuning the classification probability threshold from zero (lower left of the plot) to one (upper right of the plot). The errorbars represent the scatter in the 1,000 cross-validation trials.}
\end{figure}

\begin{figure}
\centering
\includegraphics[scale=0.66]{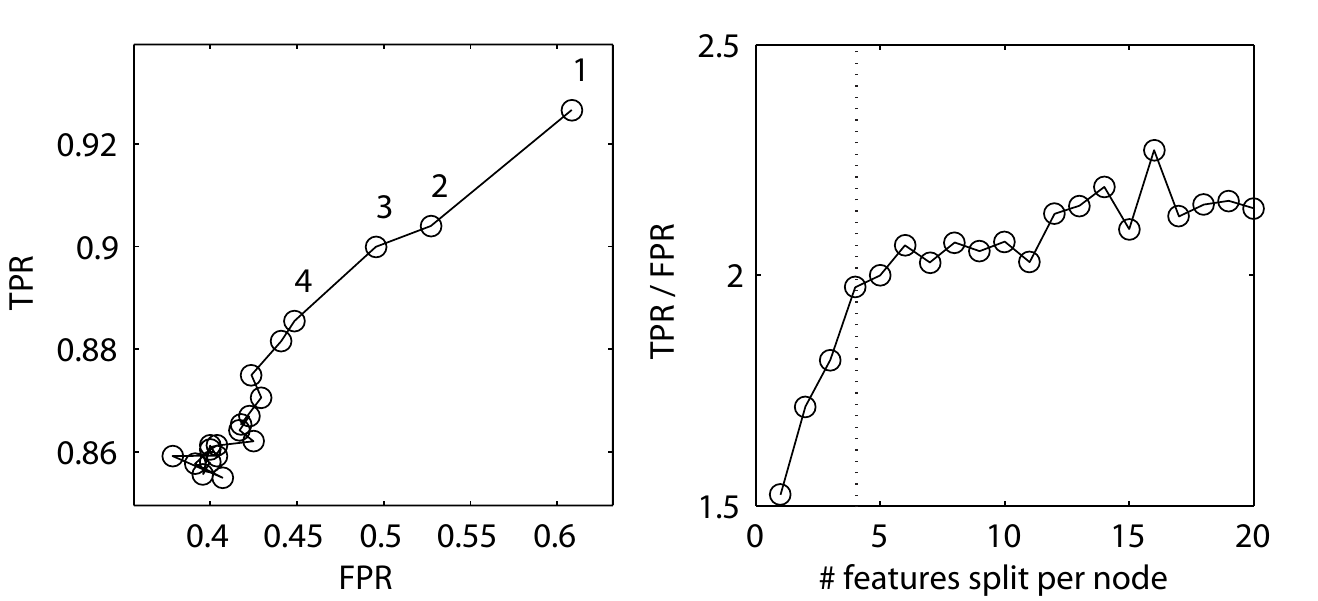}
\caption{Varying the number of features that are randomly split per node in the decision-tree-building process affects both the TPR and FPR. The values plotted here correspond to the $p>0.5$ classification threshold; each point was generated by the exact same process for which the results in Figure 12 were generated, only varying the number of features with respect to which nodes are split. In the left plot, the first four points are labeled with the number of features for that trial (for $n>4$ we omit the label). In our actual implemented model (Figure 12) we chose $n=4$ features, the value after which the TPR/FPR ratio plateaus at approximately 2, and also the value \cite{bre01} recommends, {\it i.e.}, the square-root of the total number of features (in our case, $\sqrt{20}\approx 4$).}
\vspace{10pt}
\end{figure}

Figure 9 visualizes the two-dimensional correlation coefficients for all possible pairs of the 20 lightcurve features. Overall, the correlation structure of the training sample qualitatively resembles that of the full data set, implying the training set fairly well represents the overall data set in terms of its feature-space structure. On the other hand, the \emph{distributions} ({\it e.g.}, median value, range of values) of individual features in the training set do not necessarily match the distributions in the full data set: this is evident for the several features plotted in Figure 14. An obvious example is that the full data set contains far more faint asteroids than does the training sample, even though in both cases the median magnitude (\texttt{medMag}) is positively correlated with quantities like \texttt{rmsFit} (due to Poisson noise) and \texttt{hMagRef} (since larger asteroids tend to be brighter).

\subsubsection{Classifier cross-validation}

To ascertain the trained classifier's capabilities, and to ensure that the classifier is not overfitting the training data, we perform a series of 1,000 cross-validation trials. In each trial we split each class (accurate fits and inaccurate fits) into a \emph{training} subsample (a randomly chosen\footnote{Another standard, slightly different approach is to evenly split the training data into $k$ disjoint sets (a process called $k$-folding). Also, our choice to \emph{separately} partition the two classes into training and test subsamples could be omitted.} 80\% of the class) and a \emph{test} subsample (the remaining 20\% of the class). We then train a classifier using the combined training subsamples and subsequently employ the classifier on the combined test subsamples. In each of the trials, the classifier outputs a classification probability (score) for each object in the test sample, and we track the true positive rate (TPR; fraction of accurate period fits that are correctly classified above some threshold probability) as a function of the false-positive rate (FPR; fraction of inaccurate period fits that are incorrectly classified above said threshold probability). See Figure 11 for a summary of these terms.

The results of the cross-validation are shown in Figure 12. By tuning the minimum classification probability used to threshold the classifier's output, one effectively moves along the hyperbola-shaped locus of points in TPR-vs.-FPR space seen in the plot. Several points have labels ($p=...$) indicating the corresponding threshold probability (adjacent points being separated by $\Delta p=0.05$). The errorbars in Figure 12 represent the standard deviation of the location of each point over all 1,000 trials, while the point centers are the average locations.

A classification threshold of $p>0.5$ is conventionally used when quoting single false-positive and true positive rates. In our case, this gives FPR $=0.45\pm 0.07$ with TPR $=0.89\pm 0.03$. The \emph{contamination} of positively-classified lightcurves in the cross-validation trials depends also on the actual class ratios in the sample being classified. In particular, since $\sim$$1/3$ of our known-period lightcurves are inaccurate fits (Figure 5), it follows that among all lightcurves the classifier labels as accurate fits, the contaminated fraction is $(0.45\times 1/3)/(0.89\times2/3 + 0.45\times1/3)\approx 1/5$. If instead of using the classifier we just randomly labeled some fraction of the lightcurves as accurate and the rest as inaccurate, the resulting contamination would be $1/3$ ({\it i.e.}, worse than the $1/5$ afforded by the classifier, as expected).

Several parameters can be adjusted or tuned when training a random forest classifier. First is the number of decision trees generated during the training stage. Classification accuracy typically increases with the number of trees and eventually plateaus. Most applications employ hundreds to thousands of trees; we here use 1,000 trees. Another tunable parameter is the number of randomly-selected features (out of the 20 total here considered) with respect to which nodes are split in building the decision trees. \cite{bre01} recommends using the square root of the number of features. We ran the cross-validation for all possible numbers of features with respect to which the nodes can be split ({\it i.e.}, all numbers between 1 and 20). The results are in Figure 13. We chose $n=4$ as the number of features to split, both because the classifier's performance plateaus after that value and because it follow's the recommendation of \cite{bre01} ($4\approx$$\sqrt{20}$) features.

Other parameters that can be tweaked are the maximum depth of a tree, the minimum number of samples per leaf, the minimum number of samples used in a split, and the maximum number of leaf nodes. We do not constrain any of these parameters, meaning we allow: trees of any depth, with any number of leaf nodes, leaf nodes consisting of a single sample, and splits based on the minimum of 2 samples. We note that as a result our model optimization is not comprehensive and it is possible a better classifier could be trained. However, the relatively small training sample size here is likely the limiting factor; additional data is necessary to substantially improve the classifier performance.

\begin{figure*}[h]
\centering
\includegraphics[scale=0.75]{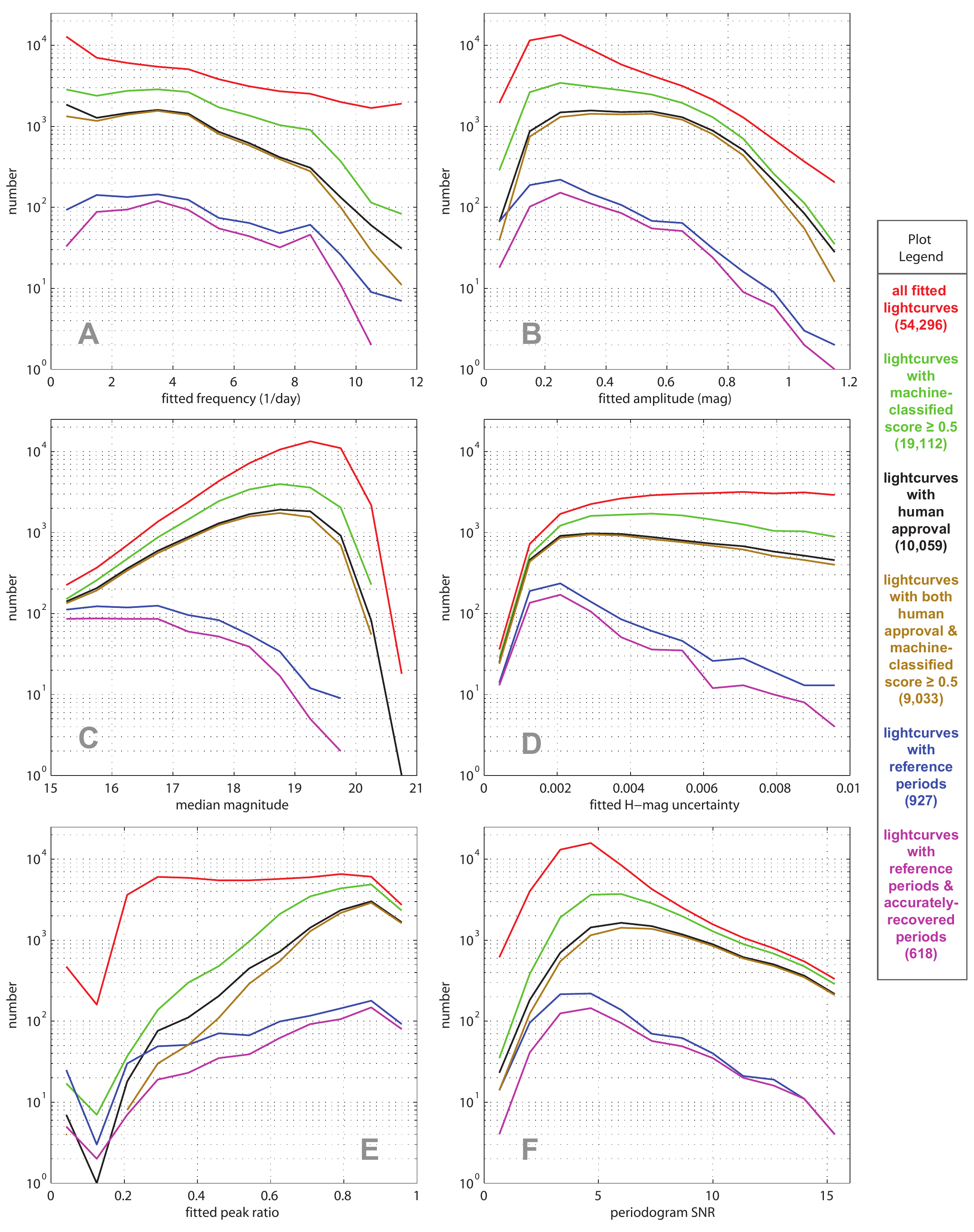}
\caption{Distributions of PTF-fitted lightcurves (and various subsets thereof) in select features/parameters. These plots are histograms with the same binning as the top rows of Figures 7 and 8. For better readability we here use line-connected bin points (rather than the stair-plot format used in, {\it e.g.}, Figure 5).}
\end{figure*}

\begin{figure*}[t]
\centering
\includegraphics[scale=0.54]{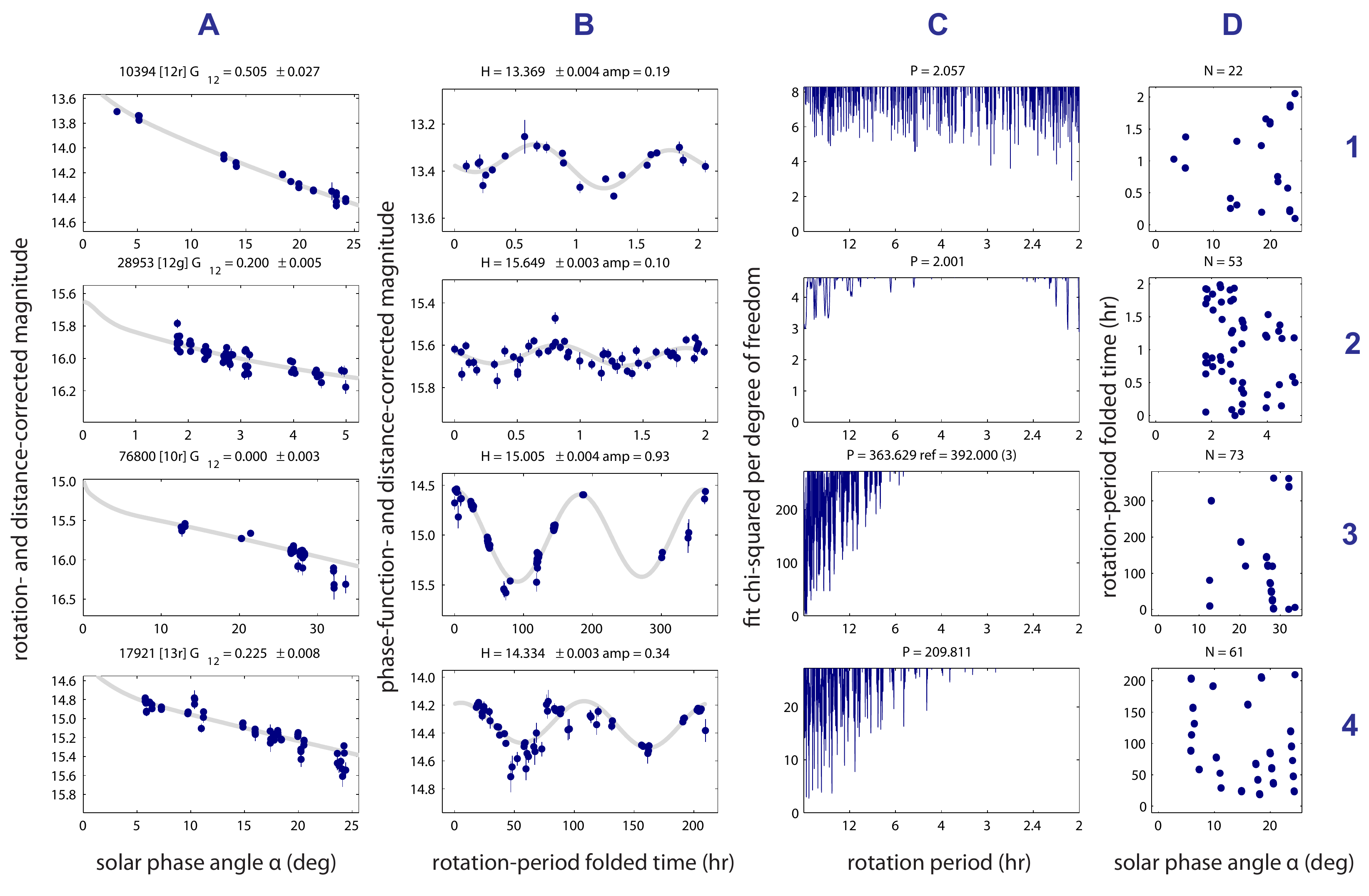}
\caption{Example lightcurves for which the machine-based and human-based reliability scores differ. \emph{Row 1}: Human approved, machine rejected ($p=0.32$). \emph{Row 2}: Human rejected, machine approved ($p=0.66$). \emph{Row 3}: Human approved, machine rejected. For this object, the fitted period differs from the known reference period of 392 hours by 7\%, hence the machine rejects it by definition. \emph{Row 4}: Human rejected, machine approved ($p=0.70$).}
\vspace{5pt}
\end{figure*}

In the bottom rows of Figure 7 and 8, we detail the dependence of the TPR and FPR on various lightcurve parameters. Averaging (marginalizing) over any of the $x$-axis quantities in these bottom-row plots (while also weighting each bin by the number of lightcurves it contains, cf. the top row of plots in Figures 7 and 8), produces precisely the TPR and FPR values of the $p=0.5$ data point in Figure 12.

In addition to the TPR and FPR estimates, cross-validation allows us to quantify the relative \emph{importance} of the features by computing the average depth in the trees at which a split was performed with respect to each feature. Those features with respect to which the training sample is consistently divided early in the building of each tree are deemed more important ({\it i.e.}, more discriminating) than those features which are split later, as the tree-building process tries to maximize the separation of the classes as early as possible by splitting features in an optimal sequence. Both Table 2 and Figure 9 list the features in order of importance.

Note that we had manually guessed several of the most important features---namely, \texttt{peakRatio}, \texttt{freqSNR} and \texttt{hMagErr}---prior to any machine-learning work via inspection of the plots in Figure 8. The numerical importance values thus agree with these initial observations, and also quantify the significance of features which would be difficult to ascertain manually. For instance, \texttt{numObsFit} appears (in Figure 8) not to be related to the fitting accuracy while \texttt{medMag} (Figure 7) \emph{does} appear related to accuracy (fainter lightcurves being less accurate), yet these two features evidently have equal importance in the classification process (cf. Table 2). Figure 9 indicates that \texttt{numObsFit} and \texttt{medMag} have quite different correlation relationships with respect to more important features. Hence, it would not be surprising if their one-dimensional distributions (in Figures 7 and 8) bear no resemblance to the multi-dimensional distributions on which the decision trees are defined and in which these two parameters apparently carry comparable weight.

\subsubsection{Machine-vetted lightcurves}

Having trained the machine classifier as described in Section 5.2.1, we use it to predict the validity of our remaining $\sim$53,000 fitted periods (of $\sim$48,000 unique asteroids) which lack quality code 3 reference periods in \cite{war09}. The automated classifier assigned positive reliability scores ($p\ge0.5$) to 19,112 of the lightcurves (35\% of the total data set). Figure 14 details the distribution of the lightcurves (raw-fitted, machine-vetted, and other subsets) with respect to some of the most important lightcurve features.

With respect to rotation period (Figure 14 panel A), the classifier rejects the largest fractions of lightcurves in the long-period ($\gtrsim$1 day) and short-period ($\lesssim$2.7 hours) bins. From Figure 7 (bottom row, leftmost column), we know that the classifier's completeness does not drop significantly for these long- and short-period objects, nor is the false positive rate higher among them. Hence we have reason to trust the classifier's heavy rejection of periods in these bins, and therefore conclude that our fitting algorithm (Section 4) is prone to erroneously fitting periods in these period extremes (as was also suggested in the known period sample in Figure 7).

Panel C shows that the mode of the apparent-magnitude (\texttt{medMag}) distribution for machine-approved lightcurves is $\sim$19 mag, as compared to the predominantly $V\lesssim 17$ mag known-period training sample. Comparing this to Figure 2 panel A shows that the limiting magnitude of reliable lightcurves is comparable to that of individual detections.

Panel E of Figure 14 shows that the raw output of our fitting process contains peak-ratio values that are uniformly-distributed above 0.2, this particular value being a hard-coded threshold that double-peaked lightcurves (at least those with amplitudes $>$0.1 mag) output by our fitting algorithm must satisfy (see Figure 3 and Section 4.1). The classifier's output clearly indicates that reliability is linearly related to the peak ratio, as was also prominently seen in Figure 8. Because Figure 8 also indicates that the classifier's true-positive and false-positive rates also relate linearly with \texttt{peakRatio}, we conclude that the slope of the \texttt{peakRatio} distribution for the machine-vetted lightcurves is likely an upper limit for the true slope.

\subsection{Manual screening}

\begin{figure}
\centering
\includegraphics[scale=0.75]{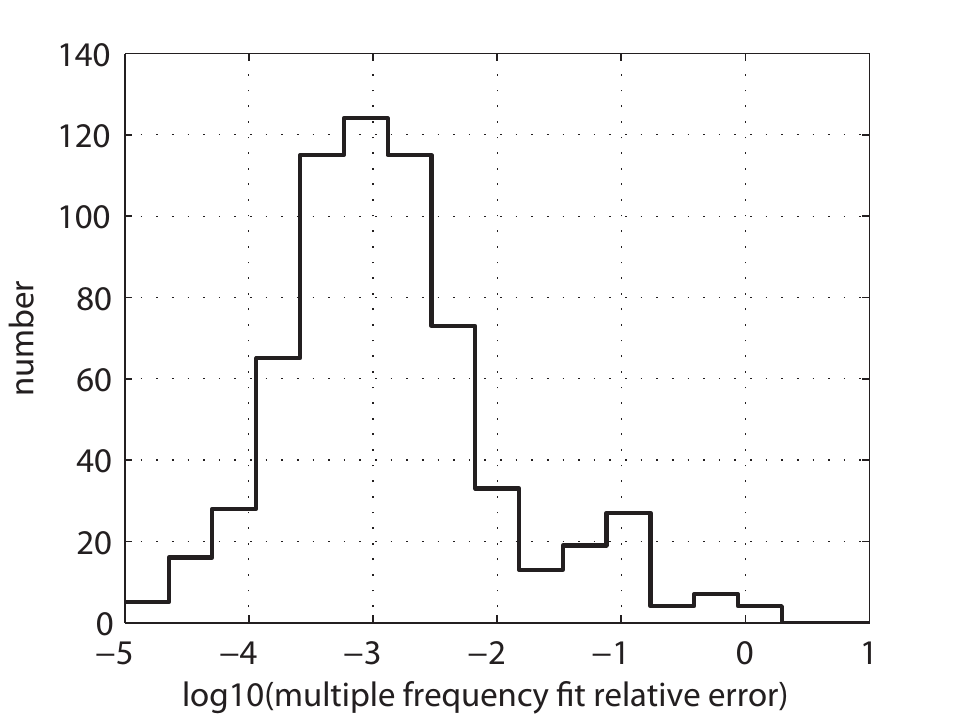}
\caption{For the 654 unique asteroids having more than one reliable lightcurve fit (either multiple oppositions and/or both $R$ and $g$ band data) we plot the log of the relative frequency error, defined as the range of the asteroid's fitted periods divided by the geometric mean of its fitted periods. Comparison with Figure 5 suggests that we can deem all cases with error $\lesssim$3\% as consistently recovered periods, and those with greater than 3\% error as inconsistent fits.}
\vspace{5pt}
\end{figure}

In addition to machine-based vetting, we manually inspected all 54,296 of the lightcurves that were output by our fitting process. A human screener first studies the ground-truth known period examples (Section 5.1) in an effort to learn to distinguish between accurate and inaccurate fits. Only the $G_{12}$ fit is considered (as was the case with the automated classifier), and for each lightcurve the screener inspects precisely the amount of information included for example in Figures 4, 6 and 10 of this paper. Specifically, for each lightcurve the screener views a row of four plots: (1) the rotation-corrected phase curve, (2) the phase-function-corrected folded rotation curve, (3) the periodogram, {\it i.e.}, the reduced $\chi^2$ plotted linearly against frequency (labeled however with the corresponding period), and (4) the rotational-phase vs. phase-angle plot. A single screener is presented with these plots through a plain-formatted webpage, allowing for efficient scrolling through the lightcurves and rapid recording of either a `reliable' or `unreliable' rating for each fitted period. In addition, all lightcurves in the known-period sample were reinserted into the screening list, with their reference periods removed. These were thus blindly assessed by the screener, independent of their formal (3\%-accuracy) classification status.

The black lines in Figure 14 plot the results of the manual screening, in which a total of 10,059 lightcurves (19\% of the total set) were deemed `reliable'. With respect to the machine-approved sample, the human-rated sample is in all cases between roughly a factor of $\sim$1 to 2 smaller in each bin relative to the features examined in Figure 14. In general the shapes of the machine-approved and human-approved distributions match fairly closely. Figure 15 shows examples of lightcurves for which the machine- and human-based classifiers differed in their rating (we focus on very short and very long fitted periods in Figure 15, but many examples exist for intermediate periods as well).

\subsection{Asteroids with multiple fitted periods}

A total of 654 unique asteroids have more than one PTF lightcurve whose fitted period was labeled as reliable by the vetting process described in the previous sections. These 654 asteroids collectively have 1,413 fits (so that the average multiplicity is $\sim$2.2 fits per asteroid) and include objects either observed in multiple oppositions and/or in both filters during one or more oppositions. Figure 16 plots the distribution of the relative error in the fitted frequencies of all such multiply-fit asteroids, this error being defined as the \emph{range} of the asteroid's fitted frequencies divided by the \emph{geometric mean} of its fitted frequencies. Just as in Figure 5 (when we compared to literature-referenced frequencies), we see a prominent mode in the histogram peaking at $\sim$0.1\% relative error, with some excess for errors greater than $\sim$3\% error. There are 63 asteroids in particular with relative errors greater than 3\%, of these only four asteroids have more than two fits. If we assume that, in the remaining 59 pairs of disagreeing periods, one of the periods is correct, then the contamination fraction of lightcurves based on the sample of multiply-fit asteroids is $\sim$$30/1413=4$\%.

\begin{figure*}
\centering
\includegraphics[scale=0.66]{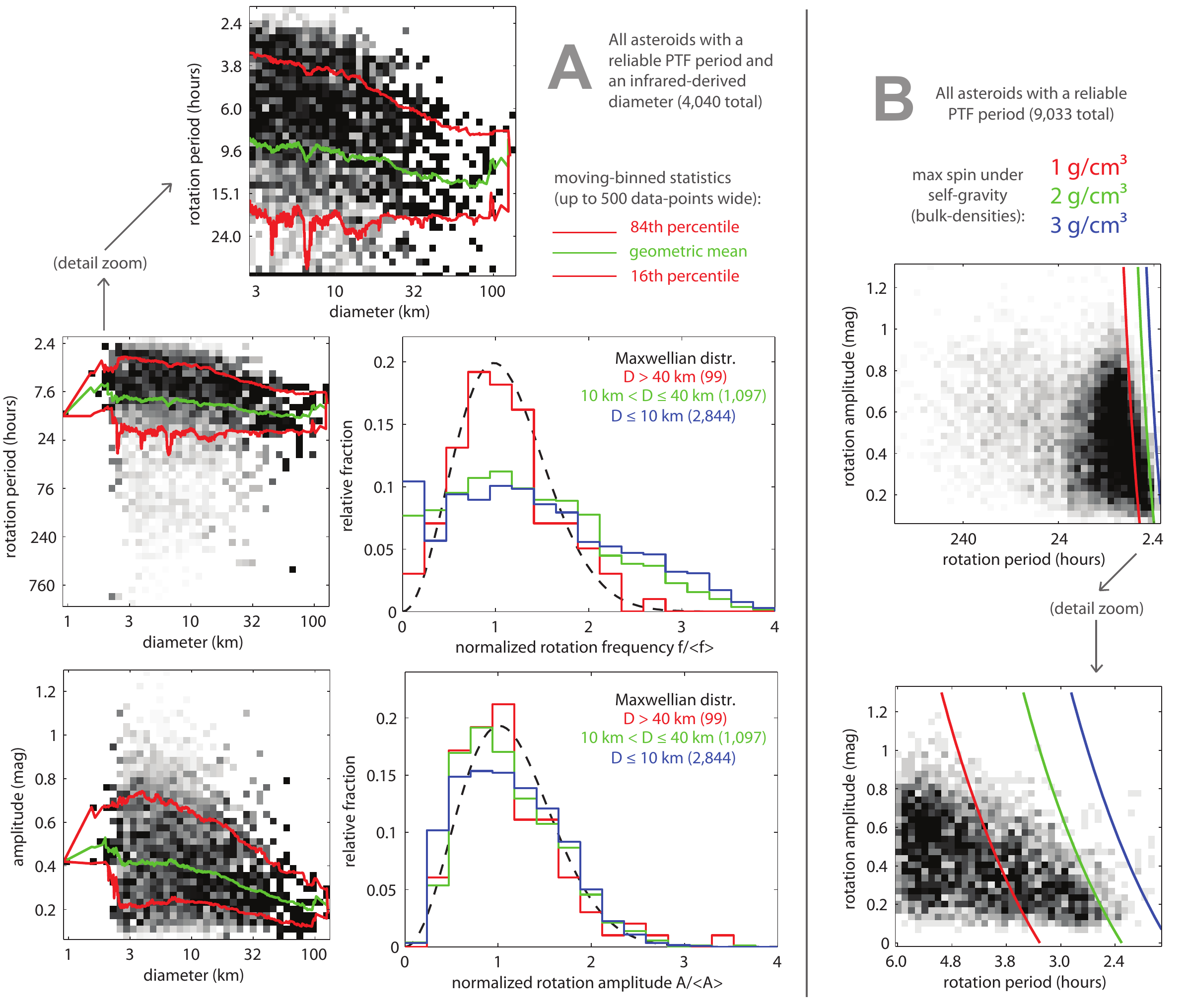}
\caption{\emph{Panel A}: Distribution of spin rate and amplitude as functions of infrared-derived diameters (see appendix for diameter data sources), including data for 4,040 of our lightcurves. The two-dimensional histograms (left side plots) are column-normalized (see text for details). \emph{Panel B}: Comparison of the period versus amplitude distribution (regular 2D histogram, \emph{not} column normalized) with max-spin-rate versus amplitude for a uniform density ellipsoid held together solely by self-gravity}.
\vspace{10pt}
\end{figure*}

\section{Preliminary lightcurve-based demographics}

In this section we perform a \emph{preliminary} analysis and interpretation of the demographic trends evident in this work's fitted lightcurve parameters. Forthcoming works and papers will more closely examine the population distributions of both rotation and phase-function parameters.

Throughout this section we repeatedly examine variation of lightcurve-derived parameters as functions of \emph{color index} and \emph{infrared-derived diameters}. In the appendix we describe the aggregation and characteristics of these two custom data sets (compiled from external sources). The color index quantifies an asteroid's probability of membership in the C-type ($p=0$) versus S-type ($p=1$) color-based clusters. Objects which in fact belong to neither C nor S groups ({\it e.g.}, V types, D types) will have color indices near $p=0.5$ provided they are in fact separated from both the C-type and S-type clusters in the 2D color spaces considered (see appendix).

There are many interesting demographic questions addressable with these lightcurve data which---in the interest of space---we do not treat in this work. For example, one could examine relationships between lightcurve parameters and orbital elements and/or family membership, proximity to resonances, and so on. We are making all of these lightcurve data available electronically (Tables 4 and 5, cf. Section 9.3) so that the community may use these data to help explore such science questions.

\subsection{Disclaimer regarding de-biasing}

The preliminary demographic analyses that follow do \emph{not} take into account fully de-biased distributions of, {\it e.g.}, spin rates, amplitudes, or phase-function parameters. The true-positive and false-positive rates given in the bottom row of plots in Figure 7 and 8 (also, the blue and violet lines in Figure 14), constitute some of the necessary ingredients for producing a fully de-biased data set, however in this work we do not attempt to compute the de-biased distributions. 

\subsection{Rotation rates and amplitudes}

In Figure 17 we reproduce several of the plots appearing in \cite{pra02} and references therein, using this work's much larger data set (characterized by at least an order of magnitude larger sample of small objects). Both spin rate and amplitude are examined for the 4,040 objects having diameter data from infrared surveys. Unlike \cite{pra02}, we are not able to individually plot each lightcurve's data (the $\sim$4,000 points would make the plot difficult to render, as well as difficult to read); hence we plot these (and other relationships later in this section) using two-dimensional histograms where the intensity of each pixel corresponds to the number of objects in that bin (darker means more, with linear scaling). Additionally, 2D histograms for which the diameter is plotted on the horizontal axis have their pixel values column-normalized, {\it i.e.}, all pixels in each column of the histogram sum to the same value. This facilitates the visual interpretation of period and amplitude variation with diameter, as the left-hand side (small-diameter end) of the plots would otherwise saturate the plot.

Following \cite{pra02}, we include the geometric mean rotation frequency as computed from a running bin centered on each object. The half-width of the bin centered on each object is either 250 (data points) or the object's distance from the top or bottom of the sorted diameter list, whichever is smallest. This ensures the geometric mean is not contaminated at the edges of the plot by the interior values, though it also means more noise exists in these edge statistics. The geometric mean is the more intuitive statistic for the rotation period as compared to the \emph{arithmetic} mean, since the rotation periods tend to span several orders of magnitude. In addition to the geometric mean, we plot the 16th and 84th percentile values from each running bin.

The basic observed trend regarding rotation rate is that smaller-diameter asteroids rotate faster on average. A slight increase in the rotation rate also appears for objects larger than $\sim$80 km. Binning the data into a coarser set of three diameter bins and normalizing each object's spin rate by the local geometric-mean rate, we see a progression from a near-Maxwellian distribution to a progressively non-Maxwellian distribution for smaller objects. The rotation rates of a collisionally-equilibrated population of rotating particles is known to approach that of a Maxwellian distribution ({\it e.g.}, \citealp{sal87}), which for a population of $N$ objects as a function of rotation frequency $f$ is:

\begin{equation}
n(N,f,f_\text{peak})=\frac{4Nf^2}{\sqrt{\pi}f_\text{peak}^3}\exp\left(-\frac{f^2}{f_\text{peak}^2}\right),
\end{equation}

\noindent where $n(N,f,f_\text{peak})df$ is the number of objects in the interval $(f,f+df)$ and $f_\text{peak}$ is the peak frequency ({\it i.e.} the frequency corresponding to the distribution's maximum). 

One way of testing how well a Maxwellian actually fits the data is the two-sided Kolmogorov-Smirnov (KS) test \citep{mas51}. This test compares an empirical distribution to a reference distribution ({\it e.g.}, Gaussian, Maxwellian, or another empirical sample) via a bootstrap method. In particular it computes a statistic quantifying the extent to which the cumulative distribution function differs in the two distributions being compared. In our case, we use Equation (32) to simulate a large sample ($10^5$) randomly drawn from an ideal Maxwellian distribution and compare this simulated sample against the 99-asteroid sample (of $D>40$ km) frequencies. Interestingly, this test indicates our 99 large-asteroid normalized frequencies differ from a Maxwellian at nearly the 10$\sigma$ confidence level, though this could be due in part to the lack of a proper de-biasing of the distribution (cf. Section 6.1)

All of these trends---including the qualitative resemblance of a Maxwellian but its formal disagreement---were noted previously by \cite{pra02}. At the time their $D<10$ km size bin contained data on only 231 objects, as opposed to our sample of 2,844 asteroids with $D<10$ km. Conversely, our $D>40$ km bin contains only 99 objects as compared to the $\sim$400 large asteroids they took into consideration in comparing to a Maxwellian.

\cite{ste15} recently described how collisional evolution of large asteroids should actually lead to a L\'{e}vy distribution, which has a significantly longer tail than a Maxwellian distribution having the same peak. They compared their theory to spin rates of $D\ge10$ km asteroids from the LCDB and found in general that the L\'{e}vy distribution fails to fit the spin distribution of large asteroids, suggesting that there may be a significant primordial component to the spin distribution. Potential primordial contributions to the angular momentum of asteroids were explored by \cite{harb79} and later authors; we will return to this topic in Section 

Our amplitude distribution contains an obvious observational bias (cf. Section 6.1) in that amplitudes less than $\sim$0.1--0.2 mag are generally ill-fit by our modeling procedure (cf. Figure 7) and thus significantly underrepresented in our sample of reliable lightcurves considered here. Nonetheless, we see a clear trend of smaller asteroids exhibiting larger rotational amplitudes, consistent with the idea that larger bodies have sufficient surface gravity to redistribute any loose mass to a more spherical shape.

As we have done for the normalized frequency distribution, we plot diameter-binned normalized amplitudes against a Maxwellian distribution, this time merely to guide the eye as opposed to validating any hypothetical physical interpretation. The fact that the normalized amplitude distributions do not deviate too drastically from the Maxwellian shape at smaller diameters indicates that the spread in the amplitude distribution is proportional to its mean value, a basic property of the Maxwellian distribution, hence the good agreement. \cite{car10} provides a recent analysis of asteroid rotation amplitudes, and highlighted a similar increase in both the amplitude's mean and spread with decreasing diameter.

\begin{figure*}
\centering
\includegraphics[scale=0.66]{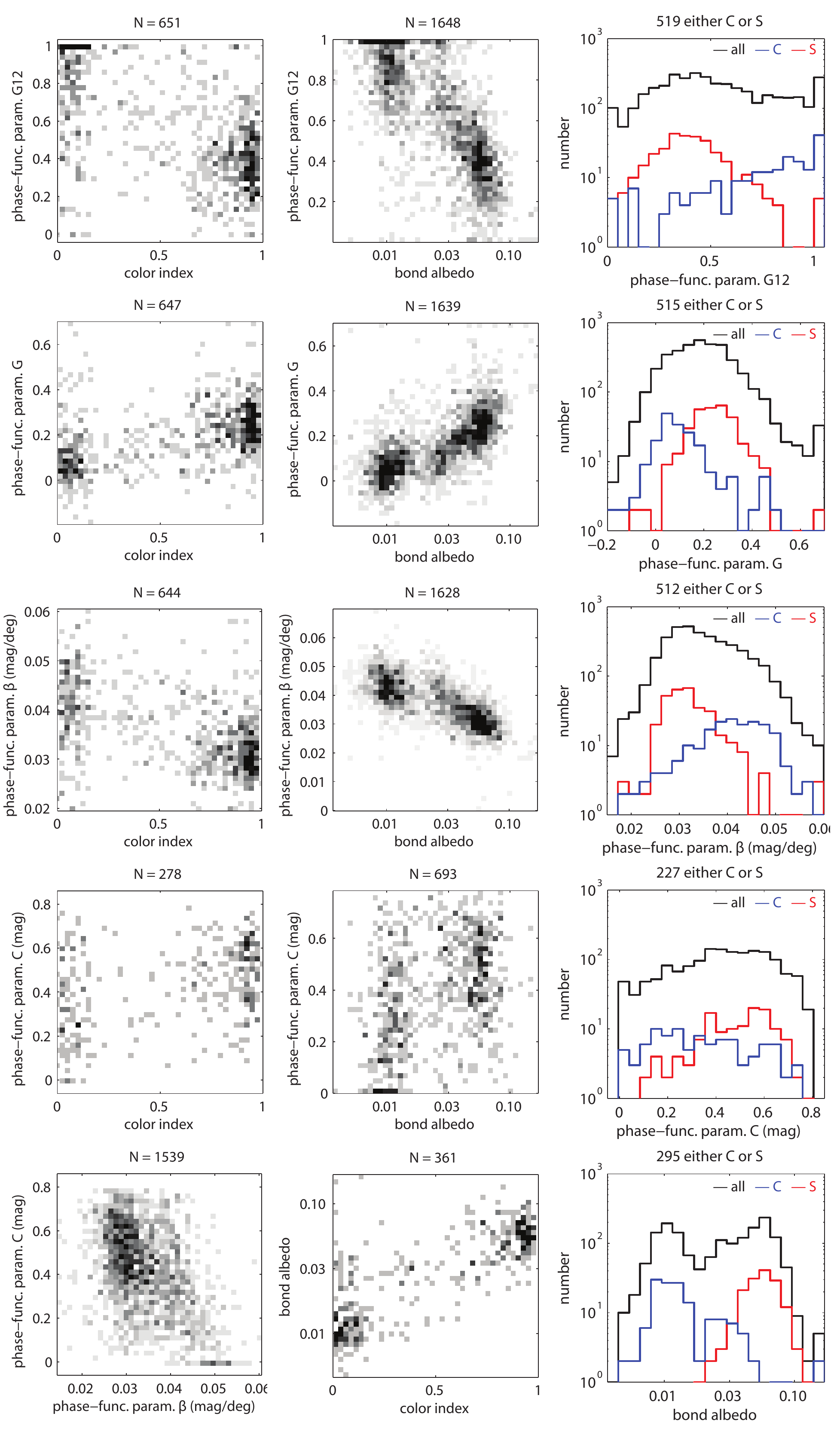}
\caption{Various fitted phase-function parameters plotted against color index and bold albedo (two-dimensional histograms; the total number of lightcurves in each plot is stated above it as $N=\ldots$). In the right column of plots, one-dimensional distributions with the color-index classified objects plotted separately. In the right column of 1D histograms, C and S types are defined as objects with color indices less than 0.25 and greater than 0.75, respectively.}
\end{figure*}

Panel B of Figure 17 shows the distribution in period-vs.-amplitude space, in which we can plot \emph{all} 9,033 lightcurves, including those lacking a diameter estimate. Contours representing the maximal spin rate of a body held-together solely by self-gravity of certain uniform densities are overplotted. Our data as a whole do not appear to populate the region beyond the $\sim$2 g/cm$^3$ contour. Later in this section we will re-examine this behavior separately for the two major taxonomic classes.

\begin{figure}
\centering
\includegraphics[scale=0.64]{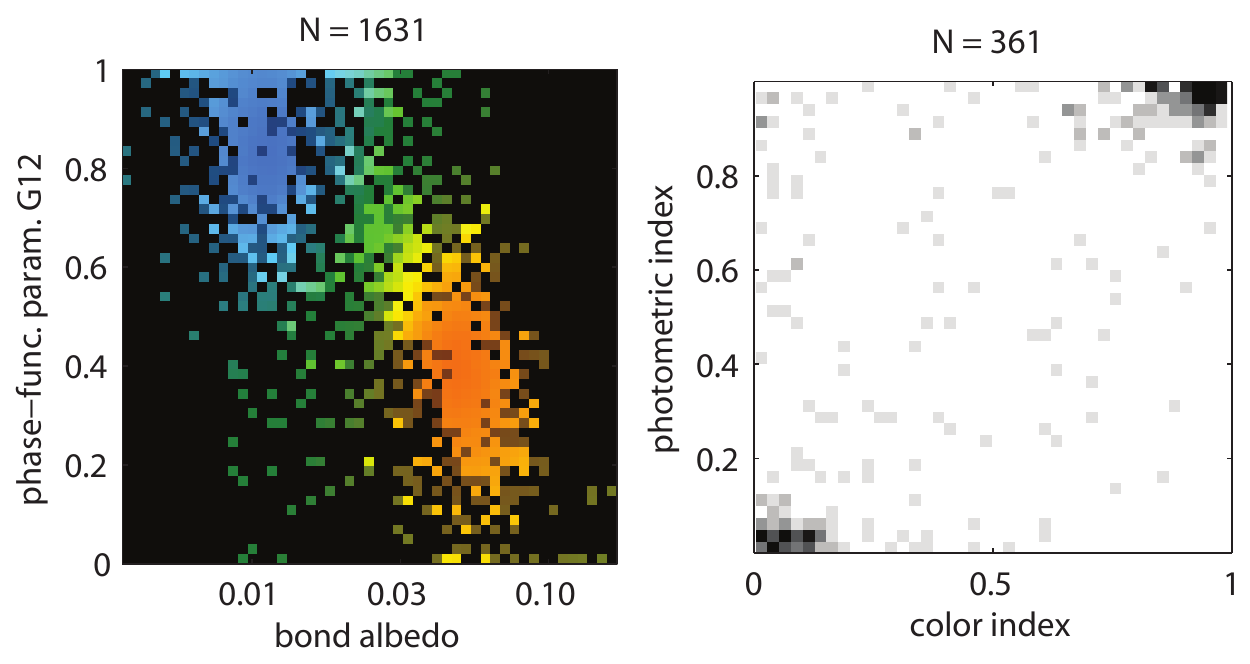}
\caption{\emph{Left}: We perform the same clustering analysis used in defining the color index (see appendix), this time on the $G_{12}$ versus $A_\text{bond}$ distribution, which contains 1,631 PTF lightcurves all of which have IR-derived diameters and reliable phase functions. The output of this clustering analysis is the \emph{photometric index}, which analogous to the color index is a number between 0 (C type) and 1 (S type) quantifying to the class membership of each constituent asteroid data point. \emph{Right}: Correlation between the color index and our photometric index, a comparison which can be made for 361 objects. Note that most data are in the lower left and upper right corners}.
\vspace{10pt}
\end{figure}

\begin{figure}
\centering
\includegraphics[scale=0.69]{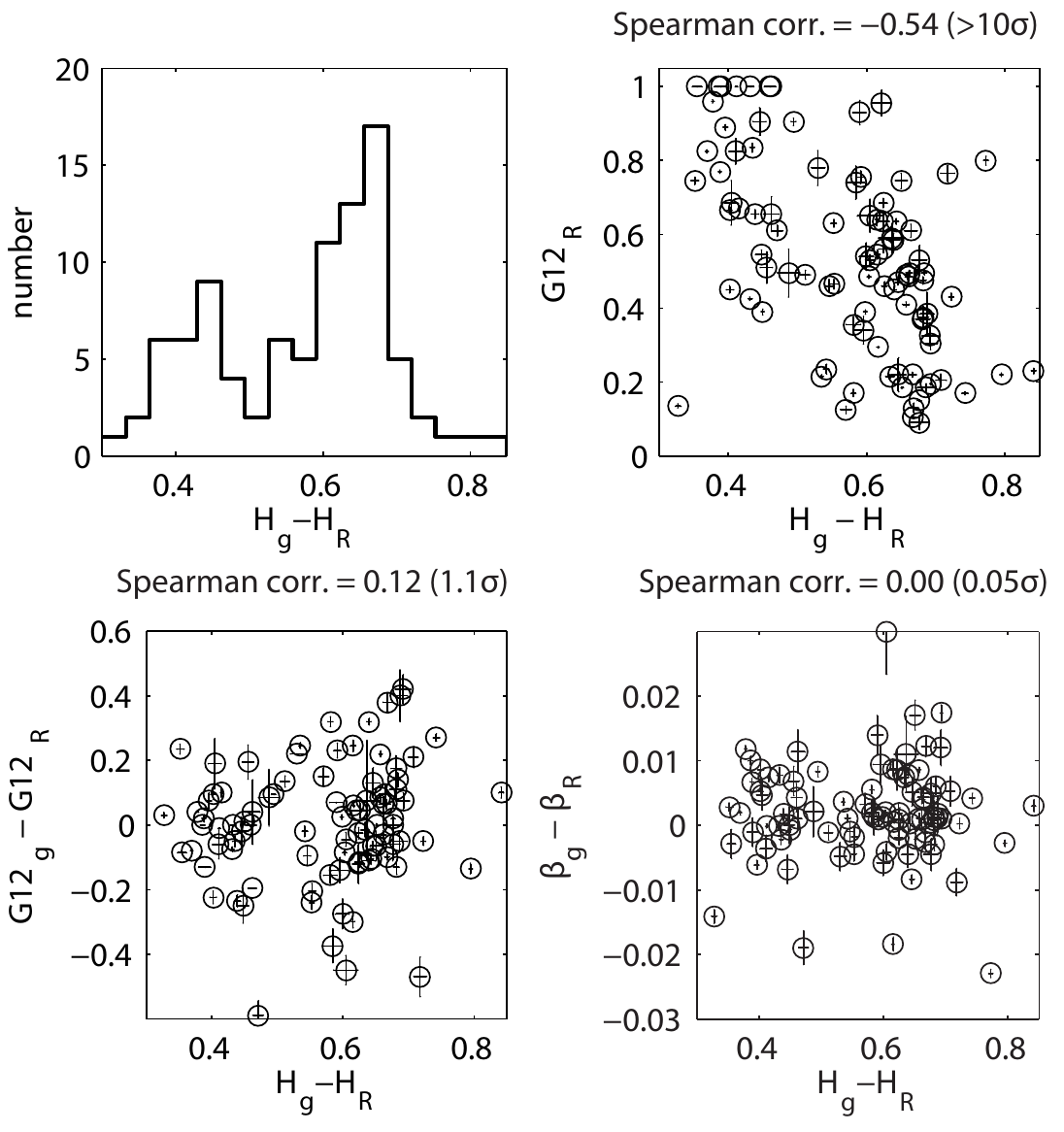}
\caption{For the 92 asteroids with both $R$-band and $g$-band lightcurve fits from the same opposition, we use the resulting difference in the absolute magnitudes $H_g-H_R$ as a proxy for taxonomy. This color distribution is qualitatively bimodal (top left), and the correlation with $G_{12}$ is very robust (top right). We detect no significant difference in the $G_{12}$ and/or $\beta$ parameters between the two bands, both in the sample as a whole, and as a function of the $H_g-H_R$ color.}
\end{figure}

\begin{figure}
\centering
\includegraphics[scale=0.59]{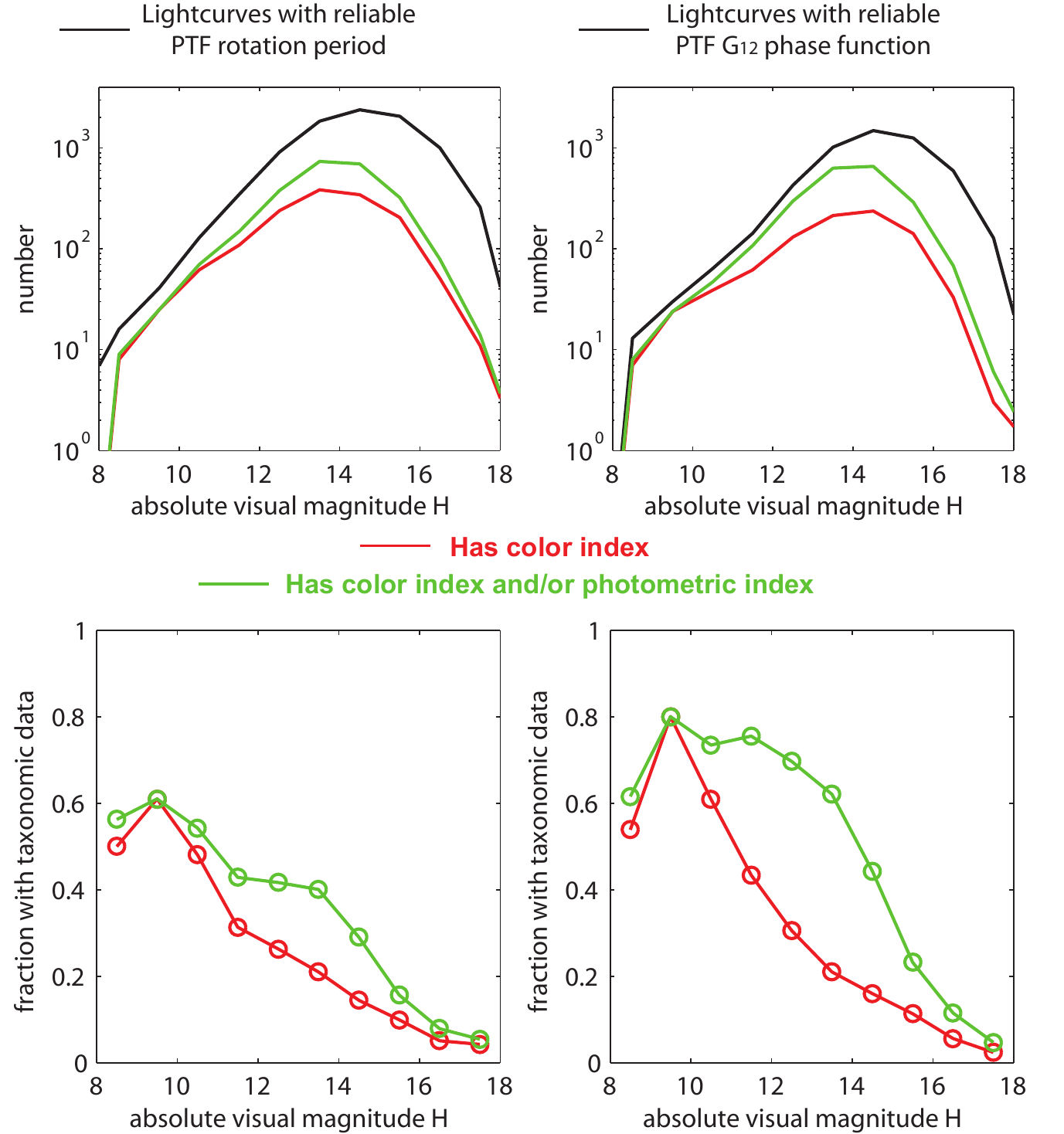}
\caption{Added completeness from supplementing the color index with the photometric index among asteroids having PTF lightcurves. Both indices are a proxy for the taxonomic type. The left- and right-hand plots apply separately to the subset labeled by the black line above each column.}
\vspace{10pt}
\end{figure}

\subsection{Phase-functions and bond albedos}

We consider any of the 54,296 fitted PTF lightcurves to have a reliably-fit phase function if \emph{both} of the following conditions are satisfied:

\begin{enumerate}

\item{The lightcurve is one of the 9,033 having a reliable period fit, \emph{or} its fitted amplitude (for the $G_{12}$ model) is less than 0.1 mag (the latter is true for 1,939 lightcurves, only 39 of which have reliable periods)}
\item{The lightcurve is fit using data from at least five phase-angle bins of width $\Delta \alpha=3$ deg. These five bins need not be contiguous, and they need not include phase angles in the region where opposition surges are typically measured ({\it i.e.}, $\alpha\lesssim 10$ deg)}

\end{enumerate}

\noindent The above two criteria are met by 3,902 out of the 54,296 PTF lightcurves. Of these, 1,648 have an infrared-based diameter available, 651 have a color index available, and 361 have both a diameter and color index.

Figure 18 details the distributions of the fitted phase parameters $G_{12}$, $G$, $\beta$ and $C$ against the color index, bond albedo, and in 1D histograms with color-based taxonomic subsets. Though the phase parameters are all correlated with color index and with bond albedo, none of the 1D phase-parameter distributions (right column of plots) exhibit bimodality alone, whereas the bond albedo (bottom right plot) does show significant bimodality. The red and blue histograms consist of all asteroids having color metric either less than 0.25 (C types) or greater than 0.75 (S types). The $G$ and $(\beta,C)$ phase parameters are only plotted for those lightcurves which also have a $G_{12}$ solution. Not every lightcurve produced a solution for all three of the phase-function models, hence the sample sizes for the $G$ and $(\beta,C)$ models include a slightly reduced number of data points.

We reiterate our statement from Section 3.2.1 that the bond albedo $A_\text{bond}$ is a more fundamental ({\it i.e.}, intensive rather than extensive) property than is the geometric albedo $p_V$, hence our focus on $A_\text{bond}$ here. The bond albedo is computed using Equation (8) together with Equation (15), and makes use of our PTF-derived absolute magnitudes---$H$ from the $G_{12}$ fit in particular---as well as the phase integral $q$ of Equation (8), also computed directly from the $G_{12}$ fit for $\phi$. In particular,

\begin{figure*}
\centering
\includegraphics[scale=0.69]{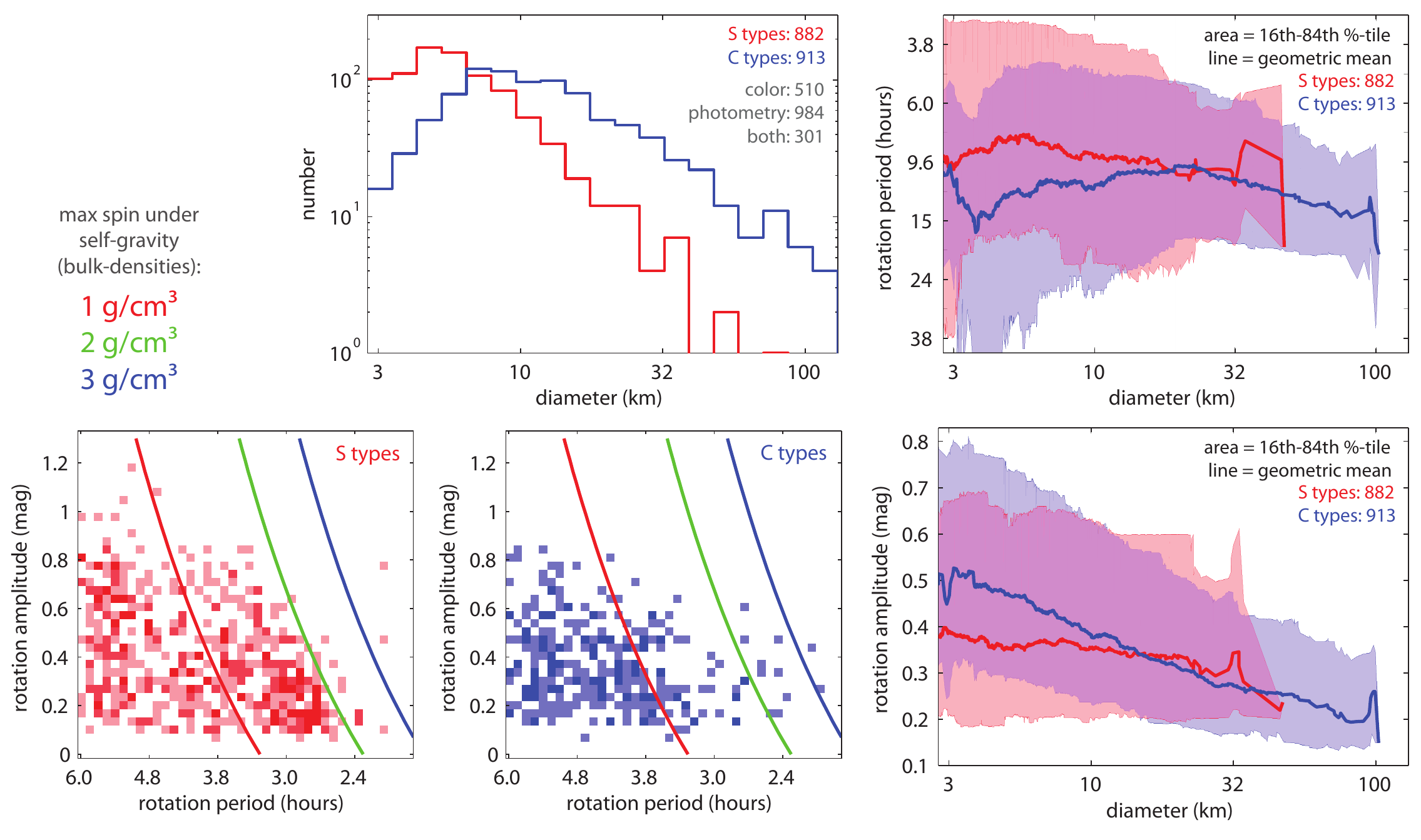}
\caption{Taxonomic dependence on spin rate and amplitude, also versus diameter, using the union of the color-index and photometric-index based C/S taxonomy.}
\end{figure*}

\begin{equation}
q(G_{12}) =\left\{
\begin{split}
&0.2707-0.236 G_{12}\;\;\;\text{if}\;G_{12}<0.2;\\
&0.2344-0.054 G_{12}\;\;\;\text{otherwise}.
\end{split}
\right.
\end{equation}

\subsubsection{Taxonomy from lightcurve data}

We use the distribution of bond albedo versus $G_{12}$ to define another taxonomic metric analogous to the color index. In particular, we apply the same clustering analysis to this distribution as we did for the seven 2D color distributions in the appendix. This procedure assigns to every object in the $A_\text{bond}$-vs.-$G_{12}$ diagram a probability of membership in each of two clusters (color coded blue and orange in Figure 19). The cluster centers are fit by the algorithm, and the output class probability of a given data point relates to its distance from these cluster centers. Probabilities near 0 represent likely C-type class membership, while probabilities near 1 represent likely S-type membership. We refer to this new metric as the \emph{photometric index}; it complements the color index as another proxy for taxonomy. There are 361 asteroids with both a photometric index and color index available (Figure 19 right plot); the two indices are clearly correlated ($\rho_\text{Spearman}=0.73$, $>$$10\sigma$ significance). Note that asteroids only have a defined photometric index if they have an infrared-derived diameter available, so that $A_\text{bond}$ is defined.

\subsubsection{Wavelength dependence}

Observational evidence for the reddening of asteroid colors with increasing phase angle is discussed by \cite{san12} and references therein. Color variation with phase angle can be equivalently stated as variation of the phase function with wavelength. Asteroids which have PTF lightcurves in both of the survey's filters ($R$ and $g$ band) allow us to investigate this phenomenon. We note however that \citep{san12} describe phase reddening as being more pronounced at longer wavelengths ($>$0.9 $\mu$m) and larger phase angles ($\alpha>30$ deg), such that \emph{a priori} we should not expect a very pronounced effect (if any) in the visible band PTF data.

Similar to the complication associated with comparing spin amplitudes from multiple oppositions (Section 3.1.1), an asteroid's mean color can potentially change if the spin axis varies with respect to our line-of-sight from year to year. Hence, we choose not to compare $R$-band and $g$-band phase-function fits from different oppositions. Aside from this constraint, we adopt the same two reliability selection criteria stated in Section 6.3, with a slight modification of requirement \#2: here we allow \emph{four} or more phase-angle bins of width $\Delta\alpha=3$ deg, as opposed to the previous sections' five-bin requirement, because of the small sample size.

There are 92 asteroids with both $R$-band and $g$-band phase-function fits acquired during the same opposition that meet the above criteria. For each asteroid we difference the $R$-band $G_{12}$ value from the $g$-band $G_{12}$ value. The mean of this difference is $-0.004_{-0.14}^{+0.19}$, indicating (for the whole sample) no significant non-zero difference between the two bands' $G_{12}$ values. Likewise, for $\beta$, we compute a difference of $0.002_{-0.003}^{+0.008}$, also consistent with zero difference between the bands.

Since these fits provide absolute magnitudes in each band ({\it i.e.}, $H_g$ and $H_R$) we compute the color $H_g-H_R$ for the 92-asteroid sample. Figure 20 shows that the distribution of this color is bimodal, suggesting it is a viable proxy for taxonomy. This is further supported by the strong correlation between $H_g-H_R$ and the $R$-band $G_{12}$ value. No correlation is seen however between $H_g-H_R$ and the difference between the two bands' $G_{12}$ value or $\beta$ values.

\subsection{Spins and amplitudes vs. taxonomy}

The union of the color-index data (see appendix) and photometric-index data (Section 6.3.1) provides significantly better taxonomic coverage of the PTF lightcurves (Figure 21). With this composite taxonomic information in hand, we can repeat the spin-amplitude analyses of Section 6.2 (Figure 17), this time considering the C-type and S-type groups separately. We define objects with one or both of the indices less than 0.25 as C type and greater than 0.75 as S type. We detail the resulting 1,795-object taxonomically-classified sample in Figure 22. There were 20 asteroids with conflicting color-based and photometric-based classifications that are not included in this 1,795-object sample.

The one-dimensional histogram in Figure 22 indicates that S-type asteroids dominate the smallest objects with data in PTF while C type dominate the largest. This reflects the fact that the survey's upper and lower sensitivity limits are defined in terms of absolute magnitude $H$ (affected by albedo) rather than physical diameter, {\it i.e.}, S-type asteroids larger than $\sim$50 km will tend to saturate the PTF detector, while C-type asteroids fainter than $\sim$5 km will usually fall below the detection limit. Adding to this effect is the fact that S-types mostly occupy the inner main-belt, where they are brighter by virtue of smaller heliocentric and geocentric distances, as compared to the usually more distant C types. While the two classes have similar representation in the sample (882 S types versus 913 C types), their true population ratio also affects the relative numbers.

\begin{figure}
\centering
\includegraphics[scale=0.78]{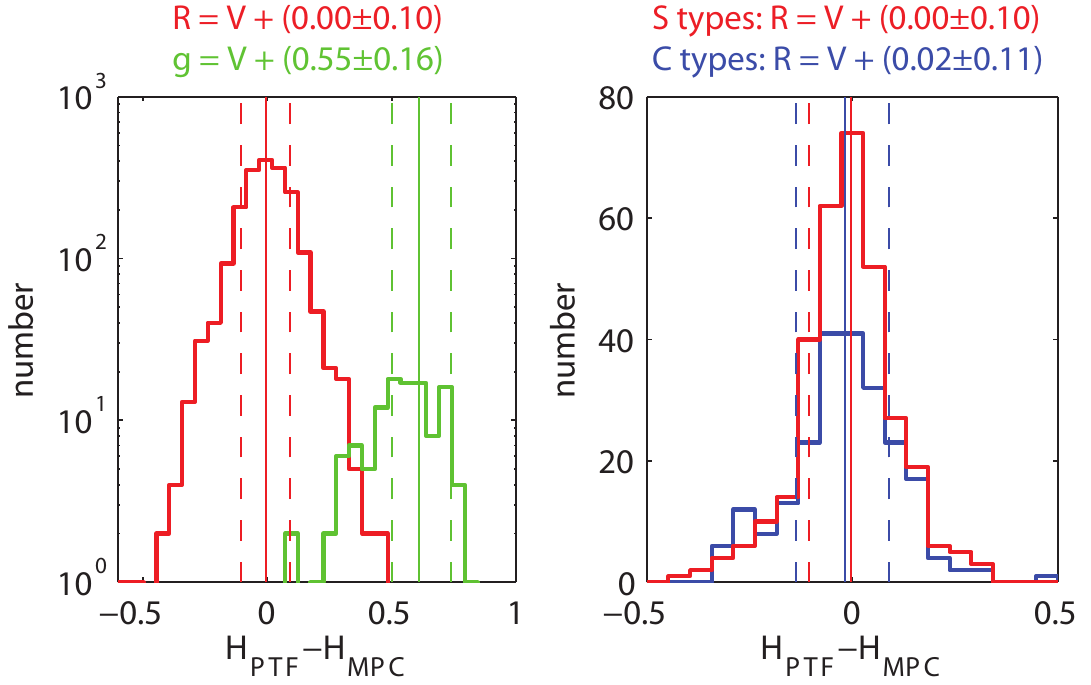}
\caption{\emph{Left}: Transformations between MPC $V$ band and the PTF $R$ and $g$ bands for asteroids, based on the difference between MPC-fitted and PTF-fitted $H$ magnitudes for asteroids whose PTF-fitted $G$ values are in the range $0.10<G<0.20$ as well as other PTF-coverage constraints (see text). \emph{Right}: $R$-band data only, with S and C types defined with either color and/or photometric indices (again using the $<$0.25 and $>$0.75 index thresholds).}
\end{figure}

The right-hand side plots in Figure 22 show rotation rate and amplitude versus diameter separately for the two taxonomic groups. Rather than plot a two-dimensional histogram as was done in Figure 17, for readability we here just plot the geometric mean and percentiles, computed by exactly the same running-bin method described in Section 6.2. The most prominent trend is that among $5\lesssim (D/\text{km}) \lesssim 20$ asteroids, C types appear to rotate slower than S-types \emph{and} have larger amplitudes than S types. Assuming both asteroid groups share the same mean angular momentum, the discrepancy could reflect the C types' ability to more efficiently redistribute material away from their spin axis, thereby increasing their moment of inertia (amplitude) while decreasing their angular rotation rate ({\it i.e.}, a simple manifestation of conservation of angular momentum).

The above-stated \emph{assumption} of a common mean angular momentum between C and S types is a merely a simple case and is neither unique nor rigorously motivated. More careful consideration of, {\it e.g.}, plausible ranges of internal tensile strengths of the two types could easily lead to more diverse scenarios wherein the two groups actually have different angular momenta and the observed spin-amplitude trends. As noted earlier (Section 6.2), large asteroids in general appear to have retained a significant primordial component in their spin distribution \citep{ste15}; it is therefore important that differences in the origin of C types and S types (accretionary, temporal and/or spatial) be taken into account along with differences in collisional evolution and differing contributions from radiative forces like YORP. Simulations of the main belt's origin, such as the Grand Tack family of models \citep{wal11}, should ultimately be modified to track particle spin evolution as well as orbits. 

\begin{figure}
\centering
\includegraphics[scale=0.57]{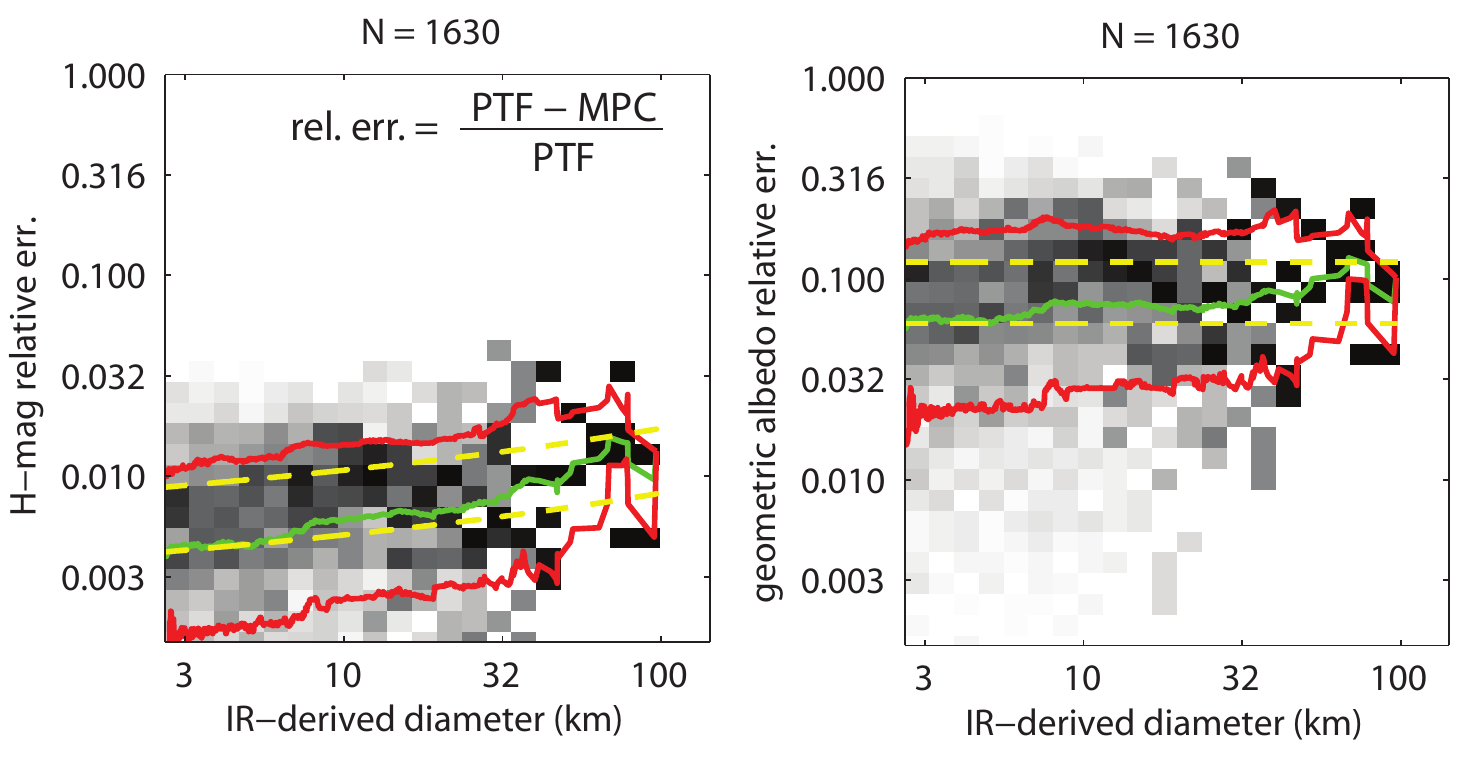}
\caption{Errors in the MPC-listed absolute magnitudes relative to the PTF $H$ values (in $R$ band and using the $G_{12}$ fit's $H$ value), only considering asteroids with IR-derived diameters. On the right is the corresponding geometric albedo relative error. Pixels in the 2D histograms shown here are column normalized. The running-bin geometric mean and 16th and 84th percentiles are shown as green and red lines. Yellow dashed lines are the mean and 84th percentile expected from the 0.1 mag transformation uncertainty alone (for 7\% geometric albedo).}
\vspace{10pt}
\end{figure}

We also reproduce the period-vs.-amplitude plot first shown in Figure 17, this time plotting separately the two taxonomic groups. The S types show a clearer cutoff at the 2 g/cm$^3$ contour line, suggesting they may in general be of greater bulk density than the C types, which show a softer boundary in this period-vs.-amplitude space, the precise location of which appears to be somewhere between 1 and 2 g/cm$^3$. Note that comparison to these density contours is only valid if the asteroids in consideration are held together mostly by self-gravity and approximated as fluids (as opposed to having significant internal cohesive or frictional resistance). These results are in general agreement with existing asteroid density estimates (\citealp{car12} and references therein). Results from an independent analysis of a smaller, more densely-sampled set of PTF asteroid lightcurves (Chang et al. in review; a study that follows closely the approach of \citealp{cha14a}) agree with the C type vs. S type rotation rate discrepancy discussed here.

\section{Comparison to MPC-generated magnitudes}

Absolute magnitudes available through the Minor Planet Center (MPC) and JPL Solar System Dynamics\footnote{\href{http://ssd.jpl.nasa.gov}{\color{blue}{http://ssd.jpl.nasa.gov}}} websites are fit using all available survey/observer-contributed photometry. These $H$ magnitudes are used in various online ephemeris tools to compute predicted $V$ magnitudes to accompany astrometric predictions. Their model assumes no rotational modulation, uses the Lumme-Bowell $G$-model (Section 3.2.2), and---with the exception of $\sim$100 large objects (nearly all with $D>30$ km)---assumes a constant $G=0.15$ for all asteroids. Our results (Figure 18 second row of plots) show that the $G=0.15$ approximation does indeed agree well with the peak of the distribution of fitted $G$ values. The PTF-fitted $G$ values obviously however show some spread and variation with taxonomy. In this section we explore the resulting differences in the absolute magnitudes $H$ and in predicted magnitudes.

\subsection{Filter transformations}

In order to compare the MPC-listed ($H_\text{MPC}$) magnitudes, which are in $V$ band, with PTF's absolute magnitudes ($H_\text{PTF}$, corresponding to the $G$-model fit) which are in either $R$ and $g$ bands, we must first compute an approximate transformation from $V$-band to each PTF band. While some transformations are given by \cite{ofe12a}, we here prefer to empirically estimate these using actual asteroid photometry from both PTF and the MPC, rather than generating them from the more general transformations of \cite{ofe12a}.

Figure 23 plots $H_\text{PTF} - H_\text{MPC}$ for asteroids whose PTF-derived $G_\text{PTF}$ is in the range $0.1<G_\text{PTF}<0.2$. By restricting the comparison to objects with fitted $G_\text{PTF}$ values close to 0.15, we in principle select $H_\text{MPC}$ magnitudes for which the MPC's $G_\text{MPC}=0.15$ assumption is actually valid (none of the asteroids in Figure 23 have MPC-listed $G$ values other than the default 0.15). Furthermore, we only consider (in Figure 23) asteroids with PTF data in at least three phase angle bins of $\Delta\alpha=3$ deg \emph{and} either a reliable period or fitted amplitude less than 0.1 mag. 

Comparing the $H_\text{MPC}$  and $H_\text{PTF}$ magnitudes for this specific subset of asteroids, we obtain approximate transformations $R=V+(0.00\pm0.10)$ and $g=V+(0.55\pm0.16)$. The $1\sigma$ uncertainties of 0.10 and 0.16 mag plausibly include a combination of the photometric calibration uncertainties of both the MPC data (coming from a variety of surveys/observers), variation in $H$ magnitude of a given asteroid between different oppositions (the MPC fits combine data possibly acquired at different viewing geometries), as well as the range of $G_\text{PTF}$ used in selecting the asteroids in this sample. Consideration of a range of $G_\text{PTF}$ values is equivalent to considering a range of asteroid colors (cf. the color-vs.-$G$ correlation seen in Figure 18). Hence the uncertainties in these transformations also encompass the variation which might otherwise be formally fit in a color term for the transformations. Such a color term for $R$ to $V$ would almost certainly be less significant than that of $g$ to $V$, as the former transformation is already zero within uncertainties. The larger uncertainty in the $g$ to $V$ transformation is likely attributable to both the smaller sample size and the fact that the $V$ bandcenter is further displaced from $g$ than from $R$, such that color variation has a more pronounced effect.

\begin{figure}
\centering
\includegraphics[scale=0.69]{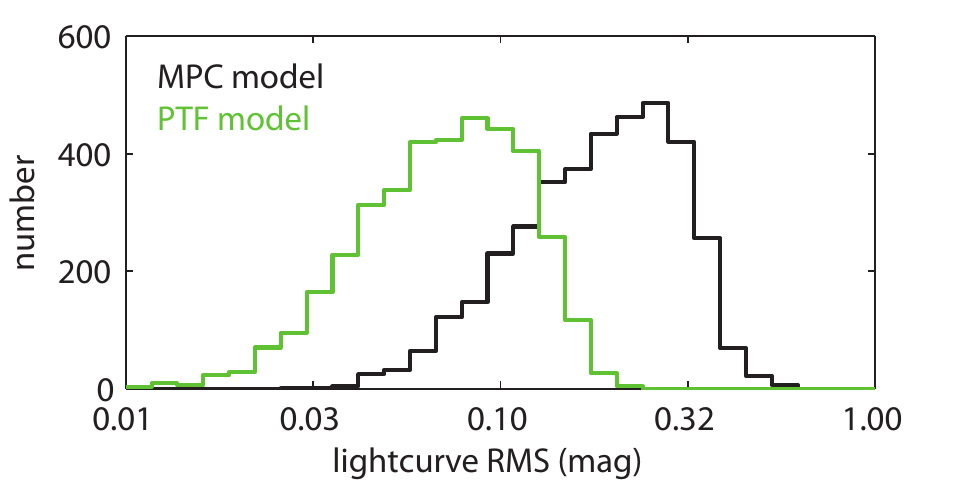}
\caption{Comparison of the root-mean-square residuals, with respect to the PTF ($H,G_{12}$) plus rotation fit and the MPC ($H,G$) fit, for all lightcurves having a reliable $R$-band PTF phase-function fit.}
\vspace{10pt}
\end{figure}

Given the above-computed transformations, and the fact that 89\% of our fitted lightcurves are in $R$ band, we proceed using only $R$-band lightcurve fits, which we compare directly against MPC magnitudes (or formally, after applying the transformation of zero). A detail of the color dependence of the $R$ to $V$ transformation appears in the right plot of Figure 23; the mean transformation differs slightly between S and C types but not at a level comparable to the uncertainty in either.

\subsection{Absolute magnitudes}

In Figure 24 we show the relative error in the MPC absolute magnitudes as compared to the PTF magnitudes, for all 1,630 lightcurves with sufficient phase angle coverage in PTF (with the five-bin phase-angle criterion). These errors should reflect not only any discrepancy due the different phase function models (PTF's $G_{12}$ versus MPC's $G$), but also variation in absolute photometric calibrations (within the MPC data internally and/or between the MPC and PTF data sets). The 0.1-mag uncertainty in the $R$ to $V$ band transformation has a prominent contribution to the errors shown here (the mean and 84th percentile of the errors expected from the 0.1-mag transformation uncertainty alone are shown as yellow dashed lines, and assume $p_V=0.07$). The green line (computed mean) and upper red line (84th percentile) indicate the errors are $\sim$1\% greater than those expected from the transformation uncertainty alone, though this increases slightly for the largest ($D>30$ km) objects. Note that many of these largest asteroids are more frequently observed by programs other than the major sky surveys; these smaller facilities tend to use smaller aperture telescopes and different absolute calibration standards, which would contribute to the error.

\subsection{Predicted apparent magnitudes}

Instead of comparing just the fitted $H$ magnitudes, for every lightcurve with a reliable PTF-fitted phase function we also compare the root-mean-square residual of all PTF data in that lightcurve with respect to both our $G_{12}$-fit-predicted $R$ magnitude and the MPC ($G=0.15$) predicted $V$ magnitude. Our fit includes more fitted parameters and obviously should result in smaller residuals; Figure 25 shows that we see a factor $\sim$3 smaller residuals in particular using the PTF fit. Note that if the 0.1-mag $R$-to-$V$ transformation uncertainty were the only significant contributor to the MPC residuals then their peak would instead be at $\sim$0.07 mag rather than $\sim$0.25 mag. Ignored rotational modulation and inaccurate phase functions move the MPC residuals distribution to higher RMS values.

The lower RMS residuals afforded by the PTF lightcurve model permit a more sensitive search for low-level transient activity ({\it e.g.}, collisional events, cometary brightening) in these asteroids. For example, \cite{cik14} perform a search for active main-belt asteroids using photometric residuals of all MPC data taken with respect to the MPC-predicted apparent $V$ magnitudes. We currently are pursuing a similar analysis using these PTF lightcurves, as a follow-up to the morphology-based search already completed with PTF \citep{was13}. A hybrid approach, wherein morphological measurements are made on stacked images of asteroids which have reliable lightcurve fits, could further reveal this kind of subtle activity.

\section{Summary}

From five years of PTF survey data we have extracted over 4 million serendipitous detections of asteroids with known orbits. We fit a photometric model to $\sim$54,000 lightcurves, each consisting of at least 20 observations acquired within a given opposition in a single filter. We adopt a second order (four-term) Fourier series for the rotation component and fit three distinct phase-function models. We assess the reliability of our retrieved rotation periods by subjecting them to both an automated classifier and manual review. Both vetting processes are trained on a sample of $\sim$800 asteroids with previously measured spin periods that also occur in our sample. We consider the intersection of the two screened samples for subsequent analysis.

Preliminary analysis (on distributions that are not de-biased) of the rotation period versus diameter confirms the previous finding that asteroids smaller than $\sim$ 40 km do not conform to a Maxwellian distribution in their normalized spin frequencies. Phase-function parameters are shown to correlate strongly with the bond albedo. None of the phase function parameters display bimodality in their measured distributions however. Together with the bond albedo, we use the phase function data to define a new taxonomic metric based solely on single-band lightcurve properties together with infrared-derived diameters ($G_{12}$ and $A_\text{bond}$). This metric complements the color-based index established previously by many visible-color and spectroscopic surveys. Combining these color- and photometry-based taxonomic indices allows us to separately examine the spin and amplitude distributions of the C-type and S-type asteroids with the largest possible sample sizes. Doing so reveals that, among small objects (5 km $<D<$ 20 km) the C types show larger amplitudes and slower spin rates. If the two populations shared a common angular momentum distribution, this could be interpreted as the two compositional types' differing tendencies to redistribute mass away from their spin axes. Comparison of the spin-amplitude distribution with contours of maximal spin rates for cohesionless bodies suggests that almost all asteroids are less dense than $\sim$2 g/cm$^3$, with C types displaying a potentially less dense upper limit of between 1--2 g/cm$^3$.

\begin{table*}
\caption{Asteroid colorimetry data sets used in computing this work's C/S color metric. These data sets are visualized in Figure 26.}
\begin{tabular}{llll}
\hline
survey name & references & data description & \# asteroids\\
\hline\hline
\multirow{2}{*}{\emph{UBV} colors} & \cite{bow78} & $U$, $B$, and $V$ broadband photometry acquired mostly at     &  \multirow{2}{*}{902} \\
                                   & \cite{ted95}  &  Lowell Observatory in the 1970s with photomultiplier tubes.        &                       \\    
\hline
\multirow{2}{*}{Eight-Color Asteroid}                  & \multirow{2}{*}{\cite{zel85}} &Photometry in eight custom filters measured with photomultipliers at Catalina and           &        \\
\multirow{2}{*}{Survey (ECAS)}              & \multirow{2}{*}{\cite{zel09}} & Steward Observatories. We compute and use the principal component color index               &    480                  \\    
                         &              & PC\#$1 = 0.771(b-v)-0.637(v-w)$. Excludes objects with PC\#1 error $>$0.3 mag.              &    \\
\hline
24-Color                     & \cite{cha79} &  Photometry in 24 interference filters measured with photomultipliers at Mauna Kea.         &   \multirow{2}{*}{262}     \\
Asteroid Survey              & \cite{cha93} &  We compute and use the mean spectral reflectance slope and first principal component.      &                     \\
\hline
Small Main-belt                   & \multirow{2}{*}{\cite{xu95}} &  \multirow{2}{*}{CCD spectroscopy (0.4--1.0$\mu$m, $R\approx 100$) conducted mostly at Kitt Peak.}       &        \\
Asteroid Spectroscopic            & \multirow{2}{*}{\cite{xu96}} &  \multirow{2}{*}{We compute and use the mean spectral reflectance slope and first principal component.}  &    305                  \\    
Survey (SMASS)                        &     &                    &    \\
\hline
Small Main-belt                   & \multirow{2}{*}{\cite{busb02}} &  \multirow{2}{*}{CCD spectroscopy (0.4--1.0$\mu$m, $R\approx 100$) conducted at Kitt Peak.}     &        \\
Asteroid Spectroscopic            & \multirow{2}{*}{\cite{bus03}} & \multirow{2}{*}{We compute and use the mean spectral reflectance slope and first principal component.}  &    1,313                  \\    
Survey II (SMASS-2)                        &     &                    &    \\
\hline
Small Solar System                   &               &  \multirow{2}{*}{CCD spectroscopy (0.5--9.0$\mu$m, $R\approx 500$) conducted at ESO (La Silla).}            &        \\
Objects Spectroscopic               & \cite{laz04} & \multirow{2}{*}{We compute and use the mean spectral reflectance slope and first principal component.}                   &    730                  \\    
Survey (S3OS2)                        &     &                    &    \\
\hline
\multirow{2}{*}{Sloan Digital Sky}      & \multirow{2}{*}{\cite{ive02}}       &  $g$,$r$,$i$, and $z$ broadband CCD photometry acquired by SDSS from 1998--2009.           &       \\
\multirow{2}{*}{Survey (SDSS)}          & \multirow{2}{*}{\cite{par08}}       &  Includes data in the Moving Object Catalog v4, supplemented with post-2007 detections         &    \multirow{2}{*}{30,518} \\
\multirow{2}{*}{$griz$ colors}          & \multirow{2}{*}{\cite{ive10}}       &  from SDSS DR10. We use the first principal component $a^*$ defined in the references.             & \\
                       &                    &  Excludes objects with $a^*$ error $>$0.05 mag or ($i-z$) error $>$0.1 mag.                          & \\
\hline
\end{tabular}
\vspace{5pt}
\end{table*}

Finally, our fitted absolute magnitudes differ from those generated by the Minor Planet Center's automated fitting procedures, though the precise discrepancy is difficult to ascertain given uncertainty in the transformation between PTF $R$-band and the MPC's $V$-band. The utility in using our model to \emph{predict} asteroid apparent magnitudes is seen in the three-fold reduction in RMS scatter about our model relative to the fiducial $G=0.15$ model that neglects rotation. This reduced scatter is an essential prerequisite for sensitive searches for cometary, collisional, and other transient activity in what would otherwise be regarded as quiescent asteroids---potentially even bright objects.

\section*{Acknowledgements}

This work uses data obtained with the 1.2-m Samuel Oschin Telescope at Palomar Observatory as part of the \emph{Palomar Transient Factory} (PTF), a scientific collaboration between the California Institute of Technology (Caltech), Columbia University, Las Cumbres Observatory Global Telescope Network, Lawrence Berkeley National Laboratory, the National Energy Research Scientific Computing Center, the University of Oxford, and the Weizmann Institute of Science (WIS).

Some data in this work (also from the 1.2-m Oschin Telescope) were obtained as part of the \emph{Intermediate Palomar Transient Factory} (iPTF) project, a collaboration between collaboration between Caltech, the Kavli Institute for the Physics and Mathematics of the Universe, Los Alamos National Laboratory, the Oskar Klein Centre, the University System of Taiwan, the University of Wisconson Milwaukee, and WIS.

A. Waszczak has been supported in part by the W.M. Keck Institute for Space Studies (KISS) at Caltech. E.O.O. is incumbent of the Arye Dissentshik career development chair and is grateful to support by grants from the Willner Family Leadership Institute Ilan Gluzman (Secaucus NJ), Israeli Ministry of Science, Israel Science Foundation, Minerva and the I-CORE Program of the Planning and Budgeting Committee and The Israel Science Foundation.

This work also makes use of data products from the \emph{Wide-Field Infrared Survey Explorer}, a joint project of the University of California Los Angeles and the Jet Propulsion Laboratory (JPL)/Caltech, funded by NASA. This work also makes use of data from \emph{NEOWISE}, which is a project of JPL/Caltech, funded by the Planetary Science Division of NASA. 

This work also makes use of data from the Sloan Digital Sky Survey (SDSS), managed by the Astrophysical Research Consortium for the Participating Institutions and funded by the Alfred P. Sloan Foundation, the Participating Institutions, the National Science Foundation, the US Department of Energy, NASA, the Japanese Monbukagakusho, the Max Planck Society, and the Higher Education Council for England.

Lastly, we thank an anonymous reviewer for helpful comments and feedback.

\newpage
\section{Appendix}

\subsection{Multi-survey visible-band color index}

The purpose of this appendix section is to introduce a one-dimensional color metric, based upon data from seven different colorimetric asteroid surveys, which quantifies an asteroid's first-order visible-band color-based taxonomy as a number between 0 (C-type endmember) and 1 (S-type endmember). Our primary motivation for doing this is to enable a uniform comparison of PTF-lightcurve-derived parameters with color spanning from the brightest/largest objects ($H\approx 8$--$9$ mag, or $D\approx125$--$80$ km diameters) down to PTF's detection limit for main-belt asteroids ($H\approx16$ mag, or $D\approx2$--$4$ km). Figure 26 panel A shows that the fraction of PTF lightcurves with color information increases by a factor of $\sim$3 among large asteroids when all seven surveys are considered, whereas for smaller objects the Sloan Digital Sky Survey's (SDSS; \citealp{yor00}; \citealp{ive02}; \citealp{par08}) moving-object catalog provides essentially all of the color information.

\begin{figure*}
\centering
\includegraphics[scale=0.59]{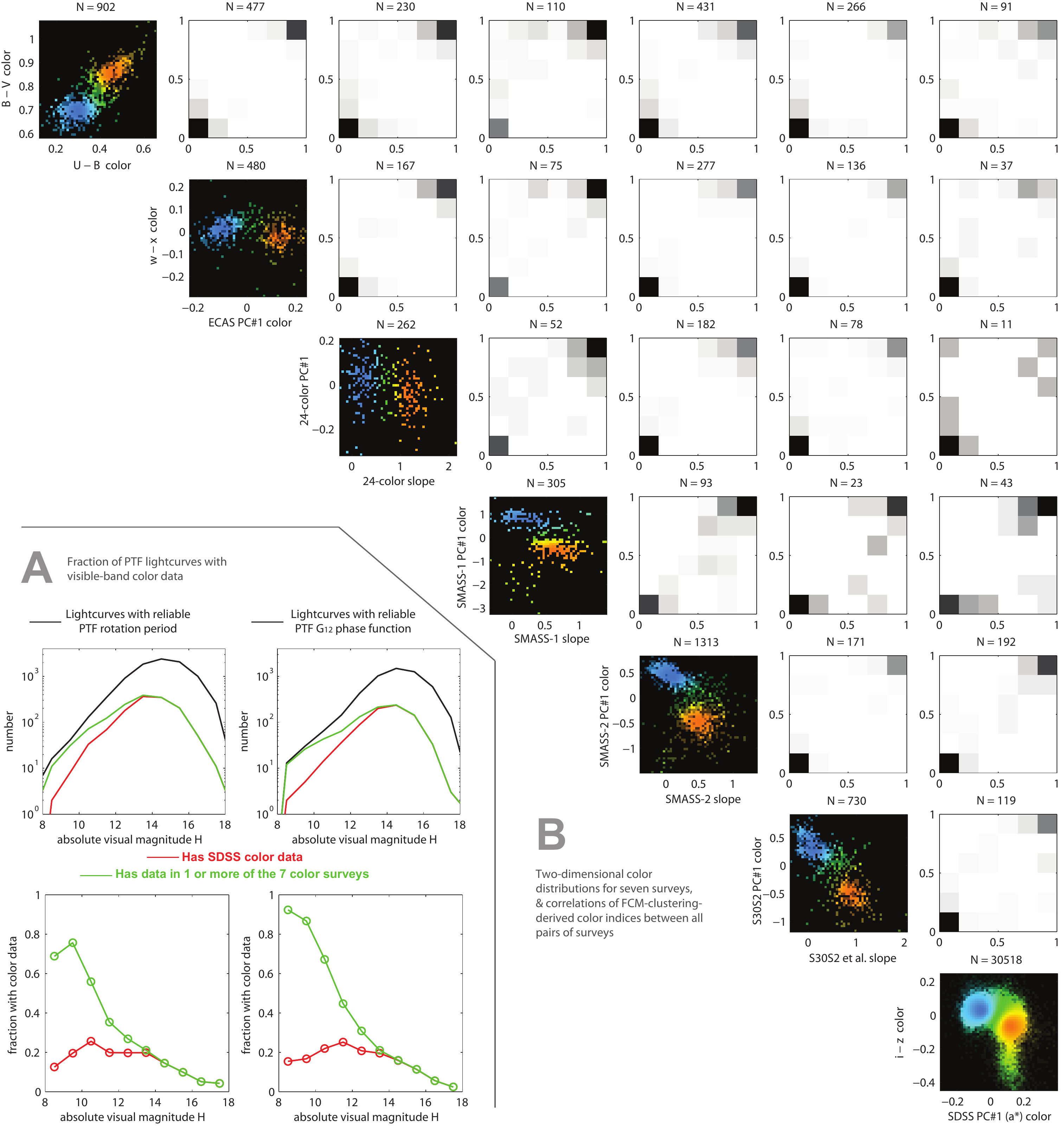}
\caption{\emph{Panel A}: Fraction of PTF lightcurves with colorimetric data available, for both the reliable-period and reliable-period-plus-$G_{12}$ sets of lightcurves. \emph{Panel B}: Two-dimensional color distributions for seven surveys, and correlations of FCM-clustering-derived classifications between all pairs of surveys.}
\vspace{5pt}
\end{figure*}

\begin{figure*}
\centering
\includegraphics[scale=0.72]{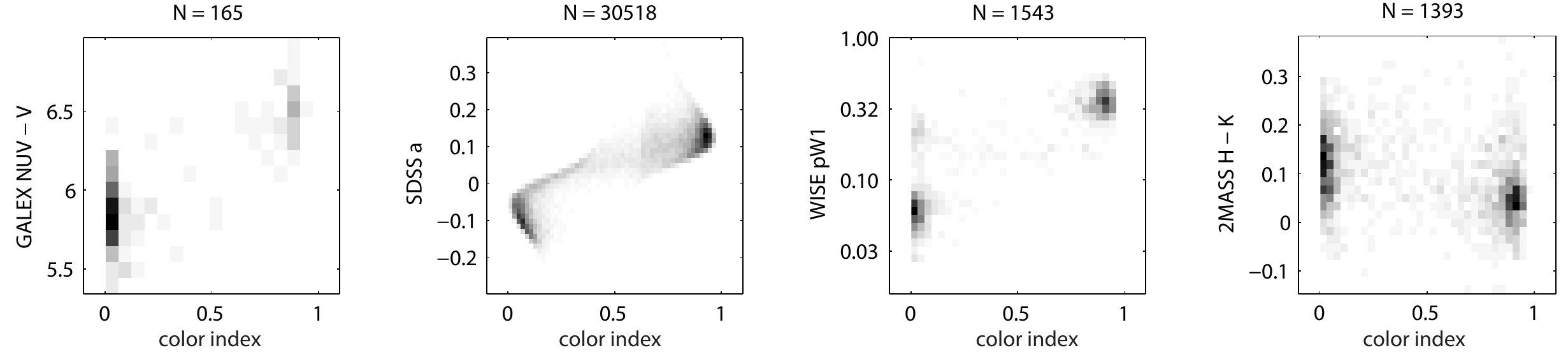}
\caption{Relationship between various asteroid surface measurements (from the UV to near-IR) and this work's visible-color-derived C/S color index. See text for descriptions of data the data sets used here, and accompanying references.}
\vspace{15pt}
\end{figure*}

\begin{figure}
\centering
\includegraphics[scale=0.85]{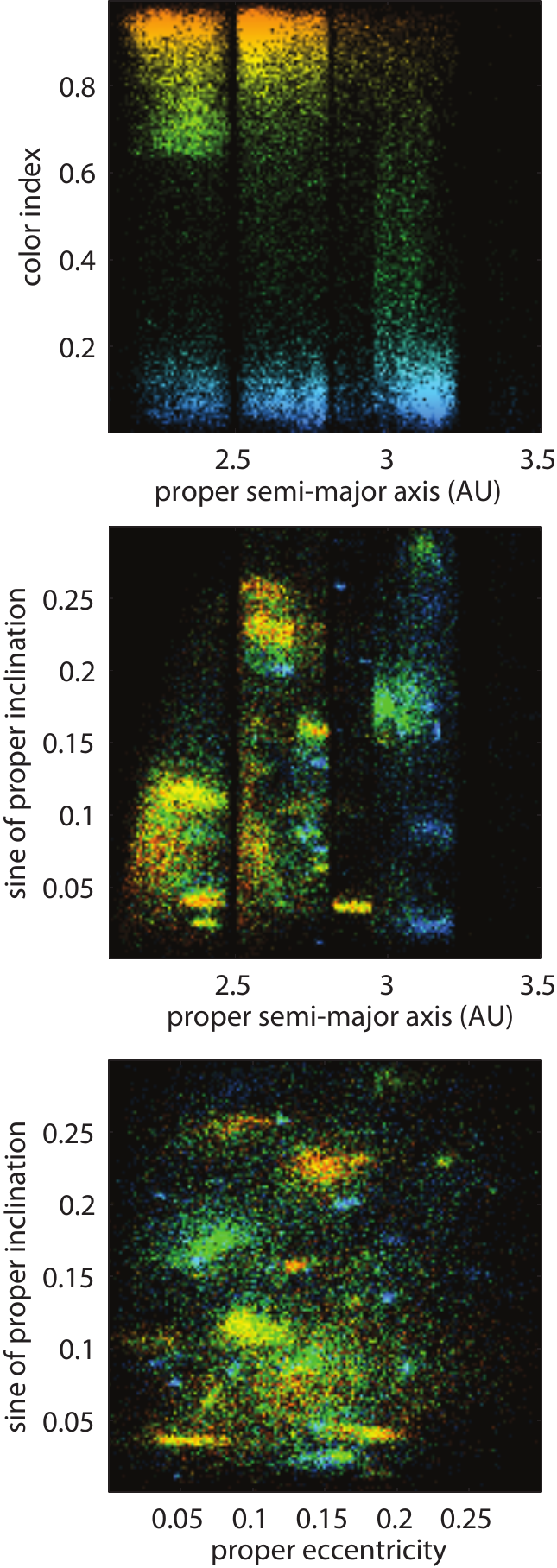}
\caption{Relationship between proper orbital elements and this work's visible-color-derived C/S color index for 30,508 asteroids.}
\end{figure}

The seven surveys we use are described in Table 3. All of these surveys contain at least two independent color measurements, and when plotting their data in these two dimensional spaces (or 2D subspaces defined by properly-chosen principal components or spectral slope parameters), the first-order C-type and S-type clusters are in all cases prominently seen (Figure 26 panel B). To each such 2D color distribution we apply a two-dimensional \emph{fuzzy c-means} (FCM) clustering algorithm (\citealp{bez81}; \citealp{chi94}). For each survey data set, FCM iteratively solves for a specified number of cluster centers (in our case, two) in $N$ dimensions (in our case one dimension) by minimizing an objective function which adaptively weights each datum according to the robustness of its membership in a given cluster. The FCM output includes computed cluster centers and, for each datum, the probability that the datum belongs to each cluster (this being related to the datum's distance from each cluster center).

In the color-distribution plots of Figure 26 panel B (the plots with black backgrounds arranged diagonally), each pixel/bin is colorized according to the average cluster-membership probability of asteroids in that pixel. Blue indicates high probability of membership in cluster 1 while orange represents high probability of membership in cluster 2.

Our color index provides a more quantitative label than that offered by popular letter-based taxonomic systems ({\it e.g.}, \citealp{bus02} and refs. therein). Several such letter-based nomenclatures were in fact defined on the basis of one or more of these seven surveys, oftentimes using a method similar to the clustering technique we use here. We identify our blue cluster with C-type asteroids and our orange cluster with S-type asteroids, though we make this association purely for connection/compatibility with the literature. This is because our computed clusters have their own unique identity/definition, formally distinct from that given in any other work. Our clusters' definitions are nonetheless completely specified/reproducible by the FCM algorithm we used to compute them.

In reducing the taxonomic classification to a single number defined by the two most prominent groups (C and S types), we lose the ability to distinguish secondary classes like V types, D types, and so on. If such a sub-group is separated from both of the two main clusters, its members will be assigned membership scores of close to 0.5. For example, in the SDSS $a^*$ vs. $i-z$ complex, the clearly-seen V-type `tail' protruding down from the S-type cluster appears mostly green in color, reflecting its intermediate classification. Likewise for the less-clearly seen D types, which in the SDSS plot lie above the S types and to the right of the C types (again in a green-colored region). The orders of magnitude lower numbers of such secondary types make them mostly irrelevant for the purpose of this analysis.

We compute the numerical uncertainty (variance) of a given asteroid's cluster-membership score in a particular survey by performing many bootstrapped trials wherein we first randomly perturb all data points by random numbers drawn from Gaussian distributions whose width are the quoted $1\sigma$ measurement ({\it i.e.}, photometric) uncertainties in each of the two dimensions, and then repeat the FCM analysis on the perturbed data. The variance in each object's reported cluster probability is then computed after a large number of bootstrap trials.

Some asteroids appear in only one of the seven surveys; for such objects the color index is simply its cluster-membership score in that particular survey. For asteroids appearing in multiple surveys, we take the variance-weighted average of the multiple membership scores (and compute that composite score's variance by summing the component variances in inverse quadrature, as usual).

The many off-diagonal plots in Figure 26 panel B compare the cluster-membership scores of all asteroids appearing in all possible survey intersections. The number of asteroids in each survey (and in the intersection of each survey pair) appears above each plot ($N=\ldots$). The survey-pair distributions are 2D-histograms where higher densities of data points correspond to black pixels/bins and low density or lack of data points is white. Evidently all possible survey combinations contain at least some asteroids (several share hundreds), and in all cases the individual taxonomic indices (on the horizontal and vertical axes) correlate strongly, confirming the consistency of the cluster membership between surveys.

In Figure 27 we illustrate some useful applications of this color index by comparing it with various asteroid surface observations. One of these quantities (SDSS $a^*$ color) was used in computing the color index, so its correlation with the clustering index is expected and thus confirmed.

\begin{figure}
\centering
\includegraphics[scale=0.59]{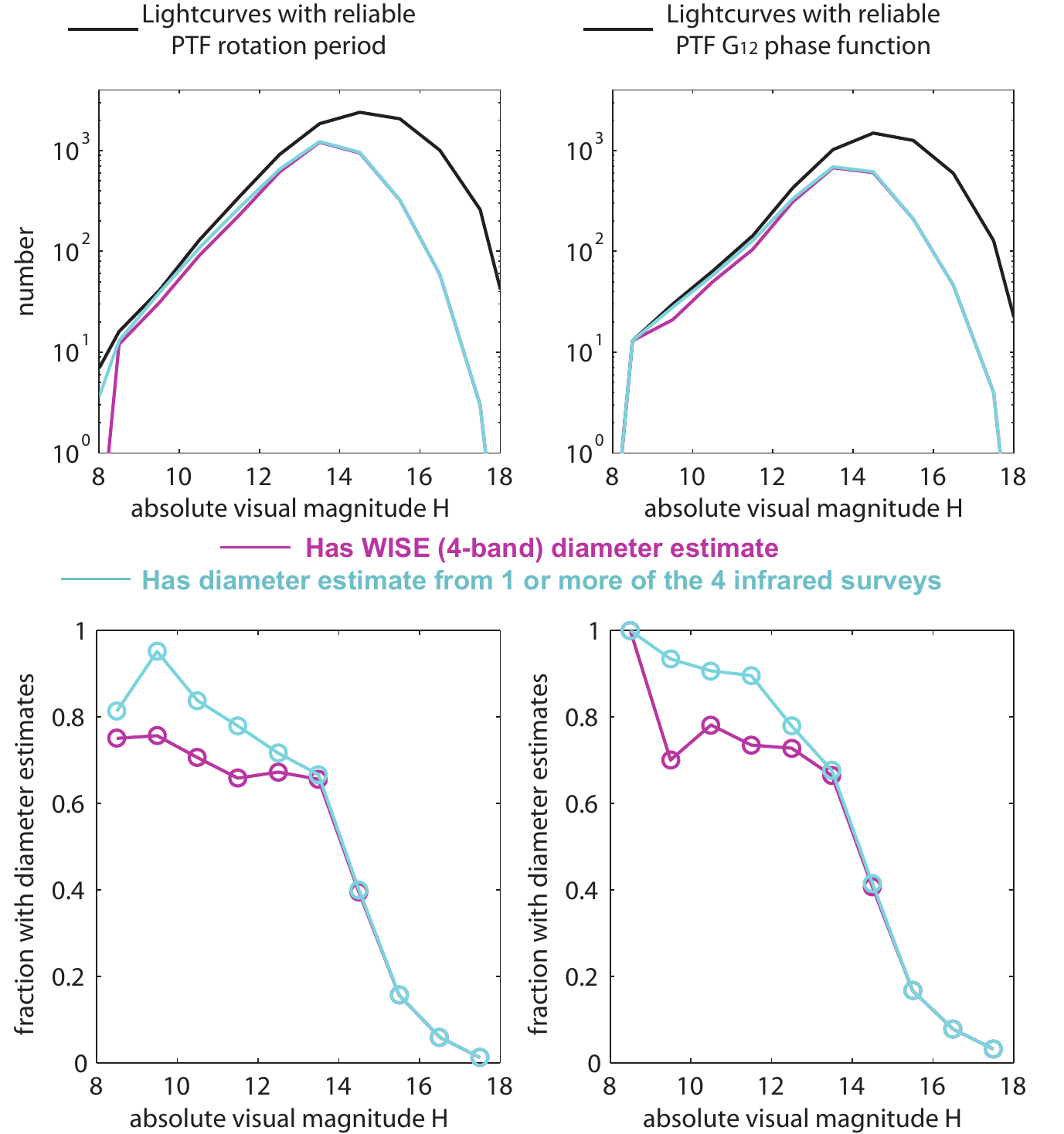}
\caption{Fraction of PTF lightcurves with thermal-IR-based diameter estimates available, for both the reliable-period and reliable-$G_{12}$ sets of lightcurves.}
\vspace{10pt}
\end{figure}

In the leftmost plot of Figure 27, asteroid photometry from \emph{GALEX}\footnote{The {\it Galaxy Evolution Explorer} ({\it GALEX}) is a NASA Small Explorer-class space telescope which from 2003--2012 conducted an imaging survey in a far-UV band (FUV, 130--190 nm) and a near-UV band (NUV, 180--280 nm). \cite{mar05} discuss the extragalactic science program; Morissey et al. (2005, 2007) discuss the on-orbit performance, survey calibration and data products. The Waszczak et al (in prep) NUV data shown here are derived from data available at \href{http://galex.stsci.edu}{\color{blue}{http://galex.stsci.edu}}.} (NUV band), compiled by Waszczak et al. (in prep), is normalized by the nominal $G=0.15$ phase-model (Section 3.2.2) predicted brightness at the time of the \emph{GALEX} observations, and the resulting NUV $-$ $V$ color evidently correlates with the visible color index. This indicates that asteroid reflectance slopes in the visible persist into the UV. 

Figure 27 also plots our color index against the $W1$-band geometric albedo derived from \emph{WISE}\footnote{The {\it Wide-field Infrared Survey Explorer} ({\it WISE}) is a NASA Medium Explorer-class space telescope which in 2010 conducted a cryogenic IR imaging survey in four bands: $W1,W2,W3$, and $W4$, centered at 3.4, 4.6, 12, and 22$\mu$m, respectively. \cite{wri10} details mission/performance; \cite{mas11} and refs. therein present preliminary asteroid data.} observations obtained during its fully cryogenic mission. We only include asteroids which were detected in both of the thermal bands ($W3$ and $W4$) and which therefore have a reliable diameter estimate. Use of this diameter in Equation (15) then permits estimation of the albedo, where the $W1$-band albedos uses the corresponding \emph{WISE} photometry ($H$ in Equation [15] being replaced with the appropriate $W1$-band absolute magnitude).

The rightmost plot in Figure 27 shows our color index's relationship to a near-infrared color from the ground-based 2MASS survey \citep{skr06}. Serendipitous asteroid detections were extracted from 2MASS by Sykes (\citeyear{syk00}, \citeyear{syk10}) and include fluxes in $J$ band (1.25 $\mu$m), $H$ band (1.65 $\mu$m---\emph{not to be confused with the absolute visible magnitude $H$, used elsewhere in this work}), and $K$ band (2.17 $\mu$m). 

Figure 28 plots our color index against proper orbital elements retrieved from the Asteroids Dynamic Site (AstDyS; \citealp{kne12}), revealing the distinct colors of dynamical families and the overall transition from S to C types with increasing semi-major axis. These are similar to the plots of \cite{par08}, which is not surprising given that the majority of the asteroids' color indices are based on SDSS data alone. Of the 32,5023 asteroids with a defined color index, there are 30,508 with proper orbital elements which are represented in Figure 28.

\subsection{Compilation of IR-derived diameters}

Similar to how we combined several surveys' colorimetric data in the previous section, here we compile thermal-infrared-derived diameter estimates from four surveys. Our aim is again to provide the largest possible sample for comparison with PTF-derived lightcurve data. Just as SDSS is the main contributor of colorimetry overall but suffers from incompleteness for large/bright asteroids, analogously \emph{WISE} provides the vast majority of IR-based diameter measurements but levels off at $\sim$80\% completeness at the bright end (Figure 29). We thus supplement \emph{WISE} with diameter data from the \emph{Infrared Astronomical Satellite} (\emph{IRAS}; \citealp{mat86}, \citealp{ted02a}), the \emph{Mid-Course Space Experiment} (\emph{MSX}; \citealp{ted02b}), and AKARI \citep{usu11}. \cite{usu14} compares several of these different data sets in terms of coverage and accuracy. As we did when defining the color index, asteroids occurring in multiple IR surveys are assigned the variance-weighted average diameter.

Regarding the WISE data in particular, we again use only those diameters which resulted from a thermal fit constrained by fluxes in all four WISE bands during the cryogenic mission. Furthermore, we use the latest (revised) diameter estimates published by \cite{mas14}, which adopted an improved thermal modeling technique first discussed by \cite{gra12}.

\subsection{Lightcurve data tables}

The online version of this article includes two electronic tables containing the derived lightcurve parameters and the individual photometric observations in each lightcurve. Tables 4 and 5 describe the columns and formatting of these tables, which include data on all reliable-period lightcurves as well as those having amplitudes less than 0.1 mag and sampling in five or more 3-deg-wide phase-angle bins (which have reliable $G_{12}$ fits). Using these tables one can produce plots of the PTF lightcurves we have analyzed in this work.

\begin{table*}
\centering
\caption{ Parameters describing PTF lightcurves with a reliable period or phase function. Byte-by-byte Description of file: \texttt{ptf\_asteroid\_lc\_parameters.txt}}
\begin{tabular}{lllll}
\hline
\hline
   Bytes & Format & Units & Label & Explanations\\
\hline
   1-  4& I4    & ---   &         &Lightcurve ID number$^1$\\
   6- 11& I6    & ---   &    &Asteroid number (IAU designation)\\
  13- 14& I2    & yr    &    &Last two digits of opposition year\\
      16& I1    & ---   &    &Photometric band: 1 = Gunn-$g$, 2 = Mould-$R$\\
  18- 20& I3    & ---   &    &Number of observations in the lightcurve\\
  22- 26& F5.2  & mag   &    &Median apparent magnitude\\
  28- 37& F10.5 & day   & $t_\text{min}$   &Time (MJD) of first observation\\
  39- 48& F10.5 & day   & $t_\text{max}$     &Time (MJD) of final observation\\
  50- 54& F5.2  & deg   &  $\alpha_\text{min}$   &Minimum-observed phase angle\\
  56- 60& F5.2  & deg   &  $\alpha_\text{max}$   &Maximum-observed phase angle\\
  62- 63& I2    & ---   &     &Number of sampled phase-angle bins of 3-deg width\\
  65- 68& F4.2  & ---   & $p$  &Reliability score from machine classifier: 0=bad, 1=good \\
      70& I1    & ---   &   &Manually-assigned reliability flag: 0=bad, 1=good\\
      72& I1    & ---   &   &Period reliability flag: 0=bad, 1=good (product of two previous columns)\\
  74- 79& F6.3  & mag   &  $H$       &Absolute magnitude from $G_{12}$ fit\\
  81- 85& F5.3  & mag   &       &Uncertainty in absolute magnitude from $G_{12}$ fit\\
  87- 91& F5.3  & ---   &  $G_{12}$     &Phase-function parameter $G_{12}$\\
  92- 98& F6.3  & ---   &     &Uncertainty in $G_{12}$$^2$\\
 100-105& F6.3  & ---   &  $G$       &Phase-function parameter $G$\\
 107-113& F7.4  &mag/deg&  $\beta$    &Phase-function parameter $\beta$\\
 115-119& F6.3  & mag   &  $C$       &Phase-function parameter $C$\\
 121-124& F4.2  & mag   &       &Amplitude from $G_{12}$ fit (max $-$ min)\\
 126-134& F9.4  & hr    &  $P$  &Period from $G_{12}$ fit\\
 136-144& F9.4  & hr    &    &Period uncertainty from $G_{12}$ fi\\
 146-152& F7.4  & mag   &  $A_{11}$     &Fourier coefficient $A_{1,1}$ from G12 fit\\
 154-160& F7.4  & mag   &  $A_{12}$     &Fourier coefficient $A_{1,2}$ from G12 fit\\
 162-168& F7.4  & mag   &  $A_{21}$     &Fourier coefficient $A_{2,1}$ from G12 fit\\
 170-176& F7.4  & mag   &  $A_{22}$     &Fourier coefficient $A_{2,2}$ from G12 fit\\
 178-181& F4.2  & ---   &    &Ratio of the two peak heights in folded rotation curve$^3$\\
 183-186& F4.2  & ---   &  $\chi^2_\text{red}$    &Reduced chi-squared of the fit\\
 188-192& F5.3  & mag   &    &"Cosmic error" (see Section 4.1)\\
 194-198& F5.3  & mag   &     &Root-mean-square residual of observations w.r.t the fit\\
 200-206& F7.3  & hr    &    &Reference period (from http://sbn.psi.edu/pds/resource/lc)\\
 208-213& F6.2  & km    &  $D$       &Diameter derived from thermal IR data$^4$\\
 215-218& F4.2  & km    &     &Uncertainty in diameter\\
 220-224& F5.3  & ---   &  $A_\text{bond}$   &Bond albedo$^5$\\
 226-231& F6.4  & ---   &  &Uncertainty in bond albedo\\
 233-236& F4.2  & ---   &    &Color-based taxonomic index: 0=C-type, 1=S-type\\
 238-241& F4.2  & ---   &    &Photometry-based taxonomic index: 0=C-type, 1=S-type\\
\hline
\end{tabular}
\smallskip
\\$^1$ID number labels individual observations in Table 5.
\\$^2$Set to $-1$ if larger than the interval tested in grid search
\\$^3$Set to 0 if there is only one maximum in the folded lightcurve
\\$^4$References for the IR diameters are given in the text (appendix)
\\$^5$Bond albedo only computed for objects with reliable $G_{12}$ and available diameter
\bigskip
\end{table*}

\begin{table*}
\centering
\caption{ Parameters describing PTF lightcurves with a reliable period or phase function. Byte-by-byte Description of file: \texttt{ptf\_asteroid\_lc\_observations.txt}}
\begin{tabular}{lllll}
\hline
   Bytes & Format & Units & Label & Explanations\\
\hline
   1-  4 &I4     &    &           & Lightcurve ID number$^1$\\
   6- 15 &F10.5  &day &     $\tau$   &  Light-time-corrected observation epoch\\
  17- 26 &F10.7  &AU  &     $r$      & Heliocentric distance\\
  28- 37 &F10.7  &AU  &    $\Delta$  & Geocentric distance\\
  39- 43 &F5.2   &deg &    $\alpha$  & Solar phase angle\\
  45- 50 &F6.3   &mag &    $R$ or $g$    & Apparent magnitude$^2$\\
  52- 56 &F5.3   &mag &           & Uncertainty in apparent magnitude\\
  58- 62 &F5.3   &mag &       & Uncertainty in apparent magnitude with cosmic-error\\
  64- 69 &F6.3   &mag &      & Magnitude corrected for distance and $G_{12}$ phase function\\
  71- 76 &F6.3   &mag &      & Magnitude corrected for distance and rotation ($G_{12}$ fit)\\
  78- 83 &F6.3   &mag &      & Residual with respect to the $G_{12}$ fit\\
  85- 89 &F4.1   &    &      & Rotational phase from 0 to 1 ($G_{12}$ fit)\\
\hline
\end{tabular}
\smallskip
\\$^1$ID number also corresponds to the line number in Table 4.
\\$^2$Filter/band is specified in Table 4.
\bigskip
\end{table*}

\end{document}